\documentclass[twoside]{elsart_mm}

\usepackage{fancyhdr}
\headsep=\baselineskip
\usepackage{natbib}
\usepackage{amssymb}
\usepackage{amsmath}

\usepackage[expert,z-times,vw-times]{tmsfnts}

\usepackage{graphicx}
\usepackage[breaklinks,hypertex]{hyperref}  

\def\reference#1{\bibitem[jnk(1999)]{jnk}}
\def\1e{\mbox{1E\,0657--56}}
\def\araa{ARA\&A}
\def\jgr{J.\ Geophys.\ Res.}
\def\nat{Nature}
\def\physrep{Phys.\ Reports}
\def\prl{Phys.\ Rev.\ Lett.}
\def\apj        {ApJ}
\def\apjl       {ApJ}
\def\apjs{ApJ Suppl.}

\def\aap{A\&A}
\def\aaps{A\&A Suppl.}
\def\mnras      {MNRAS}
\def\pasj{PASJ}
\def\pasp{PASP}
\def\einstein   {\emph{Einstein}}
\def\asca       {\emph{ASCA}}
\def\rosat      {\emph{ROSAT}}
\def\chandra    {\emph{Chandra}}
\def\xmm        {\emph{XMM}}

\def\am         {$^\prime$}
\def\as         {$^{\prime\prime}$}
\def\msun       {$M_{\odot}$}

\def\cmcube     {cm$^{-3}$}
\def\kms        {km$\;$s$^{-1}$}
\def\sax        {\emph{SAX}}
\def\hseventy   {$H_0=70$~km$\;$s$^{-1}\,$Mpc$^{-1}$}
\def\kmsmpc     {km$\;$s$^{-1}\,$Mpc$^{-1}$}
\def\lesssim{\mathrel{\hbox{\rlap{\hbox{\lower4pt\hbox{$\sim$}}}\hbox{$<$}}}}
\def\gtrsim{\mathrel{\hbox{\rlap{\hbox{\lower4pt\hbox{$\sim$}}}\hbox{$>$}}}}
\def\lax{\lesssim}

\def\deg        {$^{\circ}$}
\def\ergs       {erg$\;$s$^{-1}$}

\submitted{Updated version of an article to appear in Physics Reports
 \hfill astro-ph/0701821\/}

\begin{document}
\begin{frontmatter}

\title{SHOCKS AND COLD FRONTS IN GALAXY CLUSTERS}

\author{Maxim Markevitch and Alexey Vikhlinin}

\address{Harvard-Smithsonian Center for Astrophysics, 60 Garden St.,
  Cambridge, MA 02138, USA}

\address{Space Research Institute, Profsoyuznaya 84/32, Moscow 117997,
  Russia}

\begin{abstract}

The currently operating X-ray imaging observatories provide us with an
exquisitely detailed view of the Megaparsec-scale plasma atmospheres in
nearby galaxy clusters.  At $z<0.05$, the \chandra's 1\as\ angular
resolution corresponds to linear resolution of less than a kiloparsec, which
is smaller than some interesting linear scales in the intracluster plasma.
This enables us to study the previously unseen hydrodynamic phenomena in
clusters: classic bow shocks driven by the infalling subclusters, and the
unanticipated ``cold fronts,'' or sharp contact discontinuities between
regions of gas with different entropies.  The ubiquitous cold fronts are
found in mergers as well as around the central density peaks in ``relaxed''
clusters.  They are caused by motion of cool, dense gas clouds in the
ambient higher-entropy gas. These clouds are either remnants of the
infalling subclusters, or the displaced gas from the cluster's own cool
cores.

Both shock fronts and cold fronts provide novel tools to study the
intracluster plasma on microscopic and cluster-wide scales, where the dark
matter gravity, thermal pressure, magnetic fields, and ultrarelativistic
particles are at play.  In particular, these discontinuities provide the
only way to measure the gas bulk velocities in the plane of the sky.  The
observed temperature jumps at cold fronts require that thermal conduction
across the fronts is strongly suppressed.  Furthermore, the width of the
density jump in the best-studied cold front is smaller than the Coulomb mean
free path for the plasma particles. These findings show that transport
processes in the intracluster plasma can easily be suppressed. Cold fronts
also appear less prone to hydrodynamic instabilities than expected, hinting
at the formation of a parallel magnetic field layer via magnetic draping.
This may make it difficult to mix different gas phases during a merger.  A
sharp electron temperature jump across the best-studied shock front has
shown that the electron-proton equilibration timescale is much shorter than
the collisional timescale; a faster mechanism has to be present. To our
knowledge, this test is the first of its kind for any astrophysical plasma.
We attempt a systematic review of these and other results obtained so far
(experimental and numerical), and mention some avenues for further studies.

\end{abstract}

\begin{keyword}
galaxies: clusters: general -- X-rays: galaxies: clusters -- hydrodynamics
\end{keyword}

\end{frontmatter}

\clearpage
\tableofcontents
\clearpage

\section{INTRODUCTION}

Clusters of galaxies are the most massive gravitationally bound objects in
the Universe. They include hundreds of galaxies within a radius of 1--2 Mpc
(e.g., Abell 1958).  Dispersions of the member galaxy redshifts indicate
that the gravitational potential of the cluster is much deeper than can be
created by the total mass of its galaxies, revealing the presence of
smoothly distributed dark matter (Zwicky 1937).  Its nature is still
unknown, except that it is probably cold and collisionless.  At present, its
distribution can be directly mapped using the gravitational lensing
distortion that it introduces to the images of distant background galaxies
(e.g., Bartelmann \& Schneider 2001).  In Fig.\ \ref{1e}, panels ({\em a})
and ({\em b}), we show an optical image of the field containing a relatively
distant cluster \1e, and a map of its total projected mass derived from
lensing.  Within the radius covered by the image, this cluster has a mass of
about $10^{15}$ \msun\ ($2\times 10^{48}$ g), of which only 1--3\% is
stellar mass in the member galaxies.

\begin{figure}[p]
\centering{
\noindent
\includegraphics[width=0.75\textwidth, bb=61 389 403 706]%
{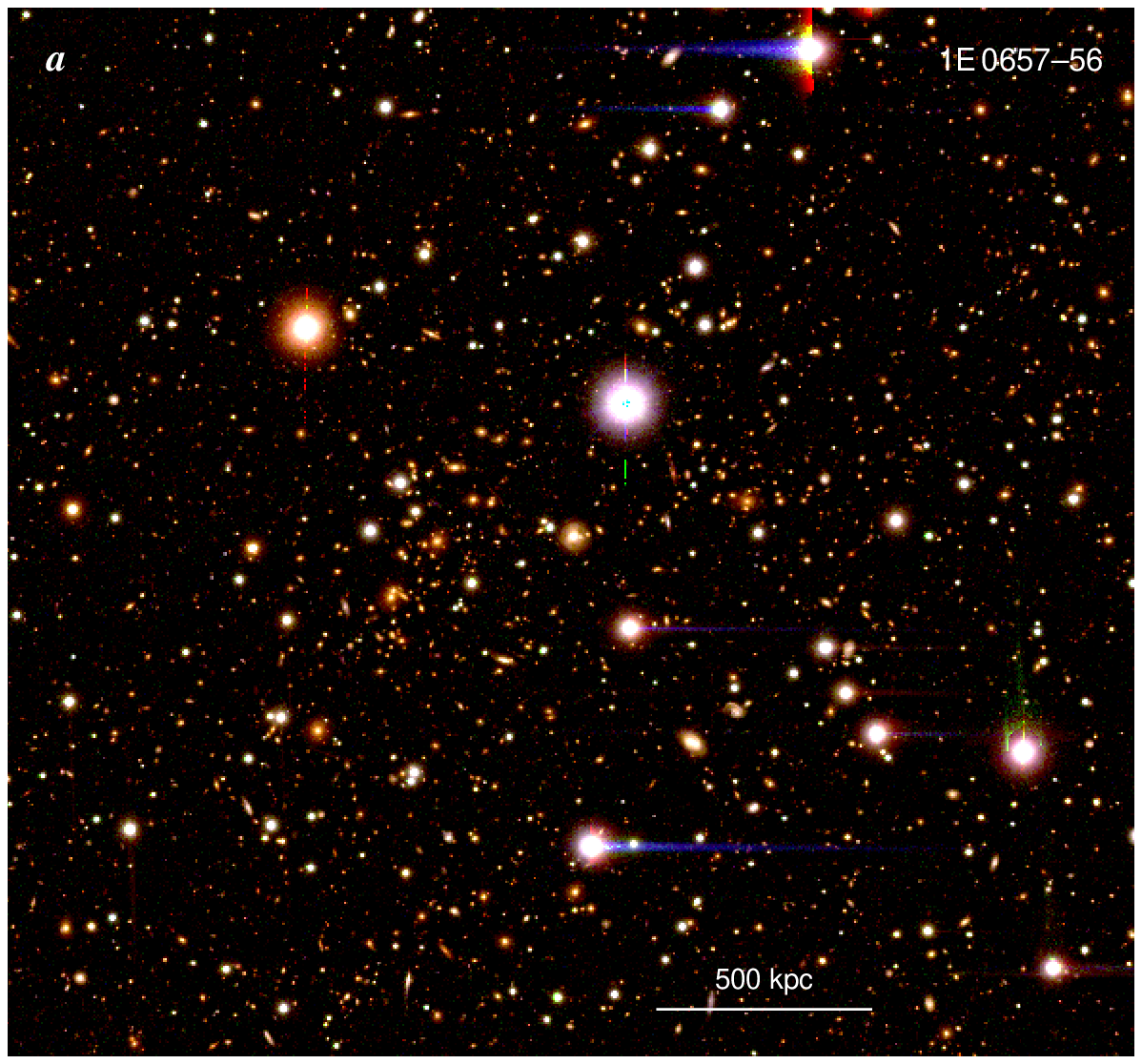}

\vspace{4mm}\hspace{0.1mm}
\includegraphics[width=0.75\textwidth, bb=61 389 403 706]%
{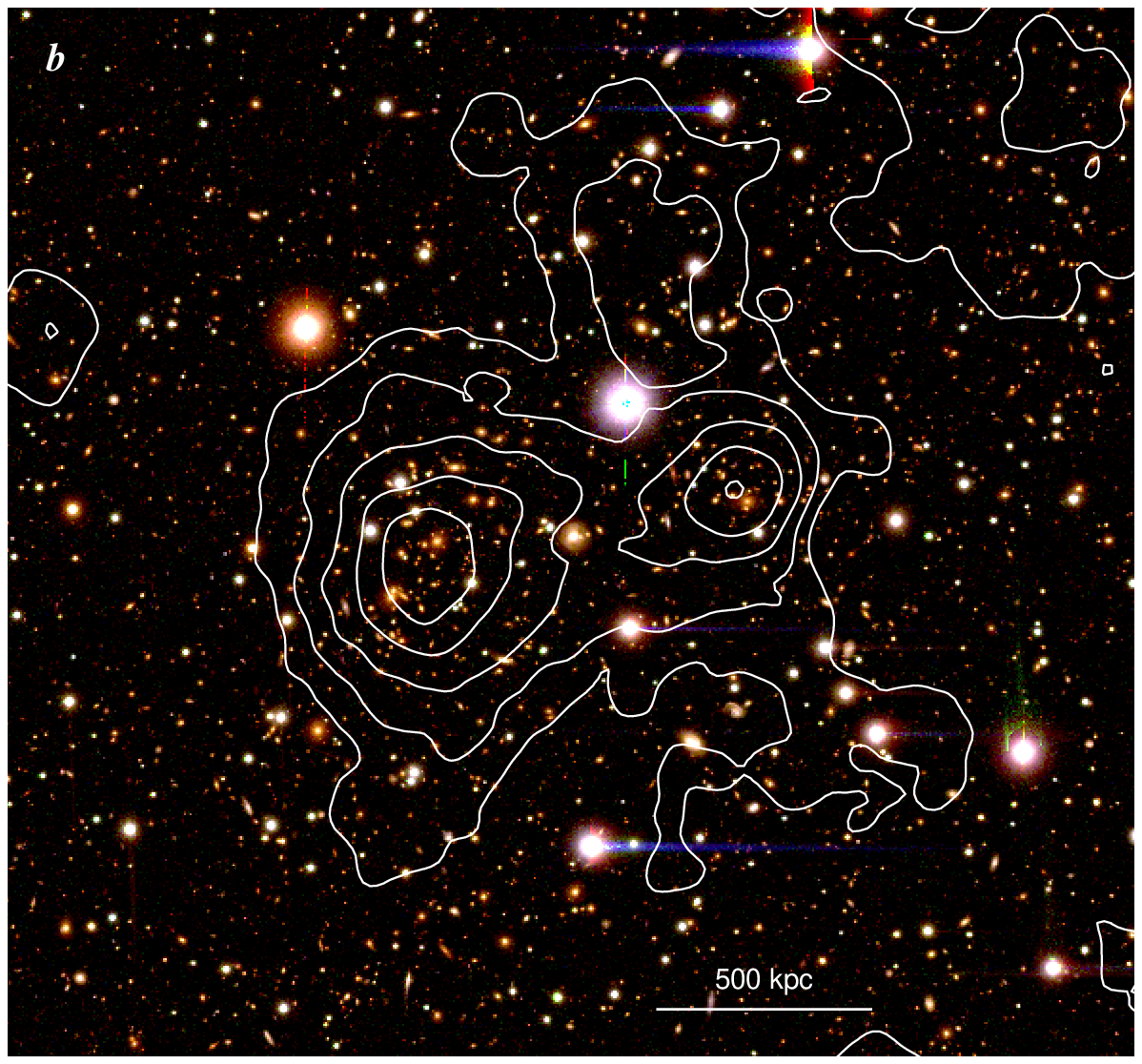}
}

\caption{({\em a}) Optical image of a field containing the \1e\ cluster
  ({\em Magellan}\/ 6.5m telescope; Clowe et al.\ 2006).  Image size is
  about 10\am\ and the ruler shows linear scale for the redshift of the
  cluster, $z=0.3$. The cluster consists of two concentrations of faint red
  galaxies in the middle.  ({\em b}) Contours show a map of projected total
  mass derived from gravitational lensing (Clowe et al.\ 2006; the lowest
  contour is noisy). The mass is dominated by dark matter.  Peaks of the two
  mass concentrations are coincident with the galaxy concentrations.}

\end{figure}
\addtocounter{figure}{-1}

\begin{figure}[p]
\centering{
\noindent
\includegraphics[width=0.75\textwidth, bb=61 389 403 706]%
{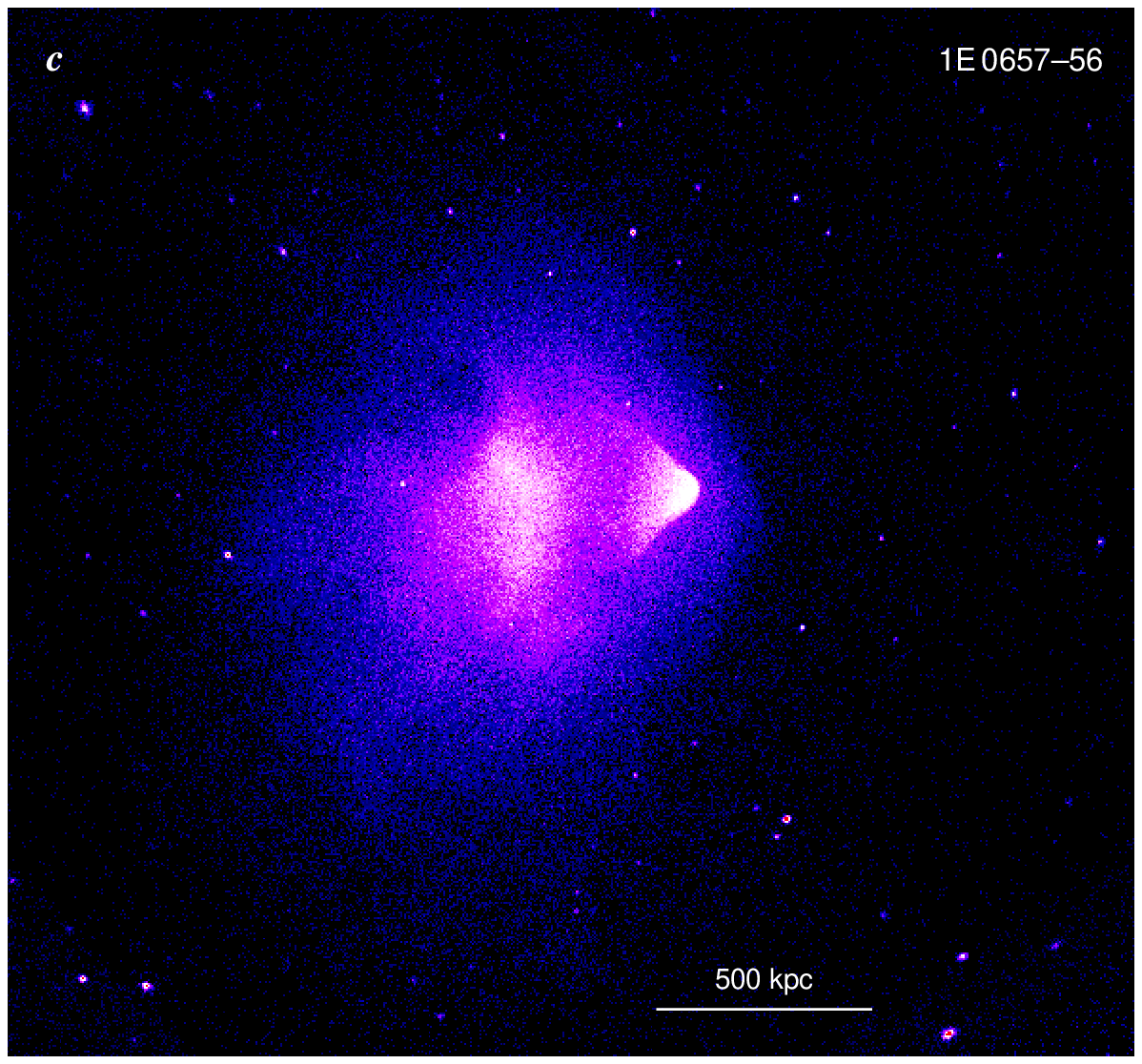}

\vspace{4mm}\hspace{0.1mm}
\includegraphics[width=0.75\textwidth]%
{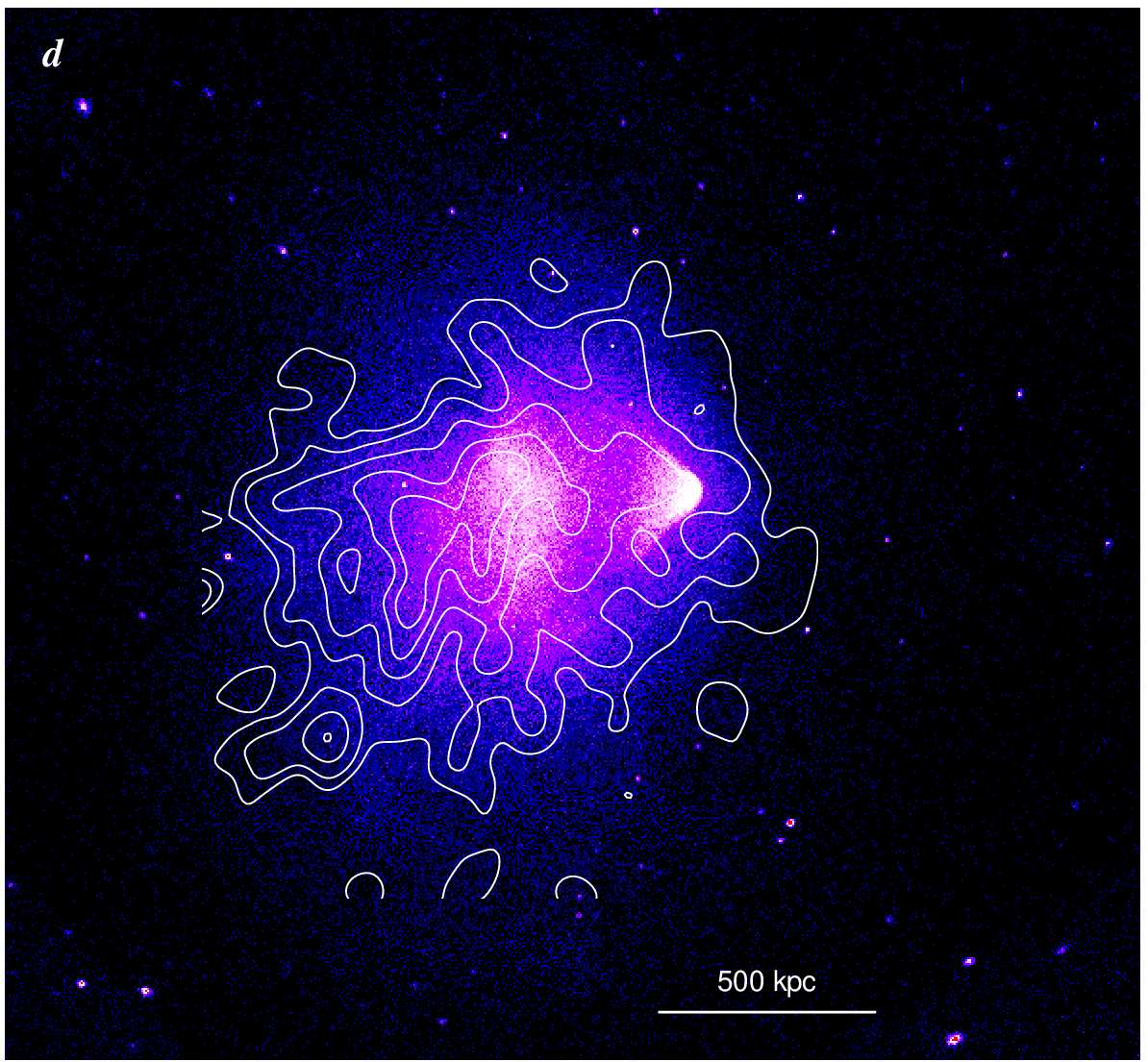}
}

\caption{---continued. ({\em c}) \chandra\ X-ray image of the same region
  of the sky containing \1e\ (M06).  Diffuse X-ray emission traces the hot
  gas.  Compact sources are mostly unrelated projected AGNs, left in the
  image to illustrate the 1\as\ angular resolution.  ({\em d}) Contours show
  the surface brightness of the diffuse radio emission (ATCA telescope, 1.3
  GHz; Liang et al.\ 2000).  The resolution is 24\as; compact sources are
  removed. The radio emission is synchrotron from the ultrarelativistic
  electrons in the cluster magnetic field, coexisting with the hot gas.}

\label{1e}
\end{figure}

\begin{figure}[t]
\centering
\noindent
\includegraphics[width=\textwidth,bb=58 587 480 726]%
{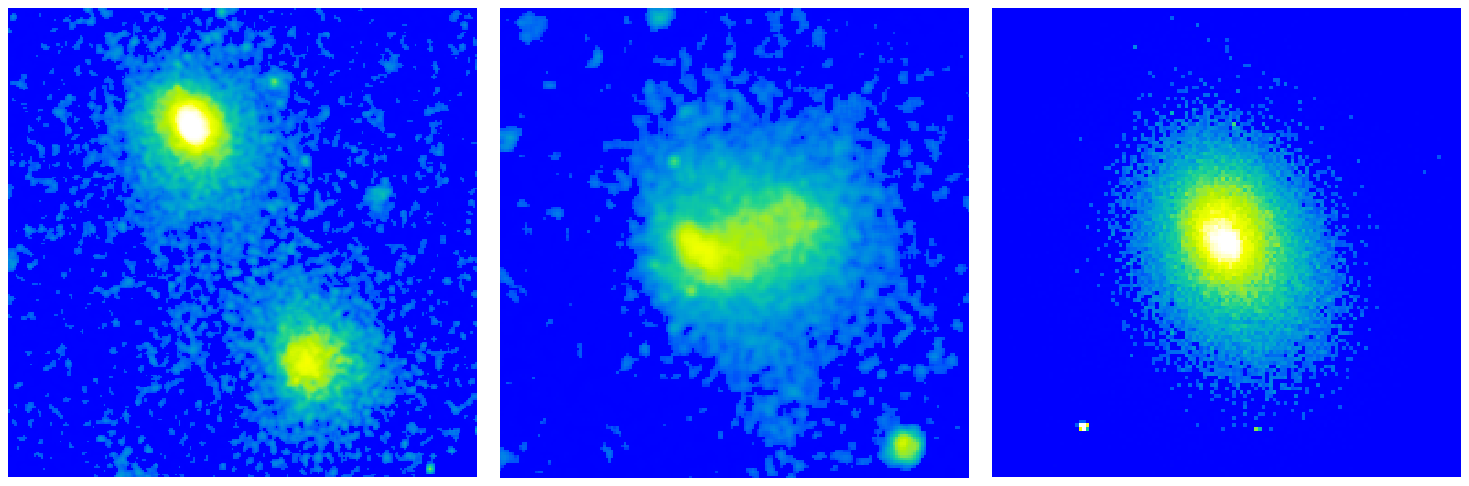}

\caption{Clusters form and grow via mergers. Panels from left to
  right show X-ray images of a pair of clusters about to merge (A399--A401),
  a system undergoing a merger (A754), and a relaxed, more massive cluster
  (A2029) that emerges in a few Gyr as a result.}

\label{merg3}
\end{figure}

X-ray observations showed that intergalactic space in clusters is filled
with hot plasma (Kellogg et al.\ 1972; Forman et al.\ 1972; Mitchell et al.\ 
1976; Serlemitsos et al.\ 1977). It is the second most massive cluster
component, and their dominant baryonic component. A theoretical and
observational review of this plasma and other cluster topics can be found in
Sarazin (1988). Here we summarize the basics which will be needed in the
sections below.

The intracluster medium (ICM) has temperatures $T_e\sim 10^7-10^8$ K ($1-10$
keV) and particle number densities steeply declining from $n\sim 10^{-2}$
\cmcube\ near the centers to $10^{-4}$ \cmcube\ in the outskirts.  It
consists of fully ionized hydrogen and helium plus traces of highly ionized
heavier elements at about a third of their solar abundances, increasing to
around solar at the centers.  It emits X-rays mostly via thermal
bremsstrahlung. At densities and temperatures typical for the ICM, the
ionization equilibrium timescale is very short. The electron-ion
equilibration timescale via Coulomb collisions is generally shorter than the
age of the cluster, so $T_e=T_i$ can be assumed in most cluster regions,
except perhaps in the low-density outskirts and at shock fronts (and
probably even there, as will be seen in \S\ref{sec:tei}).  The spectral
density of the X-ray continuum emission at energy $E$\/ from such a plasma
is
\begin{equation}
\epsilon_X \;\propto\; \overline{g}\; n^2\; T_e^{-1/2}\, e^{-E/kT_e}
\label{eq:brems}
\end{equation}
where $\overline{g}$ is the effective Gaunt factor, which includes all
continuum mechanisms and depends weakly on $E$, $T_e$\/ and ion abundances
(e.g., Gronenschild \& Mewe 1978; Rybicki \& Lightman 1979). On top of this
continuum, there is line emission, discussed, e.g., by Mewe \& Gronenschild
(1981). The timescale of radiative cooling of the ICM is generally very
long, longer than the cluster age, with the exception of small, dense
central regions. Thus, nonradiative approximation is applicable to all the
phenomena discussed in this review.

The ICM is optically thin for X-rays for all densities encountered in
clusters (except for the possible resonant scattering at energies of strong
emission lines in the dense central regions; Gilfanov, Sunyaev, \& Churazov
1987). An X-ray telescope can thus map the ICM density and electron
temperature in projection. The current X-ray imaging instruments are
sensitive mostly to X-rays with $E\approx 0.3-2$ keV. From eq.\ 
(\ref{eq:brems}), an X-ray image of a hot cluster at $E\lax kT_e$\/ is
essentially a map of the projected $n^2$.  In Fig.\ \ref{1e}{\em c}, we show
an X-ray image of \1e\ as an example.  This highly disturbed cluster has one
of the hottest and most X-ray luminous plasma halos (with
$\overline{T_e}\simeq 14$ keV and a bolometric luminosity of $10^{46}$
\ergs), and will feature in several sections below.

The electron temperature, $T_e$, averaged along the line of sight, can be
determined from the shape of the continuum component, and sometimes from the
relative intensities of emission lines, using an X-ray spectrum collected
from a spatial region of interest. The ion temperature, $T_i$, cannot be
directly measured at present.  In principle, it can be determined from
thermal broadening of the emission lines, but this requires an energy
resolution of a calorimeter.  Because of the strong dependence of
$\epsilon_X$ on $n$, it is often possible to ``deproject'' the ICM
temperature and density in three dimensions under reasonable assumptions
about the symmetry of the whole cluster or within a certain region of the
cluster.  In such a way, the mass of the hot plasma can be determined for
many clusters whose gas atmospheres are spherically symmetric. It is found
to comprise 5--15\% of the total mass, several times more than the stellar
mass in galaxies (e.g., Allen et al.\ 2002; Vikhlinin et al.\ 2006).

If a cluster is undisturbed by collisions with other clusters for a
sufficient time, its dark matter distribution should acquire a centrally
peaked, slightly ellipsoidal, symmetric shape.  After several sound crossing
times (of order $10^9$ yr), the ICM comes to hydrostatic equilibrium in the
cluster gravitational potential $\Phi$, so that the pressure of the ICM
$p$\/ and its mass density $\rho_{\rm gas}$ satisfy the equation $\nabla
p=-\rho_{\rm gas}\nabla \Phi$.  For a spherically symmetric cluster, and
assuming that the intracluster plasma can be described as ideal gas, it can
be written as
\begin{equation}
M(r)=-\frac{kT(r)\,r}{\mu m_p G}
          \left(\frac{d\ln\rho_{\rm gas}}{d\ln r} +
          \frac{d\ln T}{d\ln r}\right),
\label{eq:mass}
\end{equation}
where $M(r)$ is the total mass of the cluster enclosed within the radius
$r$, $T(r)$ is the gas temperature at that radius, and $\mu=0.6$ is the mean
atomic weight of the plasma particles. That is, by measuring radial
distributions of the gas temperature and density, one can derive the cluster
{\em total}\/ mass (Bahcall \& Sarazin 1977; Sarazin 1988).  This method of
measuring the cluster masses is independent of, and complementary to, those
using galaxy velocity dispersions and gravitational lensing.  Unlike the
lensing mass measurement, it works only for clusters in equilibrium;
however, it is less affected by the line-of-sight projections.  For those
clusters where a comparison is possible, different total mass measurement
methods usually agree to within a factor of 2.

Cluster masses are interesting because the ratio of the baryonic mass (ICM
plus stars) to dark matter mass for a cluster should be close to the average
for the Universe as a whole, which enables some powerful cosmological tests
(e.g., White et al.\ 1993; Allen et al.\ 2004). Furthermore, the number
density of clusters as a function of mass and its evolution with redshift
depend sensitively on cosmological parameters, which is the basis for
another class of tests (e.g., Sunyaev 1971; Press \& Schechter 1974; Eke,
Cole, \& Frenk 1996; Henry 1997; Vikhlinin et al.\ 2003).  Hot electrons in
the ICM also introduce a distortion in the spectrum of the Cosmic Microwave
Background (CMB), which at $\lambda>1$ mm turns clusters into negative radio
``sources'' (Sunyaev \& Zeldovich 1972).  By comparing the Sunyaev-Zeldovich
decrement and the X-ray brightness and temperature, one can derive absolute
distances to the clusters and, again, use them for a cosmological test (Silk
\& White 1978).  The best estimates of the cluster baryonic and total masses
currently come from the X-ray data. To rely on them for cosmological
studies, we have to understand in detail the physical processes in the ICM,
how well the quantities required for those tests can be determined from the
X-ray images and spectra, and how valid are the underlying assumptions about
the ICM.  This has been the main motivation for the studies discussed in
this review.

Clusters form via gravitational infall and mergers of smaller mass
concentrations, as illustrated by a time sequence in Fig.\ \ref{merg3}. Such
mergers are the most energetic events in the Universe since the Big Bang,
with the total kinetic energy of the colliding subclusters reaching
$10^{65}$ ergs (Markevitch, Sarazin, \& Vikhlinin 1999a). In the course of a
merger, a significant portion of this energy, that carried by the gas, is
dissipated (on a Gyr timescale) by shocks and turbulence.  Eventually, the
gas heats to a temperature that approximately corresponds to the depth of
the newly formed gravitational potential well.

A fraction of the merger energy may be channeled into the acceleration of
ultrarelativistic particles and amplification of magnetic fields.  These
nonthermal components manifest themselves most clearly in the radio band.
Polarized radio sources located inside and behind clusters are known to
exhibit Faraday rotation, which is caused by magnetic fields in the ICM.  In
the radial range $r\sim 0.1-1$ Mpc (outside the dense central regions often
affected by the central AGN), the field strengths are in the range $B\sim
0.1-3\;\mu$G (with different measurement methods giving somewhat diverging
values; for a review see, e.g., Carilli \& Taylor 2002).  For such fields
and the typical ICM temperatures, gyroradii for thermal electrons and
protons are of order $10^{8}$ cm and $10^{10}$ cm, respectively, many orders
of magnitude smaller than the particle collisional mean free paths
($10^{21}-10^{23}$ cm). The plasma electric conductivity is very high and
the magnetic field is frozen in.  For typical ICM densities and a $1\;\mu$G
field, the Alfv\'en velocity is $v_A \sim 50$ \kms, much lower than the
typical thermal sound speeds $c_s \sim 1000$ \kms. Thus, the plasma is
``hot'' in the sense that the ratio of thermal pressure to magnetic pressure
$\beta\equiv p/p_B \approx c_s^2/v_A^2 \gg 1$, except in special places
which will be discussed in \S\ref{sec:drap}. This means, among other things,
that the magnetic pressure contribution is negligible for the hydrostatic
mass determination using eq.\ (\ref{eq:mass}).

The magnetic field is tangled, with coherence scales of order 10 kpc
(Carilli \& Taylor 2002). This suppresses collisional thermal conduction on
scales greater than this linear scale, by a large factor that depends on
the exact structure of the field (Chandran \& Cowley 1998; Narayan \&
Medvedev 2001). As a result, plasma with temperature gradients (for example,
created by a merger) comes to pressure equilibrium much faster than those
gradients dissipate (e.g., Markevitch et al.\ 2003a). Indeed, we are yet to
find a cluster without spatial temperature variations.

Merging clusters often exhibit faint radio halos, such as that shown in
Fig.\ \ref{1e}{\em d} (for a review see, e.g., Feretti 2002; radio halos are
not to be confused with the more localized radio ``relics'' of different
origin).  The radio emission at $\nu\sim 1$ GHz is produced by synchrotron
radiation of ultrarelativistic electrons with Lorentz factor $\gamma\sim
10^4$ in a microgauss magnetic field. These relativistic electrons coexist
with thermal ICM, bound to it by the magnetic field.
Their exact origin is uncertain; one possibility is acceleration by merger
turbulence (for a review see, e.g., Brunetti 2003; we will touch on this in
\S\ref{sec:halos}). Relativistic electrons also produce X-ray emission by
inverse Compton (IC) scattering of the CMB photons. A detection of such
nonthermal emission at $E>10$ keV (where thermal bremsstrahlung falls off
exponentially with energy) was reported for some clusters (e.g.,
Fusco-Femiano et al.\ 2005 and references therein). The energy density in
the relativistic electrons should be of the order of the magnetic pressure
and thus negligible compared to thermal pressure of the ICM.  However, it
was suggested that the currently unobservable relativistic protons that may
accompany them can have a significant energy density (V\"olk et al.\ 1996
and later works).

A part of our review will deal with hydrodynamic phenomena near the cluster
centers, and a brief description of these rather special regions will be
helpful. One of the models for the radial dark matter density distribution,
widely used until recently, is the King (1966) profile, $\rho(r) \propto
(1+r^2/r_c^2)^{-3/2}$. It has a flat core in the center with typical sizes
$r_c \sim 200$ kpc (see, e.g., Sarazin 1988 for a motivation for this
model).  An isothermal gas in hydrostatic equilibrium within such a
potential also has a flat density core (Cavaliere \& Fusco-Femiano 1976).
This is an adequate description of the observed gas density profiles for
about 1/3 of the clusters.  However, most clusters exhibit sharp central gas
density peaks (e.g., Jones \& Forman 1984; Peres et al.\ 1998).
Coincidentally, Navarro, Frenk \& White (1997, hereafter NFW) found that
density profiles of equilibrium clusters in their cosmological Cold Dark
Matter simulations can be approximated by a functional form $\rho(r) \propto
(r/r_s)^{-1}(1+r/r_s)^{-2}$. Its $r^{-1}$ dark matter density cusp in the
center corresponds to a finite, but sharp density peak of the gas in
equilibrium.  The NFW model is a good description for the total mass
profiles derived from the X-ray data for such centrally peaked clusters
(e.g., Markevitch et al.\ 1999b; Nevalainen et al.\ 2001; Allen, Schmidt, \&
Fabian 2001; Pointecouteau, Arnaud, \& Pratt 2005; Vikhlinin et al.\ 2006).
These clusters usually have relatively undisturbed ICM (see, e.g., the last
panel in Fig.\ \ref{merg3}) and a giant elliptical galaxy in the center (a
cD galaxy) which marks the dark matter density peak.  Within $r\sim 100$ kpc
of this peak, the ICM temperature declines sharply toward the center (e.g.,
Fukazawa et al.\ 1994; Kaastra et al.\ 2004; Sanderson, Ponman, \&
O'Sullivan 2006), while the gas density increases, along with the relative
abundance of heavy elements in the gas. This creates a rather distinct
central region of low-entropy gas. Outside this region, the radial
temperature gradient reverses and $T_e$\/ declines outward, but the entropy
continues to increase, so on the whole, the clusters are convectively
stable.  The high central gas densities correspond to X-ray radiative
cooling times shorter than the cluster ages (a few Gyr).  This gave rise to
a ``cooling flow'' scenario, in which central regions of such peaked
clusters are thermally unstable. Recent data indicate that there has to be a
process that partially compensates for the radiative cooling; for a recent
review see, e.g., Peterson \& Fabian (2006).  We will use the term ``cooling
flow'' to signify this region that encloses the observed gas density and
temperature peaks (positive and negative, respectively), without any
particular physical model in mind.

Gas density and temperature distributions in clusters have been studied
extensively by all imaging X-ray observatories (e.g., by \einstein, Forman
\& Jones 1982; Jones \& Forman 1999; \rosat, Briel, Henry, \& Boehringer
1992; Henry \& Briel 1995; Peres et al.\ 1998; Vikhlinin, Forman, \& Jones
1997, 1999; \asca, Fukazawa et al.\ 1994; Honda et al.\ 1996; Markevitch et
al.  1996a, 1998; \sax, Nevalainen et al.\ 2001; De Grandi \& Molendi 2002;
\xmm, Arnaud et al.\ 2001; Briel, Finoguenov, \& Henry 2004; Piffaretti et
al.\ 2005).  Temperature maps proved to be more difficult to obtain than
maps of the gas density, but the currently operating \xmm\ and \chandra\ 
observatories have the right combination of spectroscopic and imaging
capabilities to derive them with a good linear resolution for a large number
of clusters at a range of redshifts.  At $z<0.05$, \chandra's 1\as\ angular
resolution, the best among the X-ray observatories, corresponds to linear
scales $<1$ kpc.  This is less than the typical collisional mean free path
in the ICM or a typical galaxy size, and provides an exquisitely detailed
view of the physical processes in the cluster Megaparsec-sized gas halos.
With \chandra, we are able to see the classic bow shocks driven by infalling
subclusters, as well as ``cold fronts'' --- unexpected sharp features of a
different nature.  While \xmm\ can measure temperatures with a higher
statistical accuracy, its angular resolution is not sufficient to see these
sharp features in full detail, so our review of these two phenomena will be
based almost exclusively on the \chandra\ results.

\chandra's main detector, ACIS, is sensitive in the 0.5--8 keV energy band,
with a peak sensitivity between 1--2 keV, and has a FWHM energy resolution
of $50-150$ eV for extended sources, sufficient to disentangle emission
lines in the uncrowded cluster X-ray spectra, but far from that needed to
resolve the line Doppler widths.  A \chandra\ overview can be found, e.g.,
in Weisskopf et al.\ (2002).  Uncertainties of the cluster gas temperatures
from typical \chandra\ exposures are limited by the photon counting
statistics, while uncertainties of the gas densities are usually limited by
the assumed three-dimensional geometry of a cluster. Physical quantities
below are given for a spatially flat cosmology with $\Omega_0=0.3$,
$\Omega_\Lambda=0.7$, and \hseventy\ (unless the dependence on $H_0$ is
given explicitly via a factor $h\equiv H_0/100$ \kmsmpc).

\section{COLD FRONTS}

Those interested primarily in physics that can be learned from this recently
found phenomenon may read \S\ref{sec:merg}, which gives a general
description of cold fronts, then skip the rest of this chapter (which
discusses the various kinds of cold fronts and their origin and evolution
based on hydrodynamic simulations), and go directly to \S\ref{sec:tools}.

\begin{figure}[p]
\centering{
\noindent
\includegraphics[width=0.7\textwidth,bb=1 14 490 512,clip]%
{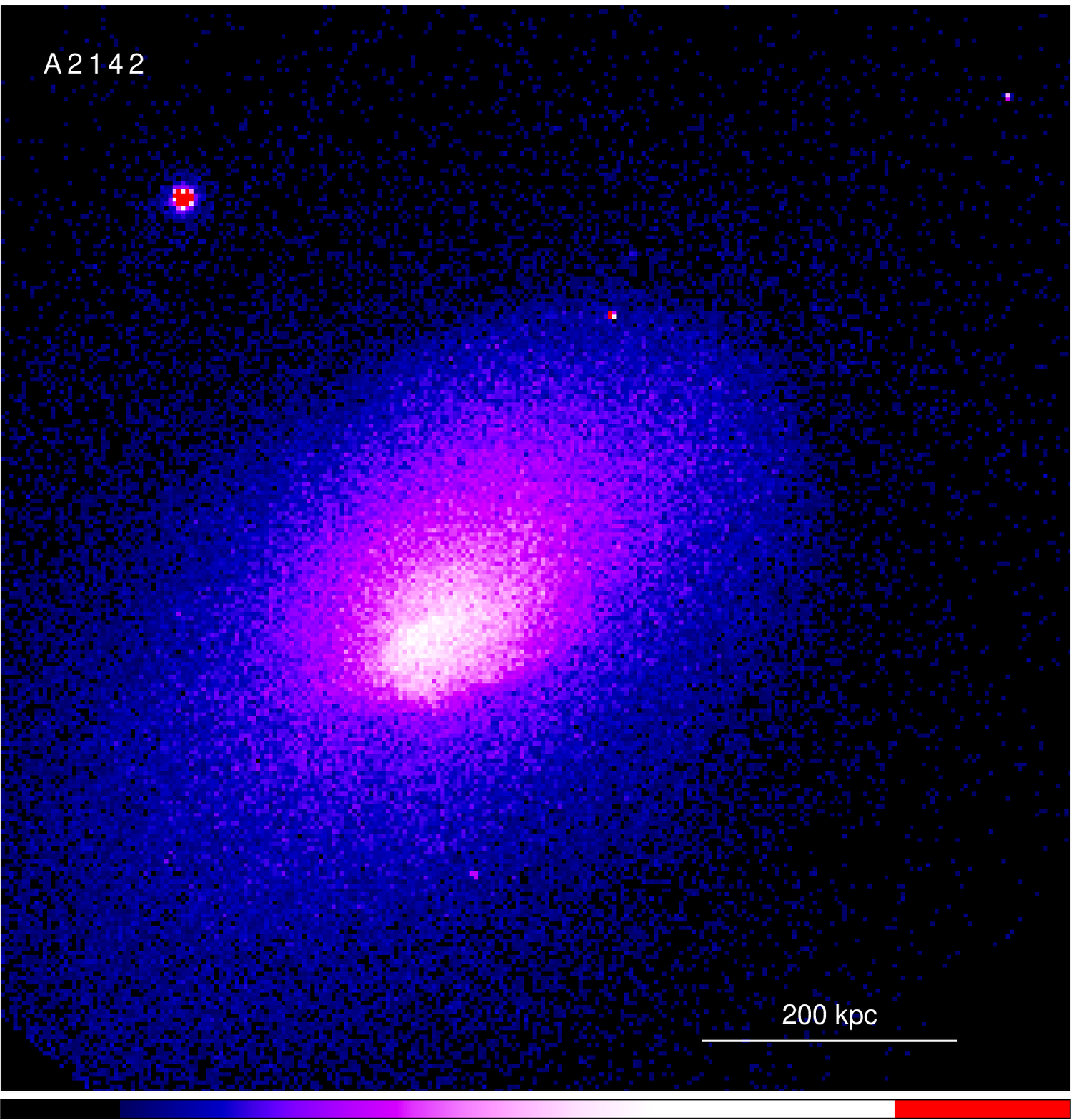}

\vspace{4mm}\hspace{0.1mm}
\includegraphics[width=0.7\textwidth,bb=1 14 490 512,clip]%
{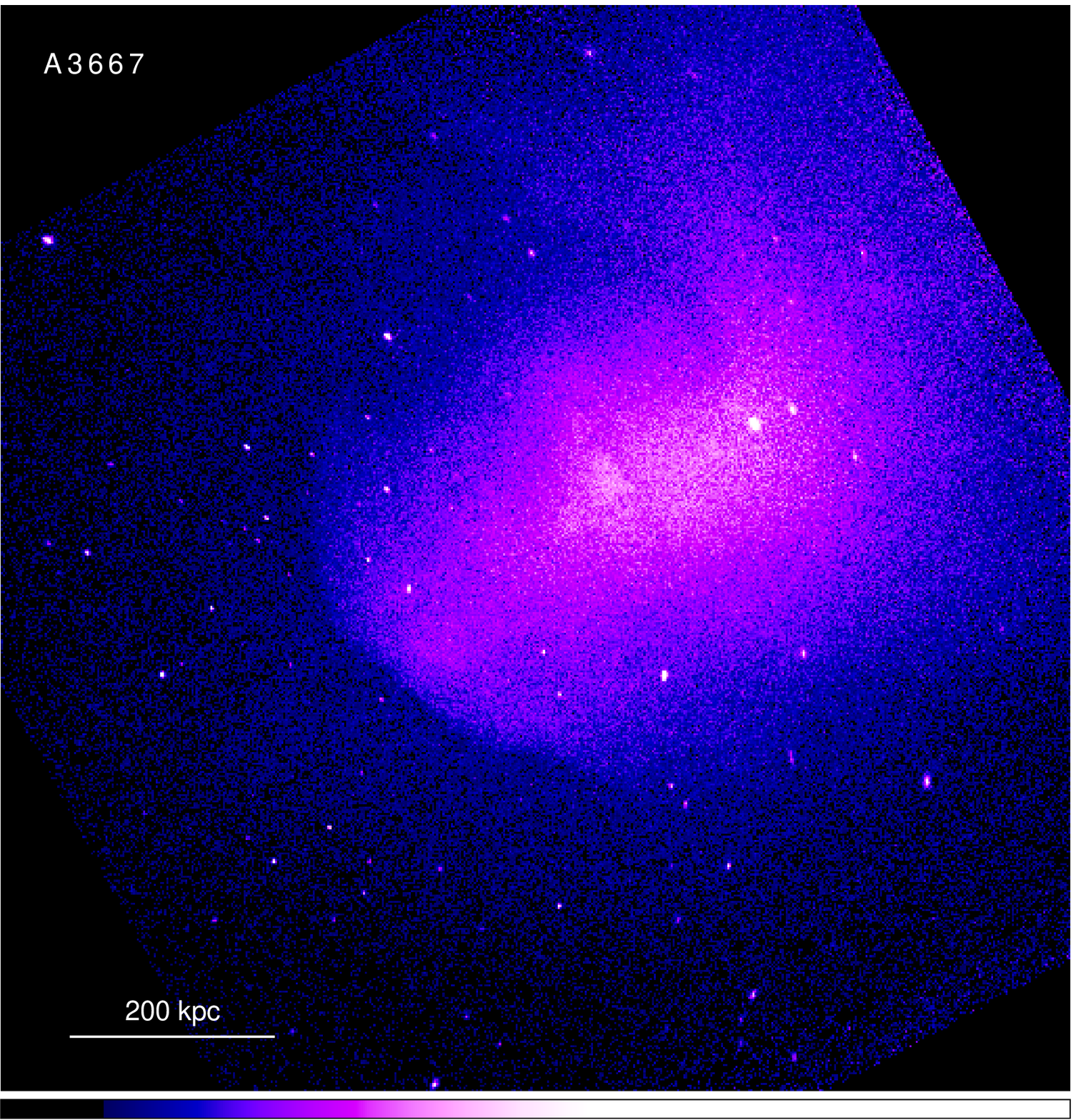}
}

\caption{\chandra\ X-ray images of clusters with the first discovered cold
  fronts, A2142 and A3667. In A2142, at least two sharp brightness edges are
  seen, between blue and black in the NW and between purple and blue south
  of center. In A3667, there is a prominent edge SE of center. Unrelated
  compact X-ray sources are not removed. (Images are created from the recent
  long \chandra\ exposures.)}

\label{2142_3667}
\end{figure}

\begin{figure}[t]
\includegraphics[width=\textwidth]{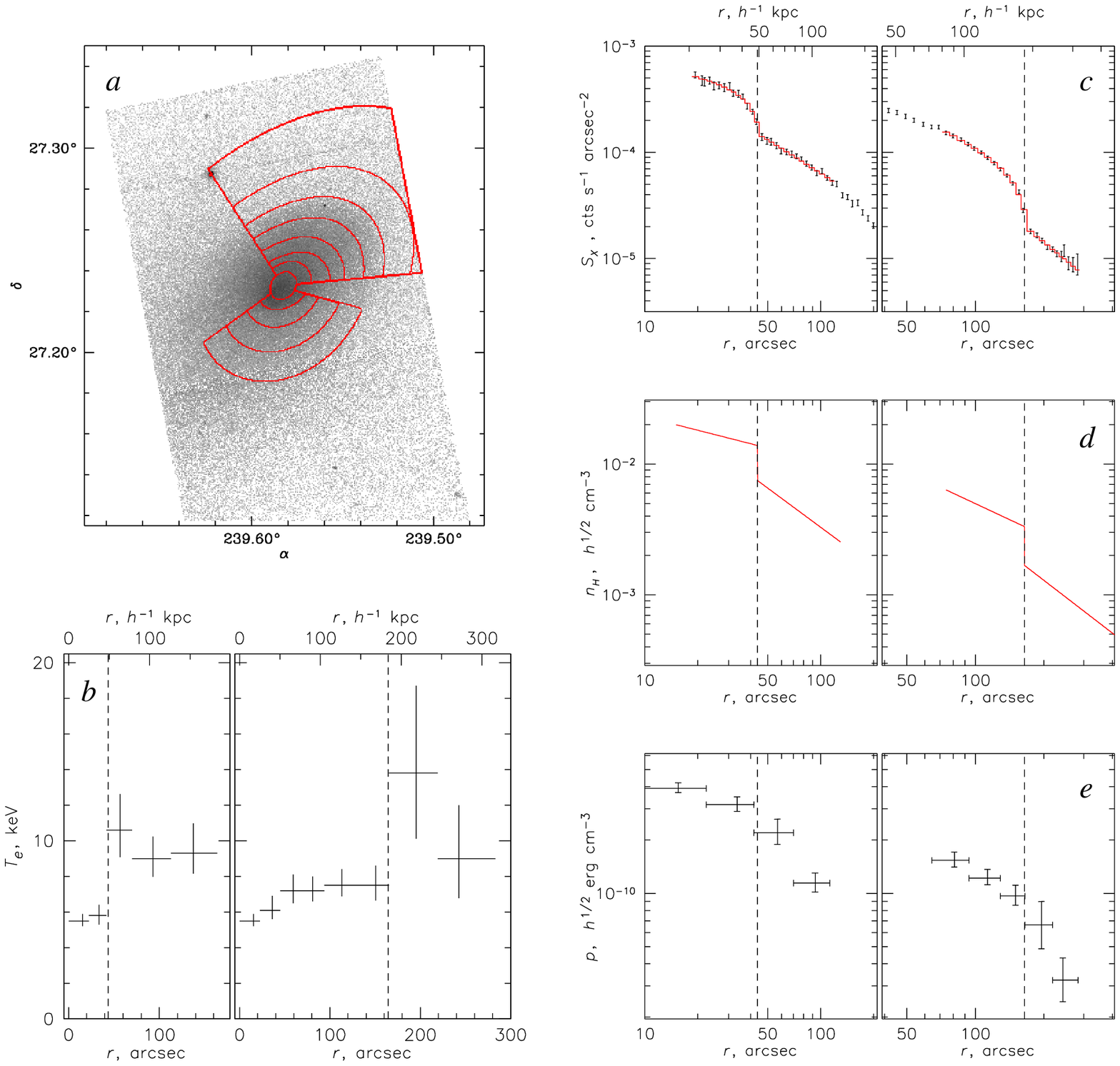}

\caption{Cold fronts in A2142 (reproduced from M00). ({\em a}) X-ray image
  with red overlays showing regions used for derivation of temperature
  profiles (panel {\em b}).  In panels ({\em b-e}), the southern edge is
  shown in the left plot and the northwestern edge is in the right plot.
  Panel ({\em c}) shows X-ray brightness profiles across the edges in the
  same sectors.  The red histogram is the brightness model that corresponds
  to the best-fit gas density model shown in panel ({\em d}). Panel ({\em
    e}) shows pressure profiles obtained from the temperature and density
  profiles.  Error bars are 90\%; vertical dashed lines show the positions
  of the density jumps.}

\label{a2142_profs}
\end{figure}

\subsection{Cold fronts in mergers}
\label{sec:merg}

Among the first \chandra\ cluster results was a discovery of ``cold fronts''
in merging clusters A2142 and A3667 (Markevitch et al.\ 2000, hereafter M00;
Vikhlinin, Markevitch, \& Murray 2001b, hereafter V01).  Figure
\ref{2142_3667} shows ACIS images of the central regions of A2142 and A3667,
which show prominent sharp X-ray brightness edges. The edge in A3667 was
previously seen in a lower-resolution \rosat\ image (Markevitch et al.\ 
1999a), and at the time, we interpreted it as a shock front, even though the
crude \asca\ temperature map did not entirely support this explanation.

If these features were shocks, the gas on the denser, downstream side of the
density jump would have to be hotter than that on the upstream side.  With
\rosat\ and \asca, we could not derive sufficiently accurate gas temperature
profiles across such edges. \chandra\ provided this capability for the first
time, so now we can easily test this hypothesis.  The \chandra\ radial X-ray
brightness and temperature profiles across the two edges in A2142 are shown
in Fig.\ \ref{a2142_profs} (from M00). They were extracted in sectors shown
in panel ({\em a}). Both brightness profiles have a characteristic shape
corresponding to a projection of an abrupt, spherical (within a certain
sector) jump of the gas density.  Best-fit radial density models of such a
shape are shown in panel ({\em d}), and their projections are overlaid on
the data as histograms in panel ({\em c}) --- they provide a very good fit.
Since there is no way of knowing the exact three-dimensional geometry of the
edge, for such fits we have to assume that the curvature of the
discontinuity surface along the line of sight is the same as in the sky
plane. To ensure the consistency with this assumption, it is important that
the radial profiles and the three-dimensional model for the gas inside the
discontinuity are centered at the center of curvature of the front, which is
often offset from the cluster center. At the same time, the model of the
outer, ``undisturbed'' gas may need to be centered elsewhere (e.g., the
cluster centroid).

Panel ({\em b}) in Fig.\ \ref{a2142_profs} shows the gas temperature
profiles across the edges. For a {\em shock discontinuity}, the
Rankine--Hugoniot jump conditions directly relate the gas density jump,
$r\equiv \rho_1/\rho_0$, and the temperature jump, $t\equiv T_1/T_0$, where
indices 0 and 1 denote quantities before and after the shock (e.g., Landau
\& Lifshitz 1959, \S89):
\begin{equation}
t=\frac{\zeta-r^{-1}}{\zeta-r}
\label{eq:t}
\end{equation}
or, conversely,
%
\begin{equation}
r^{-1}=\left[\frac{1}{4}\, \zeta^2\, (t-1)^2 +t\right]^{1/2} 
            -\frac{1}{2}\, \zeta\, (t-1),
\label{eq:r}
\end{equation}
where we denoted $\zeta \equiv (\gamma+1)/(\gamma-1)$; here $\gamma=5/3$ is
the adiabatic index for monoatomic gas.

For the observed density jump $r\sim 2$ and a presumably post-shock
temperature $T_1\sim 7.5$ keV observed inside the NW edge in A2142, one
would expect to find a $T_0\simeq 4$ keV gas in front of the shock.  This
sign of the temperature change is opposite to that observed across the edge
--- the temperature in the less dense gas outside the edge is in fact higher
than that inside (Fig.\ \ref{a2142_profs}{\em b}). The same is true for the
smaller edge in A2142, as well as the one in A3667 (V01; see also Briel,
Finoguenov, \& Henry 2004), ruling out the shock interpretation.

What are these sharp edges then? One hint is given by the gas pressure
profiles across the edges (simply the product of the best-fit density models
and the measured temperatures; Fig.\ \ref{a2142_profs}{\em e}), which show
that there is approximate pressure equilibrium across the density
discontinuity (as opposed to a large pressure jump expected in a shock
front). One also notes a smooth, comet-like shape of the NW edge in A2142,
which looks as if the ambient gas flows around it. Given this evidence, we
proposed (M00) that these features are contact discontinuities at the
boundaries of the gas clouds moving sub- or transonically through a hotter
and less dense surrounding gas --- or ``cold fronts'', as V01 have termed a
similar feature in A3667.%
\footnote{We were certainly influenced by the Burns (1998) review, which
  compared the gasdynamic phenomena in cluster mergers with ``stormy
  weather''. The term ``cold front'' has since been commonly adopted to
  denote either the discontinuity itself, or the discontinuity and the gas
  cloud behind it; which of these two is usually clear from the context.}
Strictly speaking, a contact discontinuity implies continuous pressure and
velocity between the gas phases, but a cold front often has a discontinuous
tangential velocity, when the dense gas cloud is moving.

\begin{figure}[t]
\centering
\includegraphics[width=0.5\textwidth, bb=95 145 502 718, clip]%
{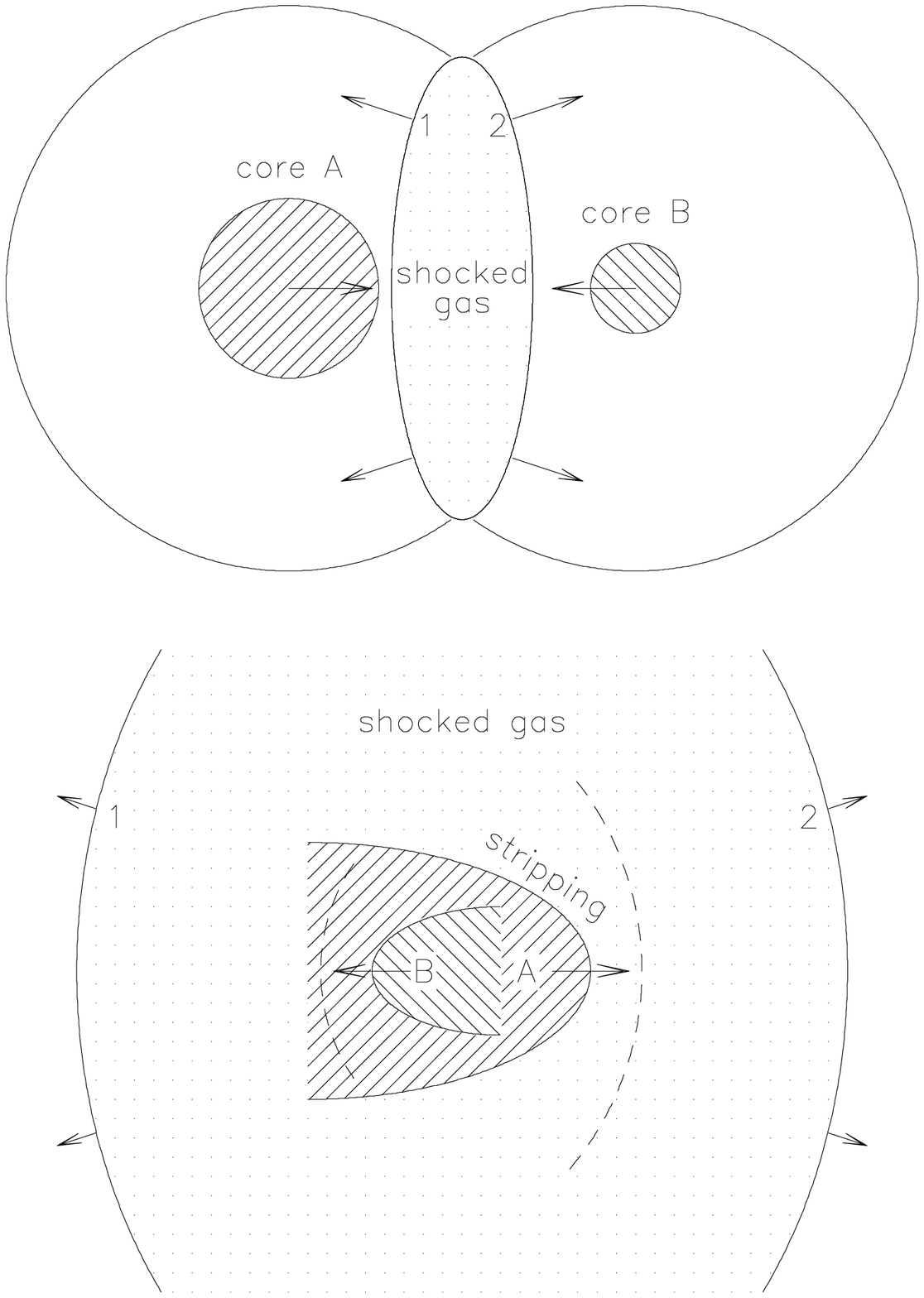}

\caption{A model for the origin of cold fronts in A2142 proposed in M00 is
  shown schematically in lower panel. The preceding stage of the merger is
  shown in upper panel. In upper panel, shaded circles depict dense cores of
  the two colliding subclusters (of course, in reality, there is a
  continuous density gradient). Shock fronts 1 and 2 in the central region
  of top panel have propagated to the cluster outskirts in lower panel,
  failing to penetrate the dense cores that continue to move through the
  shocked gas. The cores may develop additional shock fronts ahead of them,
  shown by dashed lines. See Fig.\ \ref{mathis} for a simulation
  illustrating these stages. (Reproduced from M00.)}

\label{a2142_scheme}
\end{figure} 

In the particular scenario that was envisioned for A2142 in M00, these dense
gas clouds are remnants of the cool cores of the two merging subclusters
that have survived shocks and mixing of a merger (which would have to have a
nonzero impact parameter to avoid complete destruction of the less dense NW
core). They are observed after the passage of the point of minimum
separation and presently moving apart. The hotter, rarefied gas beyond the
NW edge can be the result of shock heating of the outer atmospheres of the
two colliding subclusters, as schematically shown in Fig.\ 
\ref{a2142_scheme}. In this scenario, the less dense outer subcluster gas
has been stopped by the collision shock, while the dense cores (or, more
precisely, regions of the subclusters where the pressure exceeded that of
the shocked gas in front of them, which prevented the shock from penetrating
them) continued to move ahead through the shocked gas, pulled along by their
host dark matter clumps.

\begin{figure}[t]
\centering
\includegraphics[height=\textwidth, angle=90, bb=0 0 473 512, clip]%
{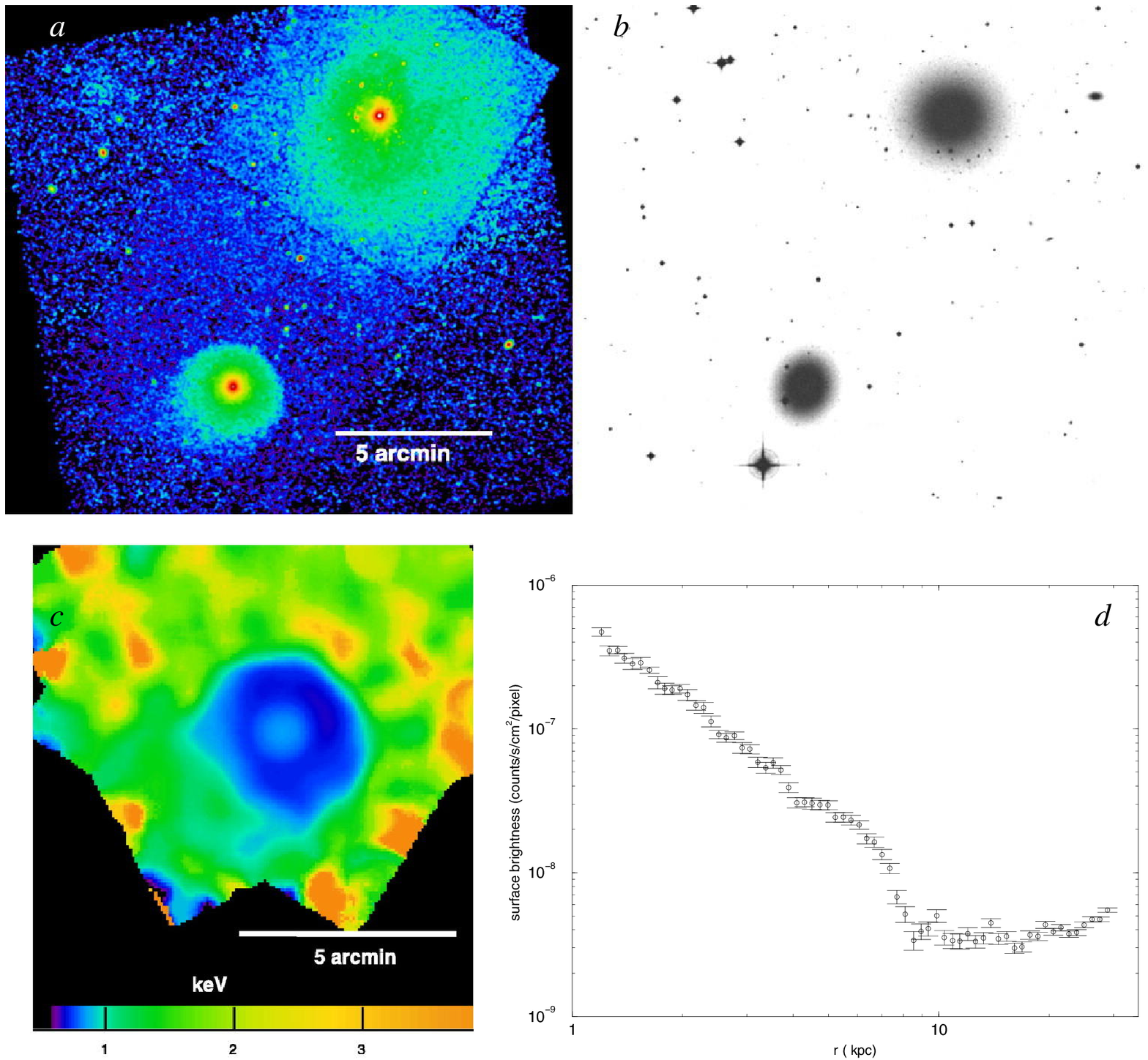}

\caption{The NGC\,1404 elliptical galaxy falling into the Fornax
  cluster. ({\em a}) \chandra\ X-ray image, showing a sharp brightness edge
  on the NW side of NGC\,1404 (the southern of the two halos). The 5\am\ bar
  corresponds to 29 kpc.  ({\em b}) Optical image (same scale as {\em a})
  showing NGC\,1404 and the central galaxy of the cluster.  ({\em c})
  \chandra\ temperature map of the NGC\,1404 region. The gas in the galaxy
  is cool.  ({\em d}) X-ray radial brightness profile in a sector across the
  edge, showing the characteristic projected spherical discontinuity shape.
  (Reproduced from Machacek et al.\ 2005.)}

\label{n1404}
\end{figure}

With the benefit of a more recent, longer \chandra\ observation of A2142,
and having seen images of numerous other clusters as well as hydrodynamic
simulations, we now think that the M00 scenario is not correct. Instead, it
seems more likely that A2142 is a cluster with a sloshing cool core (as
first pointed out by Tittley \& Henriksen 2005), a phenomenon that was
discovered later and which will be discussed in \S\ref{sec:slosh}.  However,
the physical interpretation of the X-ray edges as contact discontinuities
between moving gases of different entropies still holds --- the only
difference is the origin of the two gas phases in contact (either from
different subclusters or from different radii in the same cluster). At the
same time, the scenario proposed in M00 is realized in a number of other
merging clusters. Two particularly striking examples are the textbook merger
\1e\ and an elliptical galaxy NGC\,1401 in the Fornax cluster, although each
of these systems exhibits only one cold front.  In both objects, there is an
independent (i.e., non X-ray) evidence of a distinct infalling subcluster
that hosts the gas cloud with a cold front. In Fornax, a cold front forms at
the interface between the atmosphere of the infalling galaxy NGC\,1404 seen
in the optical image, and the hotter cluster gas (Fig.\ \ref{n1404};
Machacek et al.\ 2005). In \1e, a mass map derived from the gravitational
lensing data reveals a dark matter subcluster, which is also seen as a
concentration of galaxies in the optical image (Fig.\ \ref{1e}{\em ab};
Clowe, Gonzalez, \& Markevitch 2004; Clowe et al.\ 2006).  Its X-ray image
(Fig.\ \ref{1e}{\em c}) shows a bright ``bullet'' of gas moving westward,
apparently pulled along by the smaller dark matter subcluster.  Because of
ram pressure, the bullet lags behind the collisionless dark matter clump
(Fig.\ \ref{1e_lens}, which will be discussed in \S\ref{sec:dm}).

\begin{figure}[t]
\centering
\includegraphics[width=0.67\textwidth, bb=101 265 416 632]%
{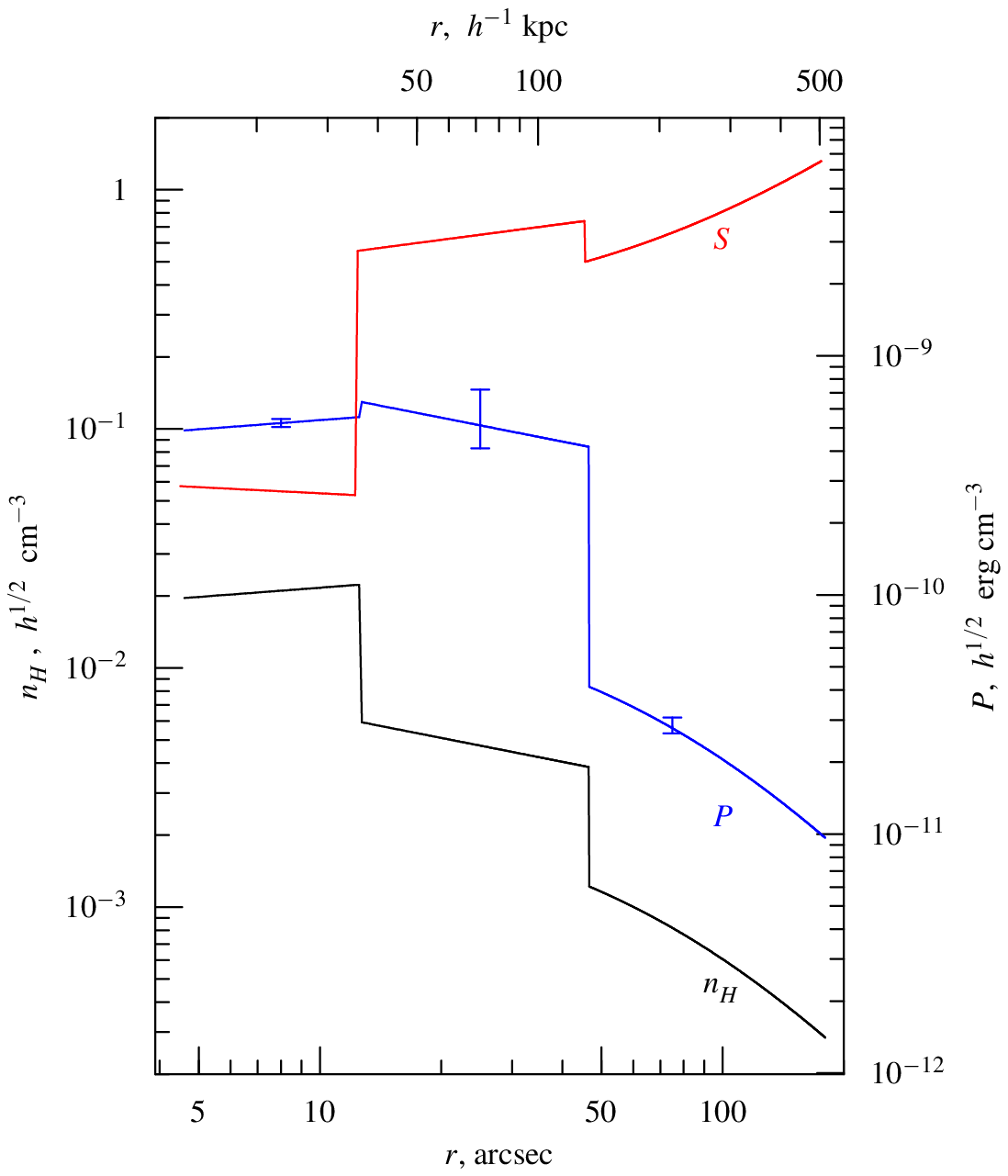}

\caption{Model radial profiles for the gas density $n$, pressure $P$\/ and
  specific ``entropy'' $S\equiv T n^{-2/3}$ for \1e\ in a sector crossing
  the bullet boundary (a cold front at $r\approx 13''$) and the shock front
  (at $r\approx 50''$), centered on the bullet.  The pressure and
  ``entropy'' profiles are simply the combinations of the best-fit density
  model and the average temperature values for each of the three regions.
  Error bars on the pressure plot correspond to temperature uncertainties.
  We omit the temperature variations inside the post-shock and bullet
  regions, so the plot can serve only as a qualitative illustration of the
  changes at shocks and cold fronts.  (Reproduced from M02, with updated
  temperatures and added entropy profile.)}

\label{1e_densprof}
\end{figure} 

The gas bullet in \1e\ has developed a sharp edge at its western side, which
is a cold front. Ahead of it is a genuine bow shock (a faint blue-black edge
in Fig.\ \ref{1e}{\em c}), confirmed by the temperature profile (Markevitch
et al.\ 2002, hereafter M02; \S\ref{sec:1e_M} below). It is instructive to
look at the radial profiles of the gas density, thermal pressure and
specific entropy derived in a narrow sector crossing both these edge
features (Fig.\ \ref{1e_densprof}).  The two discontinuities have density
jumps of similar amplitudes (a factor of 3).  As expected, the pressure has
a big jump at the shock, but is nearly continuous in comparison at the cold
front.  In broad terms, thermal pressure in the cool gas behind a moving
cold front should be in balance with the thermal plus ram pressure of the
gas flowing around it.  For a substantially supersonic motion (this shock
has $M=3$, see \S\ref{sec:1e_M}), the gas flow between the bow shock and the
driving body is very subsonic. So the ram pressure component is small
compared to thermal pressure, hence the near-continuity of thermal pressure
(a more detailed picture will be presented in \S\ref{sec:vel}).  The
entropy, on the other hand, shows only a small increase at the shock (as
expected for this relatively weak shock), but a big drop at the cold front.
This is because in the past, the bullet apparently used to be a cooling flow
(M02).  The merger brought this region of low-entropy
gas in direct contact with the high-entropy gas from the cluster outskirts.
These are the characteristic features of all cold fronts, regardless of the
exact origin of the two gas phases in contact.

In addition to the examples mentioned above, prominent if less clear-cut
cold fronts have been observed in a large number of other clusters (e.g.,
RXJ\,1720+26, Mazzotta et al.\ 2001; A2256, Sun et al.\ 2002; A85, Kempner,
Sarazin, \& Ricker 2002; A2034, Kempner, Sarazin, \& Markevitch 2003; A496,
Dupke \& White 2003; A754, Markevitch et al.\ 2003a; A2319, O'Hara, Mohr, \&
Guerrero 2004, Govoni et al.\ 2004, hereafter G04; A168, Hallman \&
Markevitch 2004; A2204, Sanders, Fabian, \& Taylor 2005).  Many of such
features have been observed at smaller linear scales in the cool dense gas
near the cluster centers, which we will separate (somewhat artificially)
into a class of their own (\S\ref{sec:slosh}).

\begin{figure}[t]
\centering
\includegraphics[width=0.95\textwidth,clip]%
{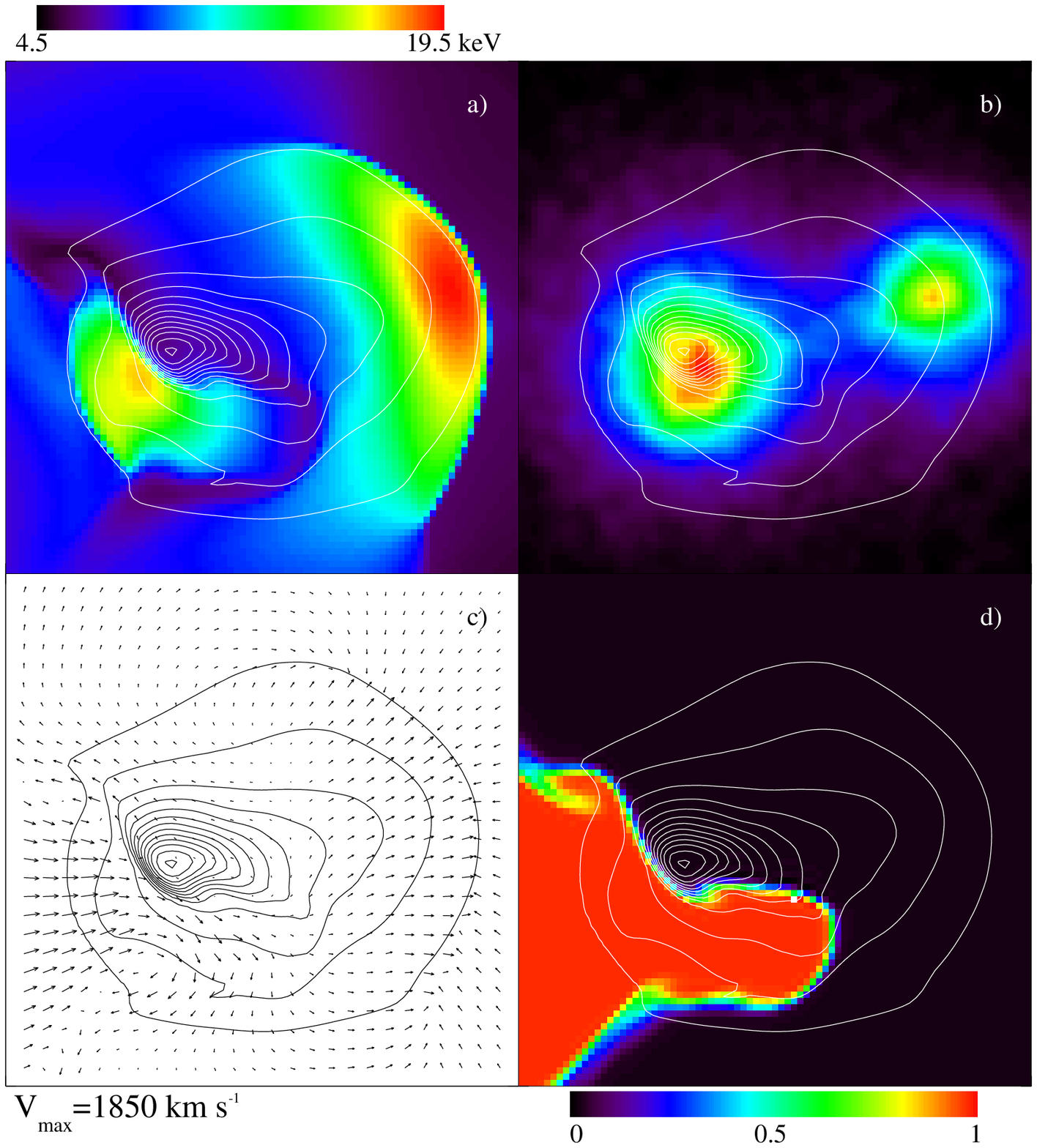}

\caption{A snapshot from a simulated off-axis merger of two
  subclusters. Colors show ({\em a}) projected gas temperature (increasing
  from black to red), ({\em b}) projected dark matter density, ({\em d})
  fraction of gas that initially belonged to each subcluster (red belonged
  to the smaller subcluster that is now on the right side and moving away
  from the collision site; black belonged to the bigger subcluster).
  Contours in all panels show X-ray brightness. Arrows in ({\em c}) show gas
  velocities.  The interface between the two gases near the brightness peak
  is a cold front. (Reproduced from Roettiger et al.\ 1998.)}

\label{roettiger}
\end{figure} 

\begin{figure}[p]
\centering

~~\includegraphics[width=0.95\textwidth]%
{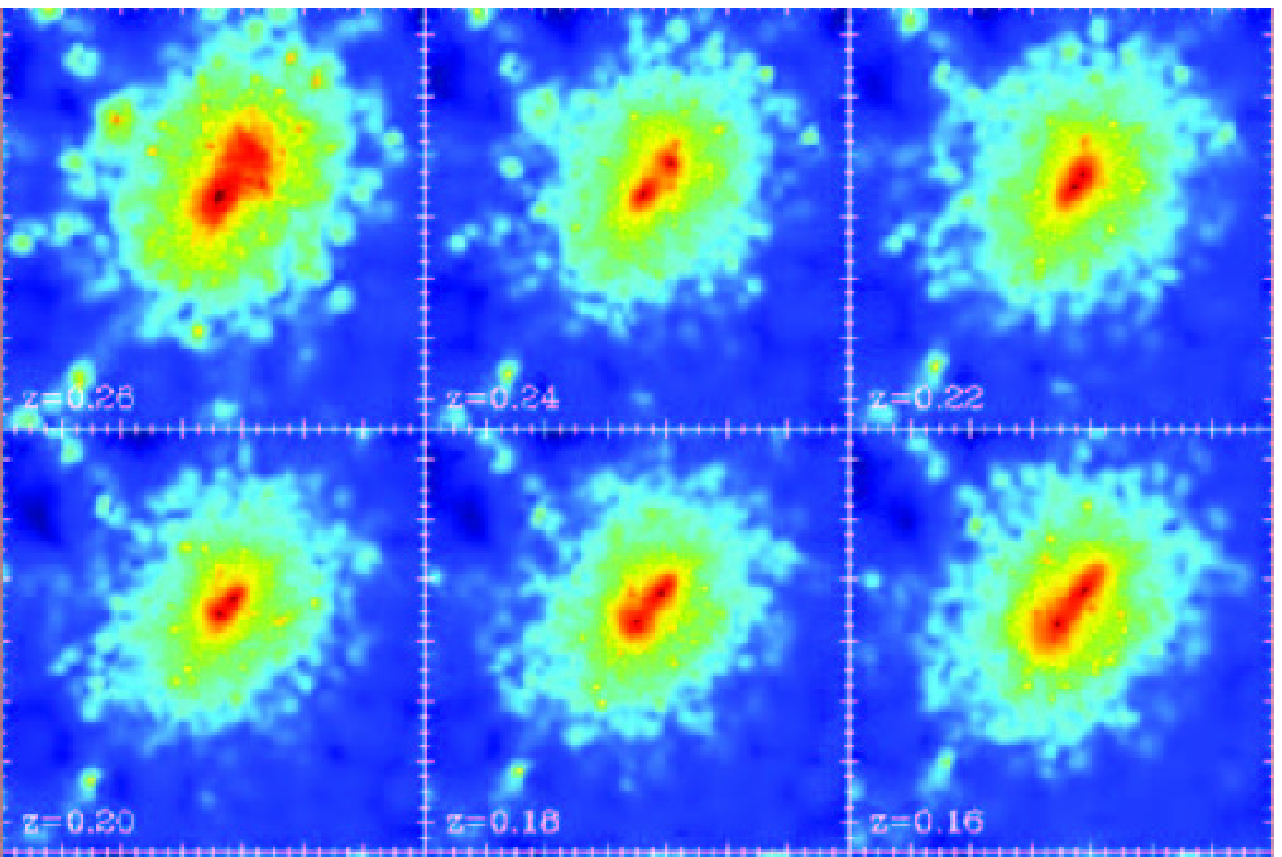}
\vspace{5mm}

\includegraphics[width=\textwidth, bb=132 288 478 489, clip]%
{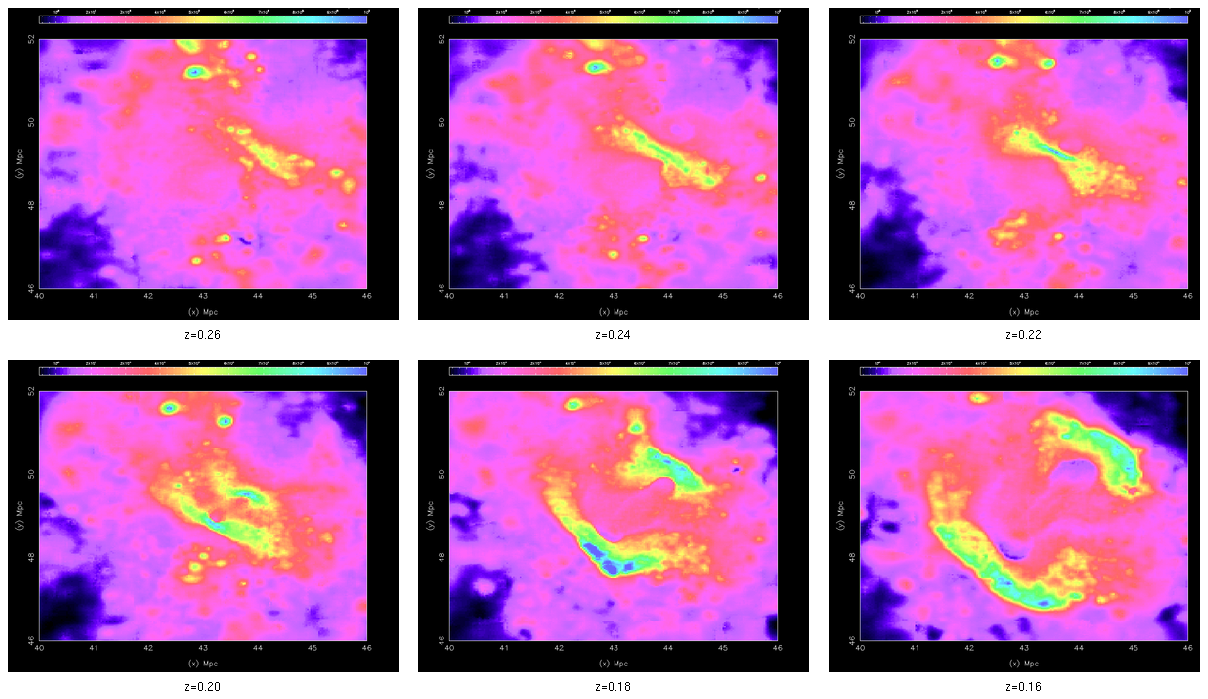}

\caption{Simulated merger of two approximately equal clusters (selected from
  a large cosmological simulation), resulting in the emergence of cold
  fronts. The clusters collide along the NW-SE direction (but not exactly
  head-on: their centers pass each other about 600 kpc apart). Upper panels
  are 14 Mpc in size and show the density of dark matter in a 500 kpc thick
  slice in the merger plane. Lower panels show the gas temperature in a
  slice; their size is 6 Mpc (at centers of the corresponding upper panels).
  Labels in each panel give the redshift of the snapshot.  $z=0.22$ is the
  moment right before core passage; there is a hot shock-heated gas strip (a
  pancake in projection) between the cores.  $z=0.20$ is right after the
  core passage; from this moment on one can see two cold fronts moving in
  the opposite directions.  These two snapshots are similar to the two
  stages shown schematically in Fig.\ \ref{a2142_scheme}.  (Reproduced from
  Mathis et al.\ 2005.)}

\label{mathis}
\end{figure} 

\begin{figure}[t]
\centering
\includegraphics[width=0.45\textwidth, bb=0 0 550 550]%
{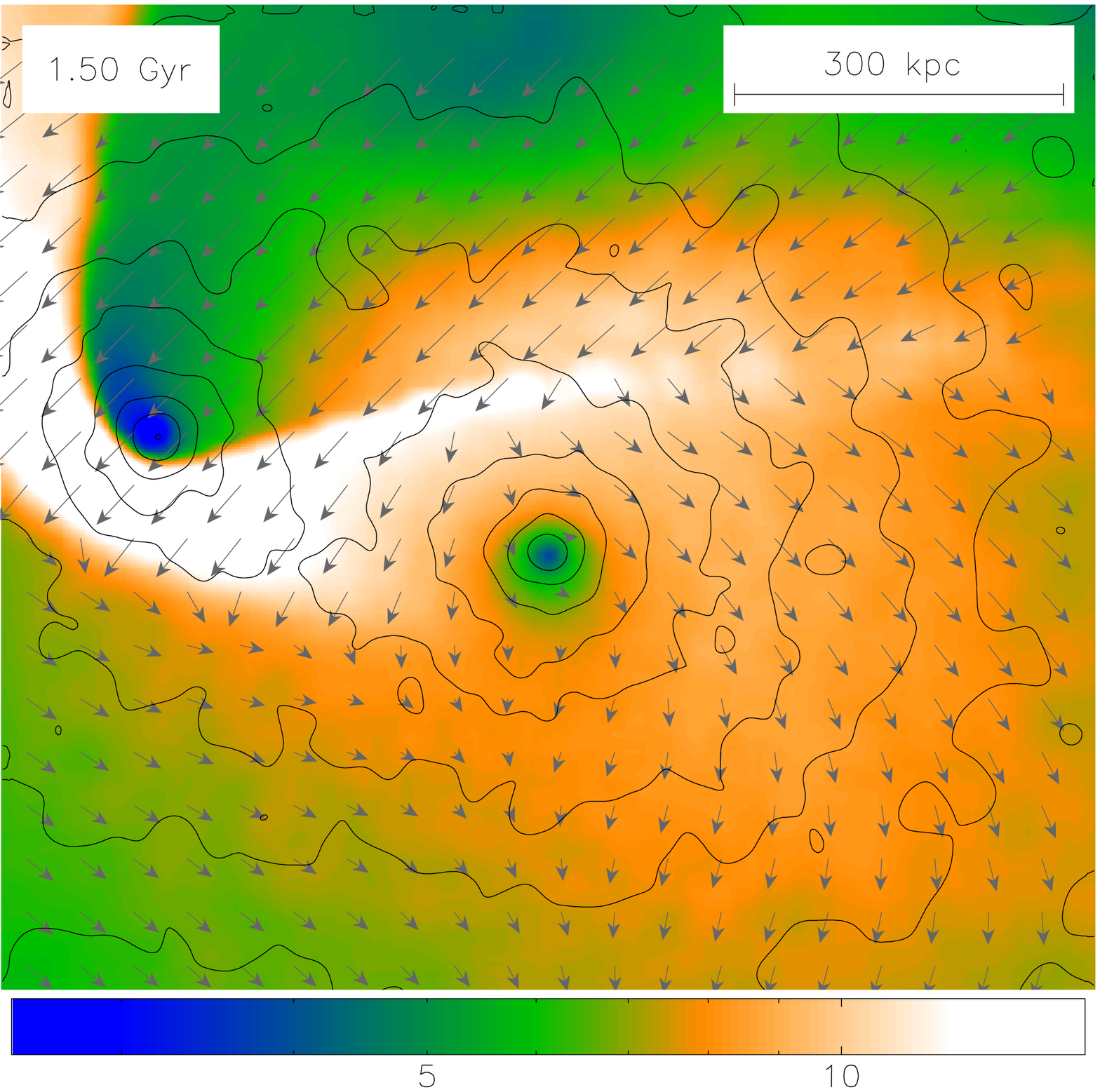}
\includegraphics[width=0.45\textwidth, bb=0 0 550 550]%
{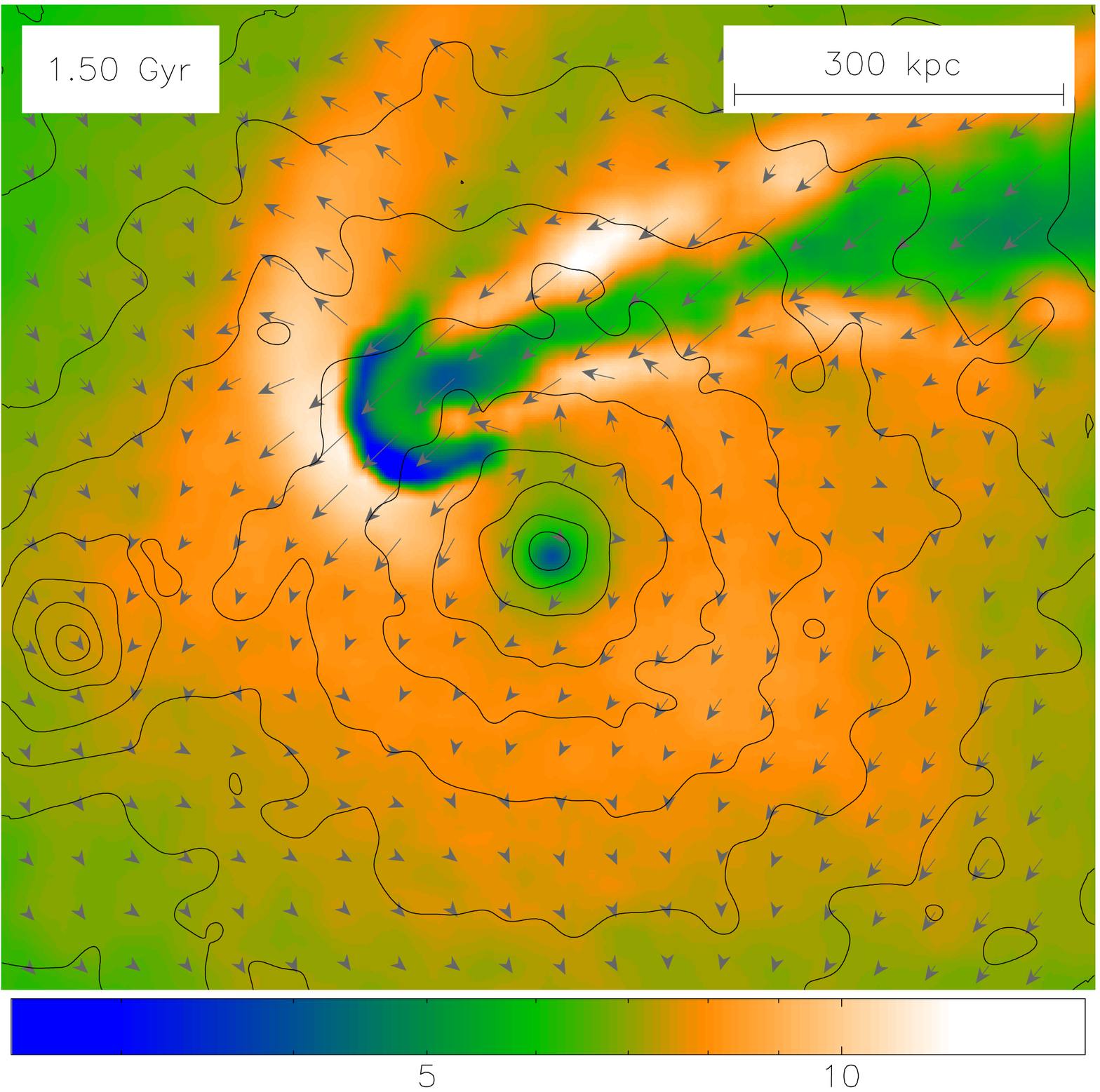}

\caption{Simulated off-axis mergers with different subcluster masses and
  trajectories. Contours show dark matter density, colors show gas
  temperature (the scale bar gives its values in keV), and arrows show the
  gas velocity field relative to the center of the bigger cluster (in the
  center of each panel). The subcluster enters from upper-right. In the left
  panel, the subcluster has retained some of its gas and developed a cold
  front, preceded by a bow shock. However, in the right panel, a less
  massive subcluster flying through denser regions has been completely
  stripped of its gas by ram pressure.  (Reproduced from A06.)}

\label{yago_strip}
\end{figure}

\subsection{Origin and evolution of merger cold fronts}
\label{sec:origin}

Since their discovery, cold fronts in merging clusters have been looked for,
and found, in hydrodynamic simulations with cosmological initial conditions
(e.g., Nagai \& Kravtsov 2003; Onuora, Kay \& Thomas 2003; Bialek, Evrard,
\& Mohr 2002; Mathis et al.\ 2005).  Several other recent works used
idealized 2D or 3D merger simulations to model the effects of ram-pressure
stripping of a substructure moving through the ICM (e.g., Heinz et al.\ 
2003; Acreman et al.\ 2003; Takizawa 2005; Ascasibar \& Markevitch 2006,
hereafter A06). In fact, these features could already be seen in earlier
simulations of idealized mergers, such as those of Roettiger, Loken, \&
Burns (1997). For example, an obvious cold front is seen in a merger
simulated by Roettiger, Stone, \& Mushotzky (1998), although they have not
yet heard of this term then and therefore concentrated on other aspects of
their result.  We present their simulated cluster in Fig.\ \ref{roettiger},
which shows maps of the gas temperature and velocity, dark matter density
and X-ray brightness for two subclusters right after a core passage. Panel
({\em d}) shows the fraction of gas that initially belonged to each
subcluster; a cold front is the boundary between the two gases that did not
mix.
The linear resolution of this and other contemporary simulations (as well as
most of the present-day ones) was limited to $\gg 1$ kpc, which is why they
could not predict that these interfaces would be so strikingly sharp in real
clusters when looked at with \chandra.  Nevertheless, keeping this
limitation in mind, we can use these and newer simulations to clarify the
origin and evolution of cold fronts.

Indeed, simulations show that when two subclusters collide, the outer
regions of their gas halos are shocked and stopped, while the lower-entropy
gas cores are often dense enough to resist the penetration of shocks and
stay attached to their host dark matter subclusters. This can be seen in a
time sequence for an interesting region selected from a cosmological
simulation (Fig.\ \ref{mathis}, from Mathis et al.\ 2005).  The upper panels
show two similar dark matter subclusters colliding and passing nearly
through each other (the pericenter passage occurs between $z=0.22$ and
$z=0.20$).  In lower panels, we see how the gas between the clusters is
first heated via compression and then by shocks, which propagate outwards
after the pericentric passage.  At $z=0.20$ and later, we see the emergence
of two cold fronts, which are the boundaries of the unstripped remnants of
the two former subcluster gas cores. This is pretty much the picture
proposed in M00 (Fig.\ \ref{a2142_scheme}) and seen in other simulations
(e.g., Nagai \& Kravtsov 2003).

\begin{figure}[t]
\centering
\includegraphics[width=\textwidth,bb=75 278 553 490,clip]%
{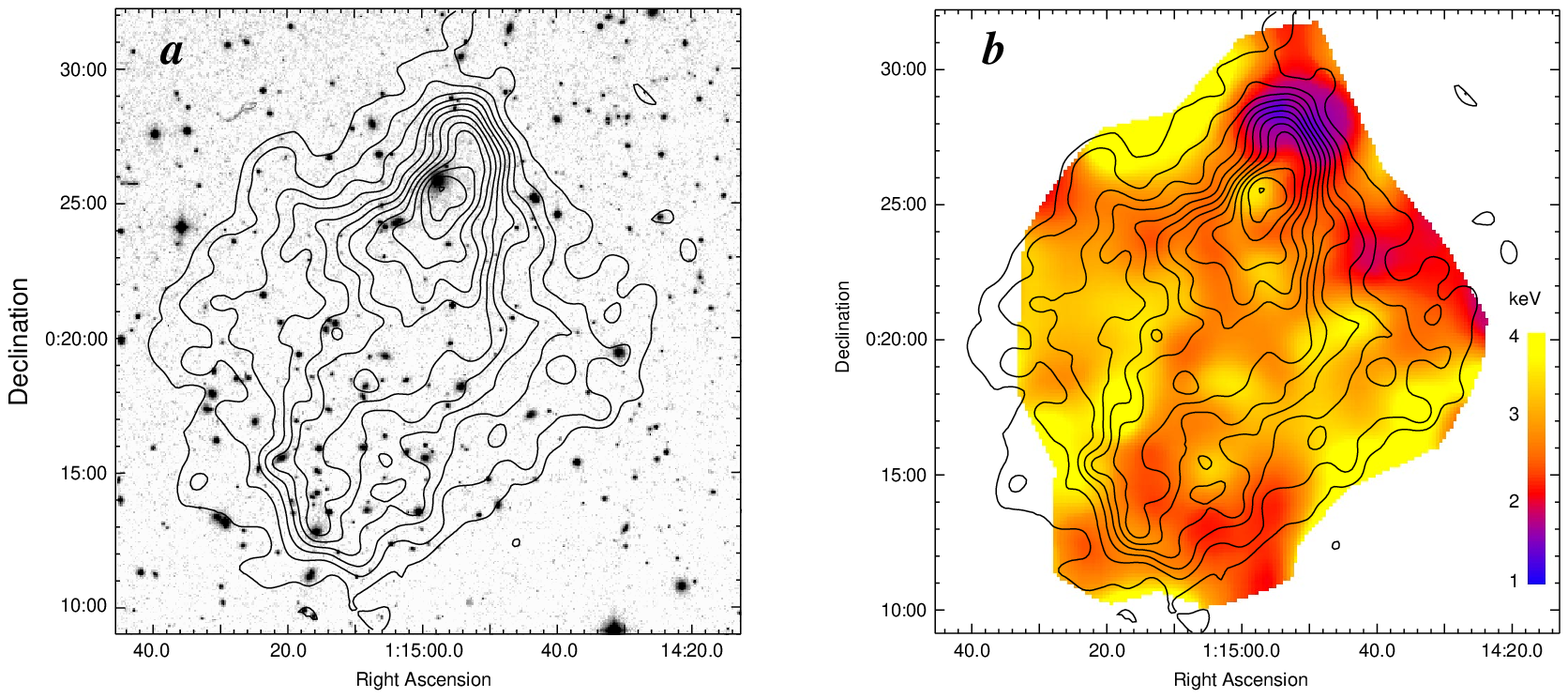}

\caption{({\em a}\/) Contours of the \chandra\ X-ray brightness map
  (smoothed) overlaid on the Palomar Digitized Sky Survey optical image of
  the merging cluster A168. ({\em b}\/) Projected temperature map (colors)
  overlaid with the image contours. The tip of the tongue in the north is a
  cold front. The cD galaxy is the likely gravitational potential minimum.
  The cold front apparently moved ahead of its host dark matter halo in a
  ram pressure slingshot. (Reproduced from Hallman \& Markevitch 2004.)}

\label{a168}
\end{figure} 

\begin{figure}[p]
\centering
\includegraphics[width=0.45\textwidth, bb=0 0 550 550]%
{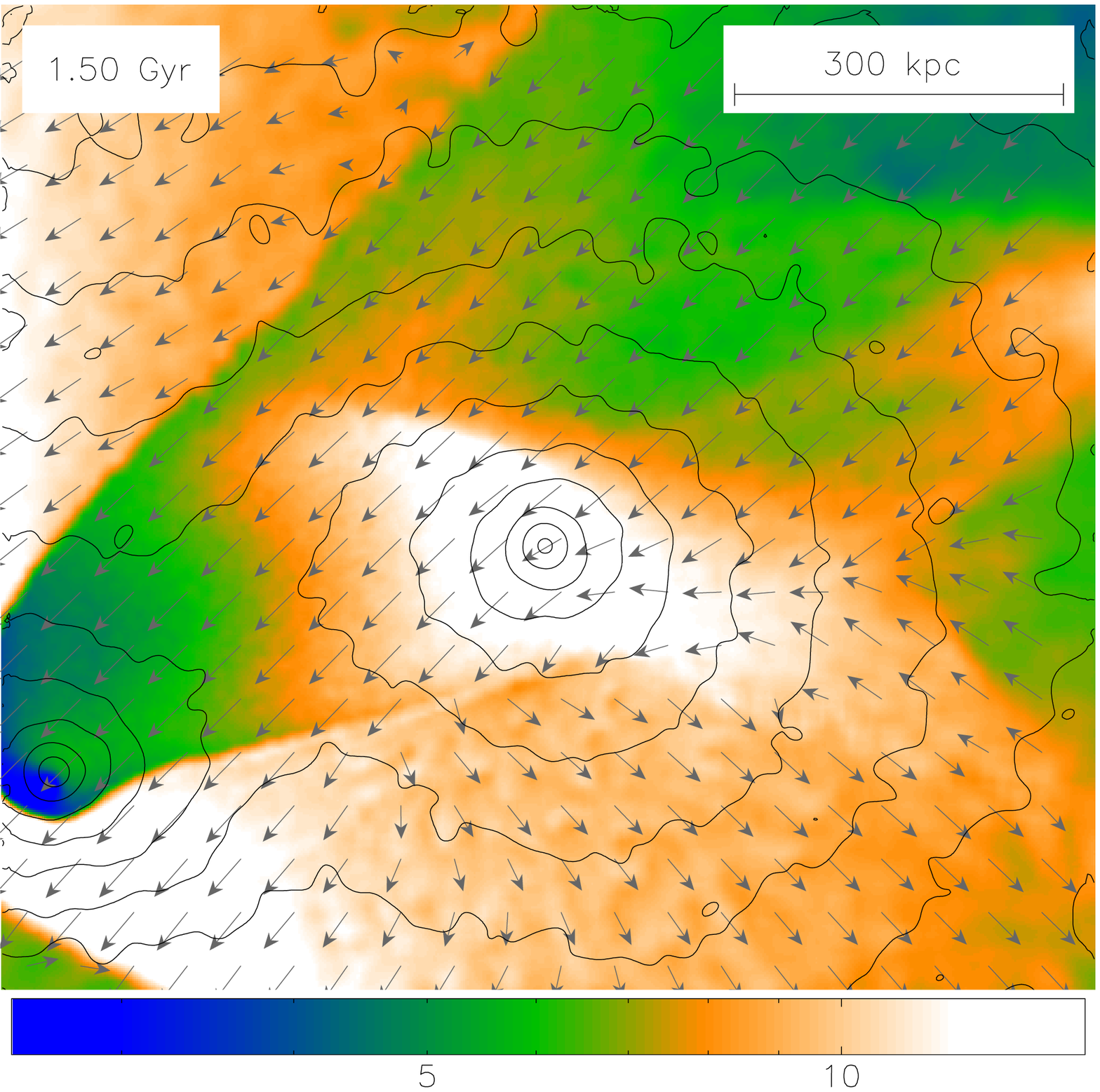}
\includegraphics[width=0.45\textwidth, bb=31 25 580 573]%
{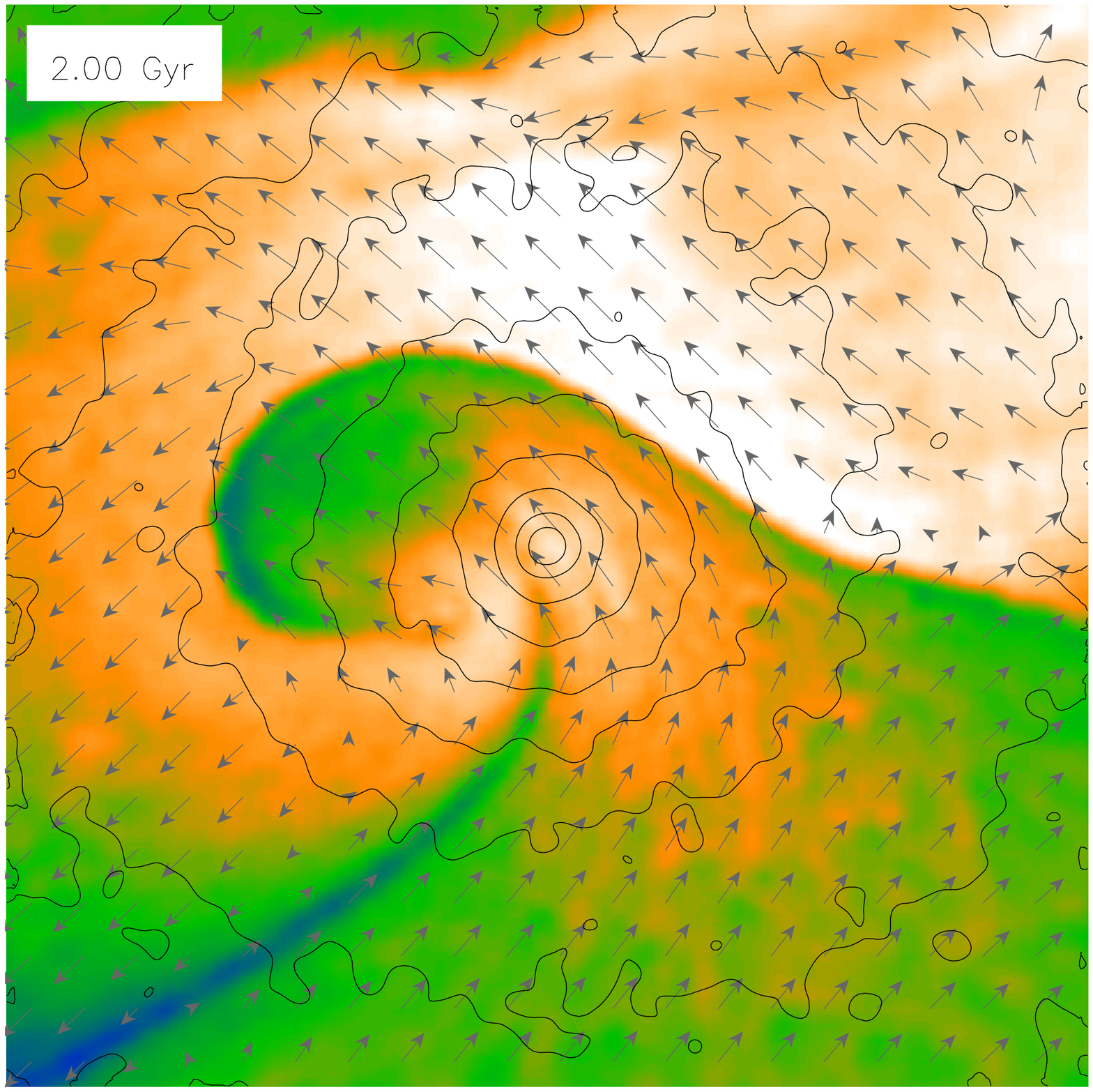}\\
\includegraphics[width=0.45\textwidth, bb=82 76 529 523]%
{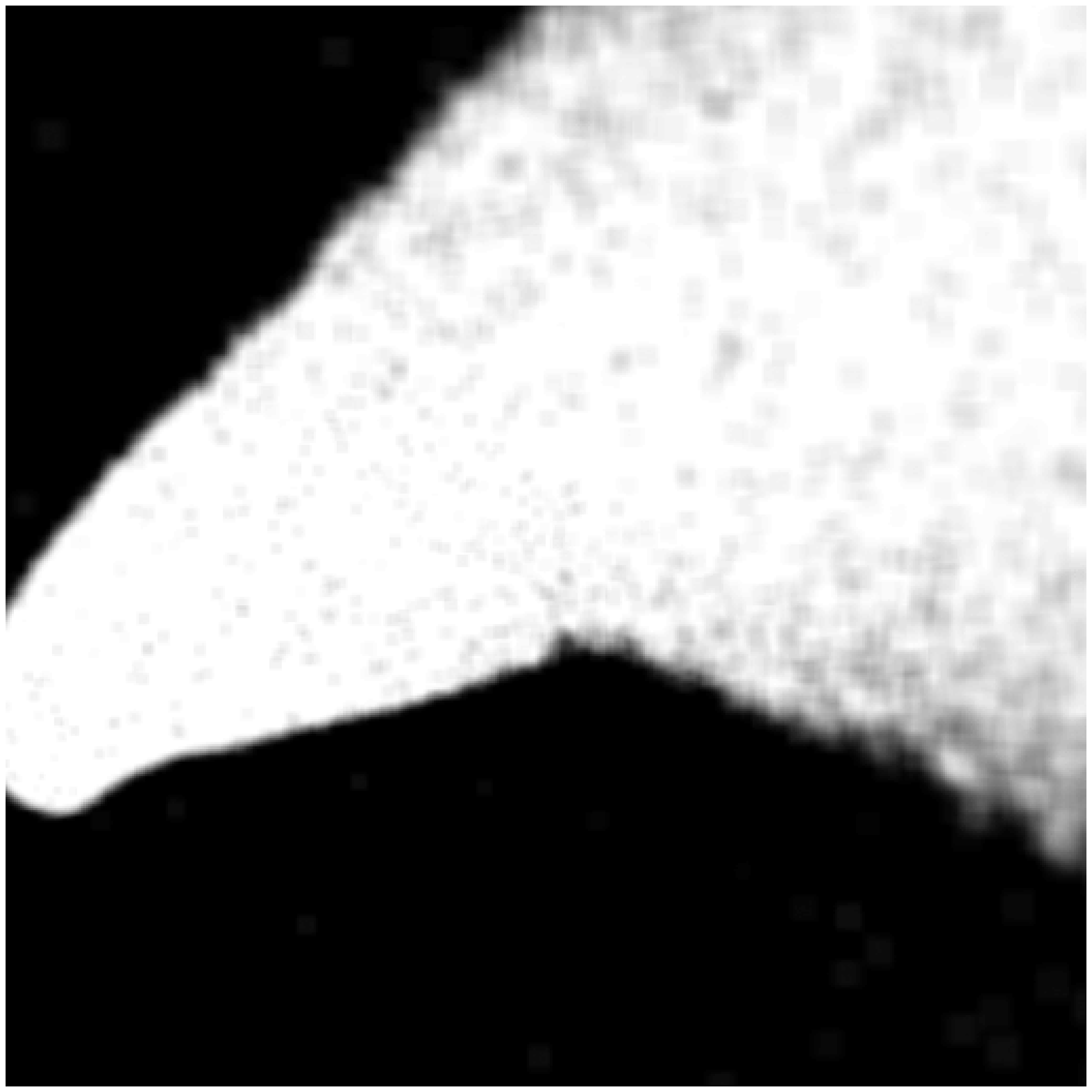}
\includegraphics[width=0.45\textwidth, bb=82 76 529 523]%
{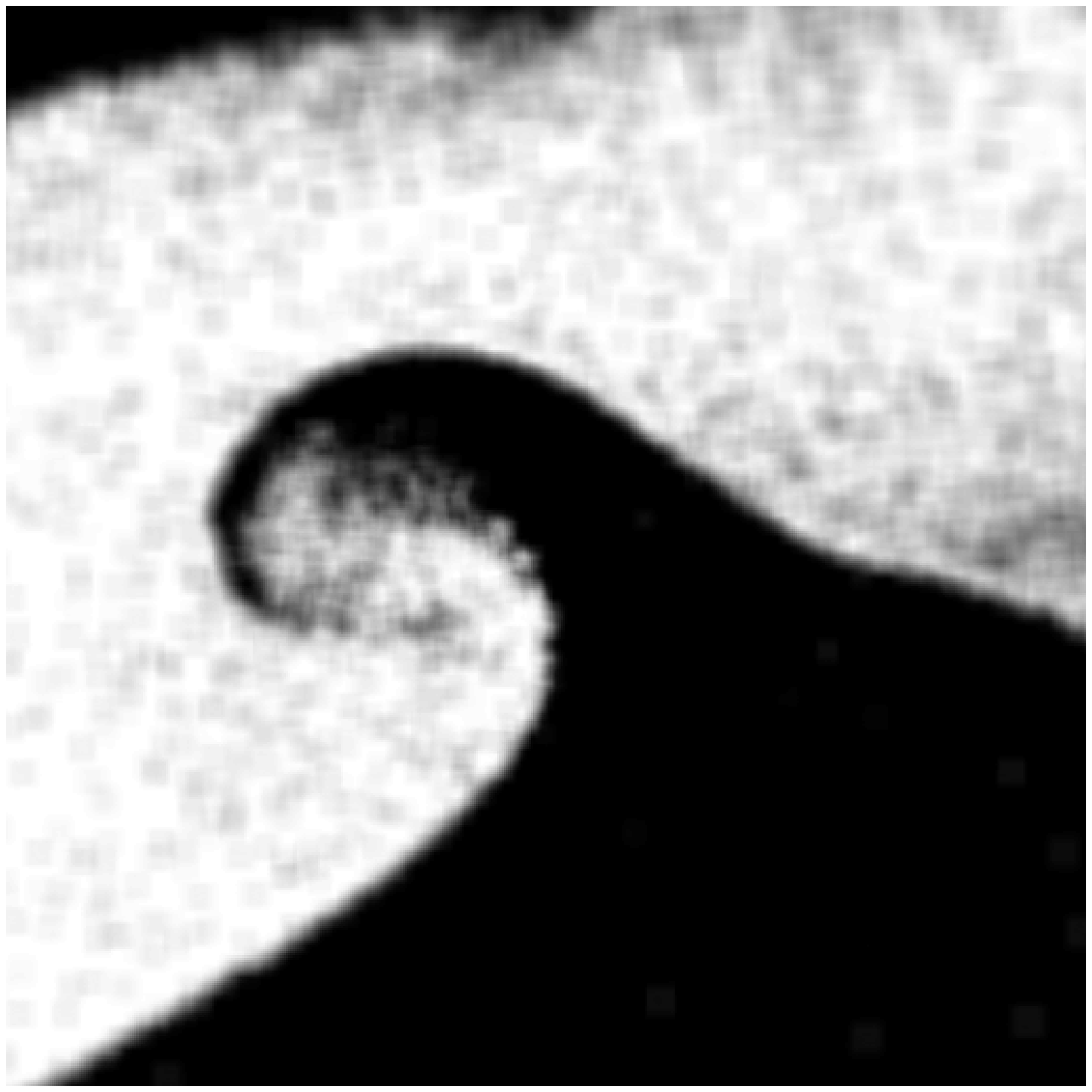}

\caption{Ram-pressure slingshot illustrated by a simulated off-axis merger
  of two subclusters. In top panels, contours show the dark matter density,
  while colors show the gas temperature, in a slice along the merger plane.
  Arrows show local gas velocities. In lower panels, black and white shows
  gas particles that initially belonged to each subcluster, for the
  corresponding upper panels. The subcluster (that entered from the
  upper-right corner) has a cool core, while the main cluster (in the
  center) was initially isothermal and did not have the usual sharp density
  peak at the center. Such initial profiles are chosen here for
  illustration, to increase the amplitude of the motions of the main central
  gas.  The main gas core is pushed back from its gravitational
  potential, but rebounds and overshoots the dark matter peak as soon as the
  ram pressure drops. (Reproduced from A06.)}

\label{yago_sling}
\end{figure}

\begin{figure}[p]
\centering
\includegraphics[width=1\textwidth,bb=65 480 573 580,clip]%
{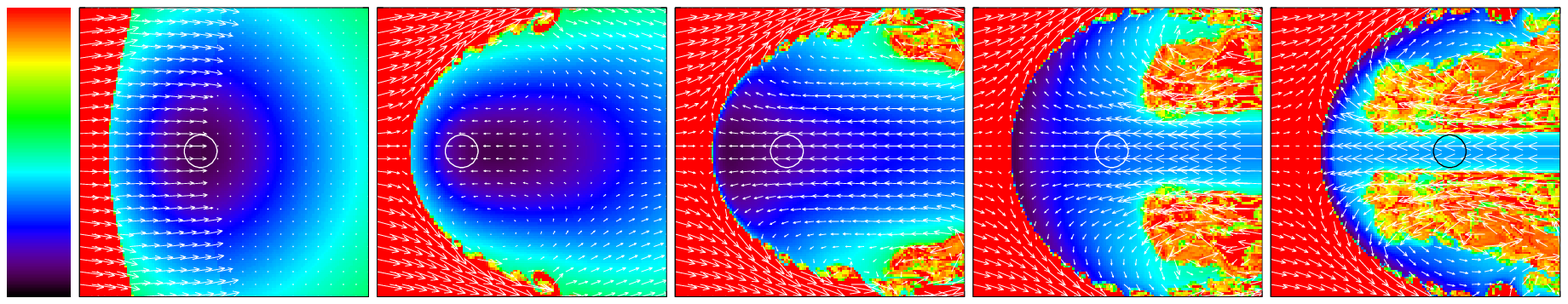}

\caption{A time sequence from a simulation of the passage of a planar shock
  wave through a cluster-like isothermal gas halo (reproduced from Heinz et
  al.\ 2003). The quantity shown is the gas specific entropy; arrows show
  gas velocities.  The shock propagates from the left. The gravitational
  potential is fixed; its minimum is shown as a circle in each panel.}

\label{heinz}
\end{figure}

\paragraph*{Ram pressure stripping}

Ram pressure of the ambient gas first pushes these gas remnants out of the
gravitational potential wells of their respective subclusters. Depending on
the depth of the well, the density of the ambient gas and the merger
velocity, the ram pressure may or may not succeed in stripping the
subcluster gas completely, as shown in Fig.\ \ref{yago_strip}. As long as it
does not succeed, the cool dense gas core is dragged along by the gravity of
the subcluster, initially slightly lagging behind its dark matter peak.  The
ambient shocked gas flows around it, separated by a sharp contact
discontinuity. This is the stage at which we observe the bullet subcluster
in \1e\ (Fig.\ \ref{1e_lens}).

\paragraph*{Ram pressure slingshot}

For a subcluster that has managed to retain its cool core through the
pericenter passage where the ram pressure was the highest, an interesting
thing happens at a later stage. As the subcluster moves away from the
pericenter and slows down, it also enters the region with a lower density of
the ambient gas, and the ram pressure on the cool cores drops very rapidly
($p_{\rm ram}=\rho v^2$).  As a result, the cool gas rebounds and overtakes
the dark matter core as if in a slingshot. The forward region of the cool
core moves away from the gravitational potential minimum which kept it at
high pressure, expands adiabatically and cools, further enhancing the
temperature contrast at the cold front (as noted by Bialek et al.\ 2002).
This is what we observe in A168 (Fig.\ \ref{a168}, from Hallman \&
Markevitch 2004) --- instead of lagging behind, a cold front in that cluster
is located ``ahead'' of the most likely center of the northern subcluster (a
giant galaxy seen in the optical image).  This process is also seen at late
stages of the Mathis et al.\ simulations (note the crescent-shaped cool
regions appearing in the last panel in Fig.\ \ref{mathis}). This ``ram
pressure slingshot'' is further illustrated in Fig.\ \ref{yago_sling} (taken
from A06), which shows a small subcluster passing near the center of a
larger cluster. The corresponding black and white panels show the gas that
initially belonged to each of the subclusters.  At first, ram pressure
exerted by the dense subcluster gas pushes the main cluster core far away
from the dark matter peak. However, as soon as the subcluster passes, that
ram pressure drops, and the main cluster gas (black) rebounds under the
effect of gravity and unbalanced thermal pressure behind the front,
overshooting the center.  Note that in both panels, the boundary of the main
cluster core is a cold front, but at the latter moment, the temperature
contrast at the front is enhanced by adiabatic expansion.  Interestingly, a
gas temperature map for A3667 obtained with \xmm\ (Heinz et al.\ 2003; Briel
et al.\ 2004), which has sufficient statistical accuracy to show the
small-scale detail, shows that the coolest gas is located right along the
cold front, suggesting that the front in A3667 is at this late,
``slingshot'' stage of its evolution.  (Another possibility is that the cool
spot in A3667 is a remnant of a cooling flow-like initial temperature
distribution.)

There may be an additional effect that helps to enhance the temperature
contrast at the cold fronts. Heinz et al.\ (2003) used idealized
two-dimensional hydrodynamic simulation to model the evolution of a contact
discontinuity between a uniform wind flowing around a cool, initially
isothermal (and therefore, with the specific entropy declining toward the
center), cluster-like gas cloud in a stationary gravitational potential of
the underlying dark matter halo. The initially planar discontinuity has
developed into a spheroidal cold front with a flow of ambient (post-shock)
gas around it (Fig.\ \ref{heinz}).  The gas halo in this simulation is first
displaced from the potential minimum along the direction of the wind (panel
2), but then the central, lowest-entropy gas rebounds, overshoots the
potential minimum (as in the ram-pressure slingshot described above) and
starts flowing toward the cold front (panel 3).  At the same time, the
ambient flow around the contact discontinuity has generated a shear layer in
which the Kelvin-Helmholtz (KH) instabilities drag the cool gas located just
under the surface of the front to the sides and away from the tip (panel 3
and later). This is in addition to the usual flattening and sideways
expansion of a dense gas sphere subjected to a wind. If occurs in real
clusters, this ``circulation'' may help the low-entropy gas located deeper
under the surface to reach the tip.

\begin{figure}[t]
\centering
\includegraphics[width=0.48\textwidth]%
{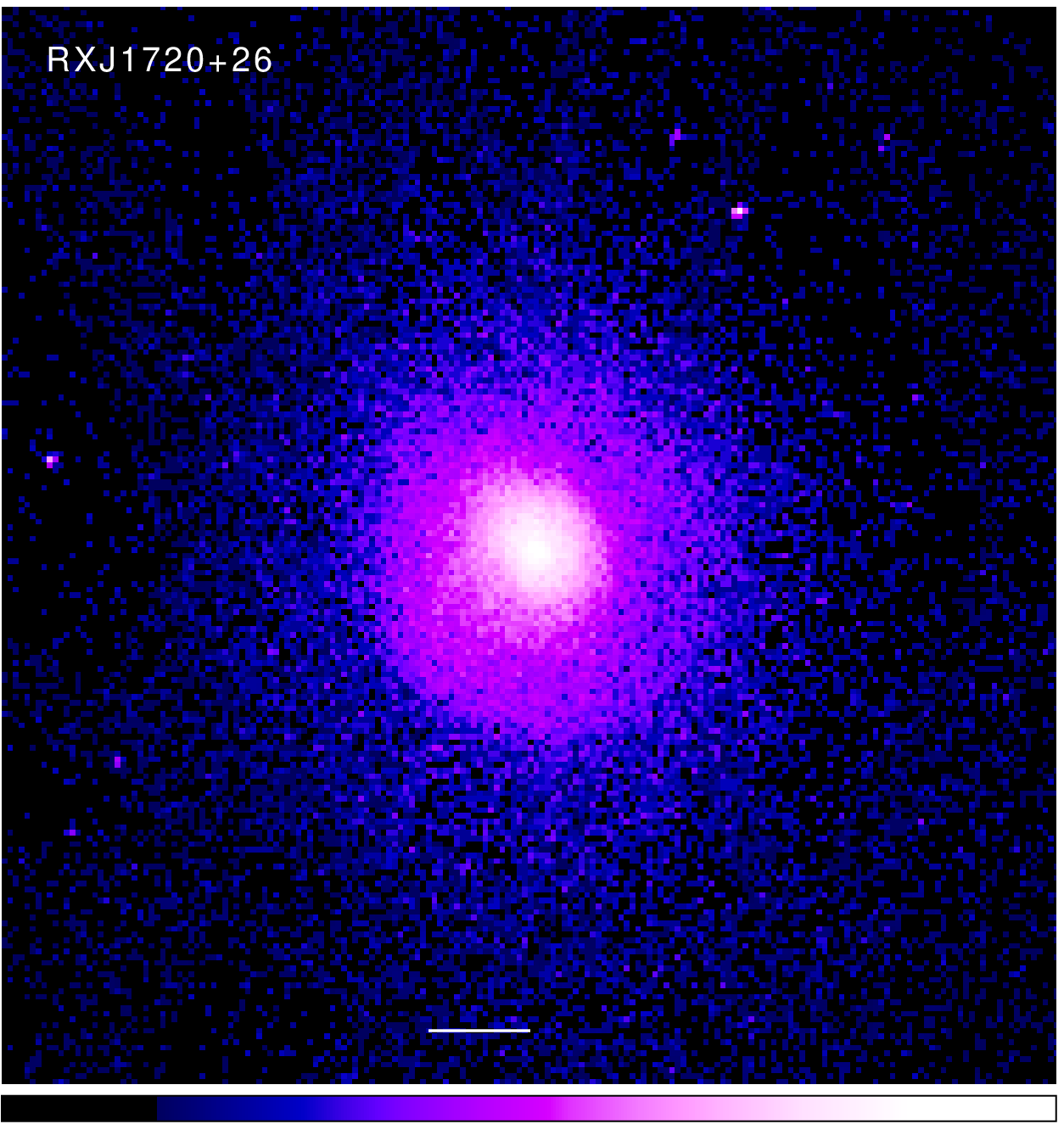}~~
\includegraphics[width=0.48\textwidth]%
{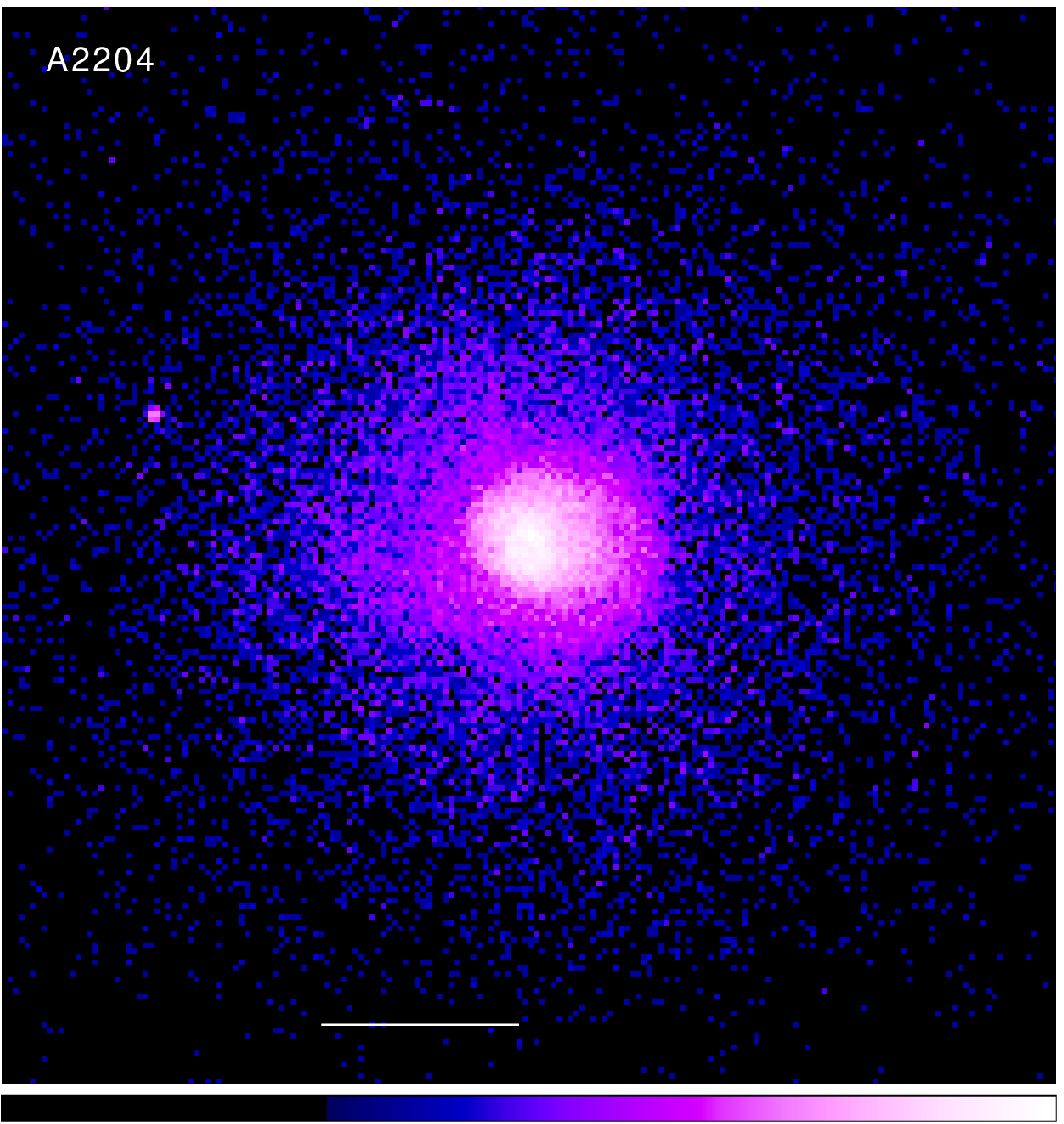}
\vspace{5mm}
\includegraphics[width=0.48\textwidth]%
{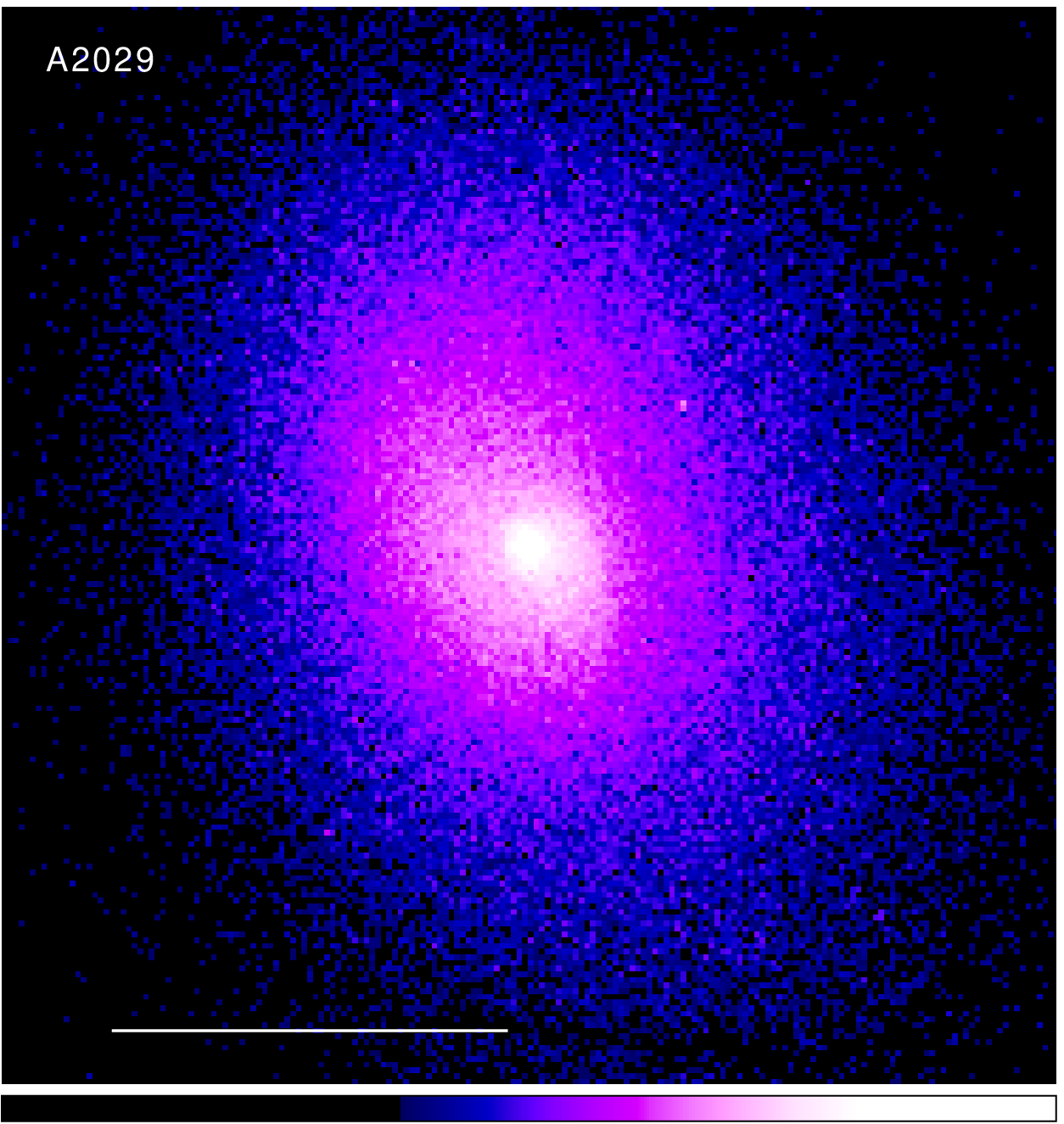}~~
\includegraphics[width=0.48\textwidth]%
{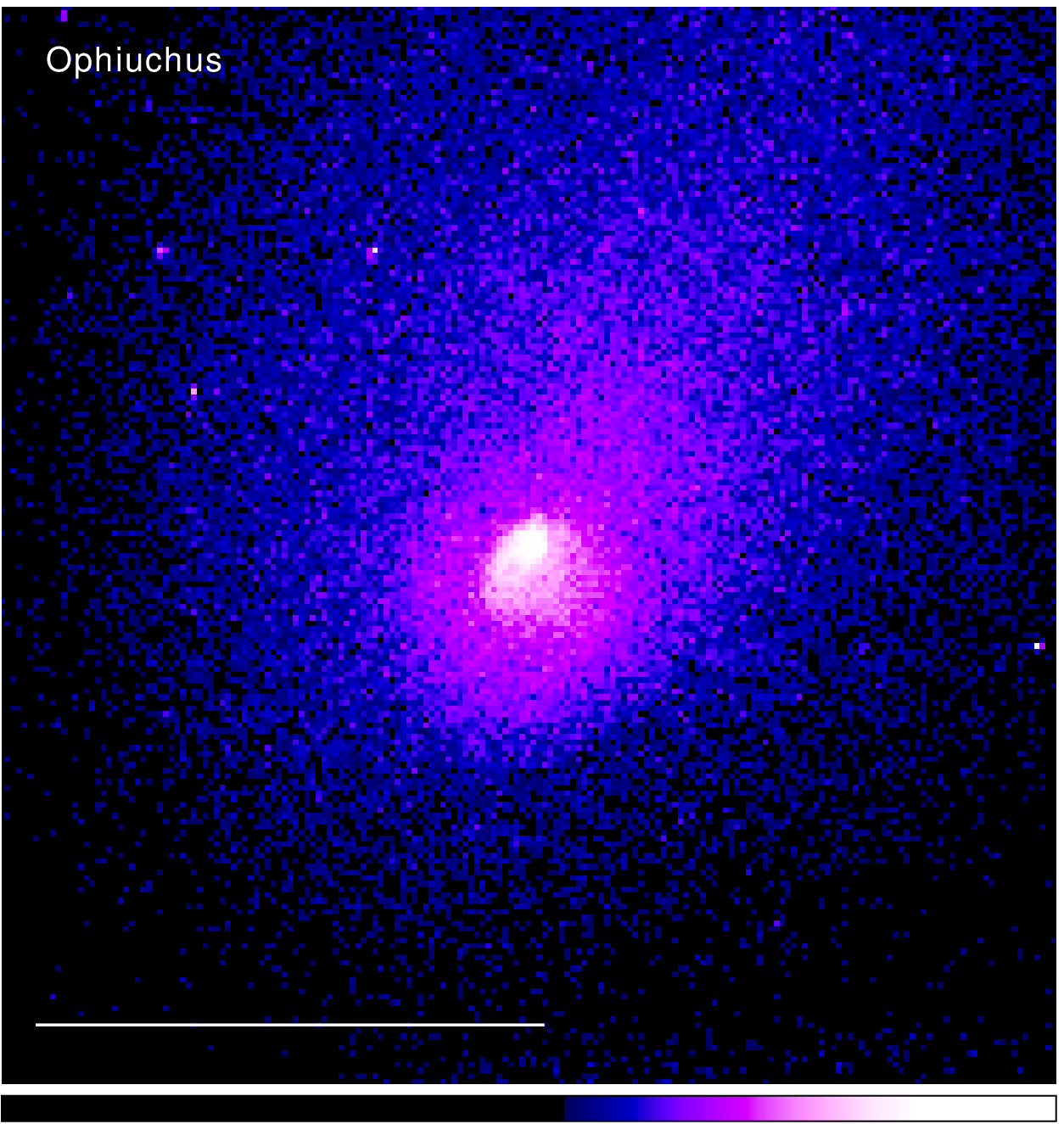}

\caption{X-ray images of several non-merging clusters exhibiting cold fronts
  inside their cool cores.  Horizontal bars are 100 kpc.  A2029 (Clarke et
  al.\ 2004) exhibits 2 edges in a spiral pattern at $r\sim 7$ kpc and 20
  kpc (between white and pink and between light and dark pink).
  RXJ\,$1720+26$ (Mazzotta et al.\ 2001) exhibits a large, $r\approx 250$
  kpc edge (pink--blue), while A2204 (Sanders et al.\ 2005) shows a spiral
  pattern consisting of at least two edges at 20 kpc and 70 kpc (white--pink,
  pink--blue).  Ophiuchus has some evidence in the outskirts of a recent
  merger; note three edges on scales around $r\sim 3$ kpc, 8 kpc and 40 kpc
  (white--pink, white--darker pink, blue--darker blue).  (Reproduced from
  A06.)}
\label{slosh_examples}
\end{figure}

\subsection{Cold fronts in cluster cool cores}
\label{sec:slosh}

When the subclusters merge, one does expect to see vigorous gas flows,
including moving remnants of the subcluster cores which give rise to cold
fronts. Surprisingly, though, cold fronts are also observed near the centers
of most ``cooling flow'' clusters, many of which are relaxed and show little
or no signs of recent merging (e.g., Mazzotta et al.\ 2001; Markevitch,
Vikhlinin, \& Mazzotta 2001, hereafter M01; Mazzotta, Edge, \& Markevitch
2003; Churazov et al.\ 2003; Dupke \& White 2003; Sanders et al.\ 2005).
These fronts are typically more subtle in terms of the density jump than
those in mergers, and occur on smaller linear scales close to the center
($r\lax 100$ kpc), with their arcs usually curved around the central gas
density peak. There are often several such arcs at different radii around
the density peak.  Cooling flow clusters by definition have a sharp
temperature decline and an accompanying density increase toward the center
(that is, a sharp decline of specific entropy).  The edges are seen inside
or on the boundaries of this cool central region.  This is a very common
variety of the cold fronts; we found them in more than a half of the cooling
flow clusters (Markevitch, Vikhlinin, \& Forman 2003b).  Given the
projection, this means that most, if not all, cooling flow clusters may have
one or several such fronts.  Some of the clusters with such fronts are shown
in Fig.\ \ref{slosh_examples}.  One of them is A2029, which on scales
$r>100-200$ kpc is the most undisturbed cluster known (e.g., Buote \& Tsai
1996).  As in mergers, cold fronts in these clusters must indicate gas
motion; however, the moving gas clearly does not belong to any infalling
subcluster.

\begin{figure}[t]
\centering
\includegraphics[width=1\textwidth,bb=59 223 553 612,clip]%
{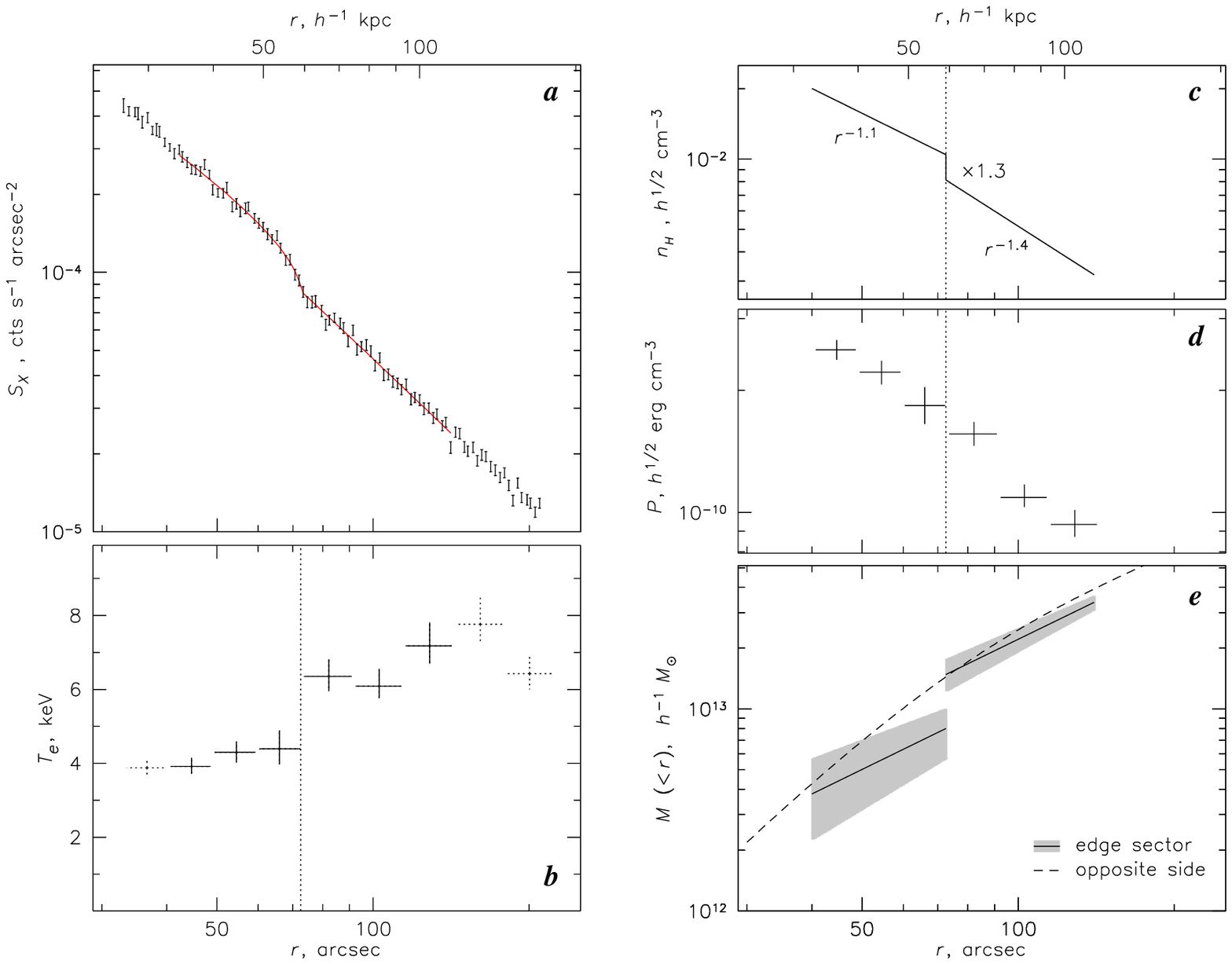}

\caption{Profiles from a sector in the A1795 cluster, centered on the
  cluster peak and containing a brightness edge (which is concentric with
  the cluster peak).  Vertical dotted lines show the edge position. ({\em
    a}) X-ray surface brightness. The red line is a projection of the
  best-fit density model shown in panel ({\em c}). ({\em b}) Gas temperature
  profile, corrected for projection.  ({\em d}) Pressure profile.  ({\em e})
  Total mass within a given radius derived using the hydrostatic equilibrium
  assumption (with an error band shown in gray).  For comparison, a fit in
  the sector opposite to the edge, where the gas distribution is continuous,
  is shown by dashed line. If the gas were indeed in hydrostatic
  equilibrium, they would show the same mass. (Reproduced from M01.)}

\label{a1795_profs}
\end{figure}

\begin{figure}[t]
\centering
\includegraphics[width=0.5\textwidth]%
{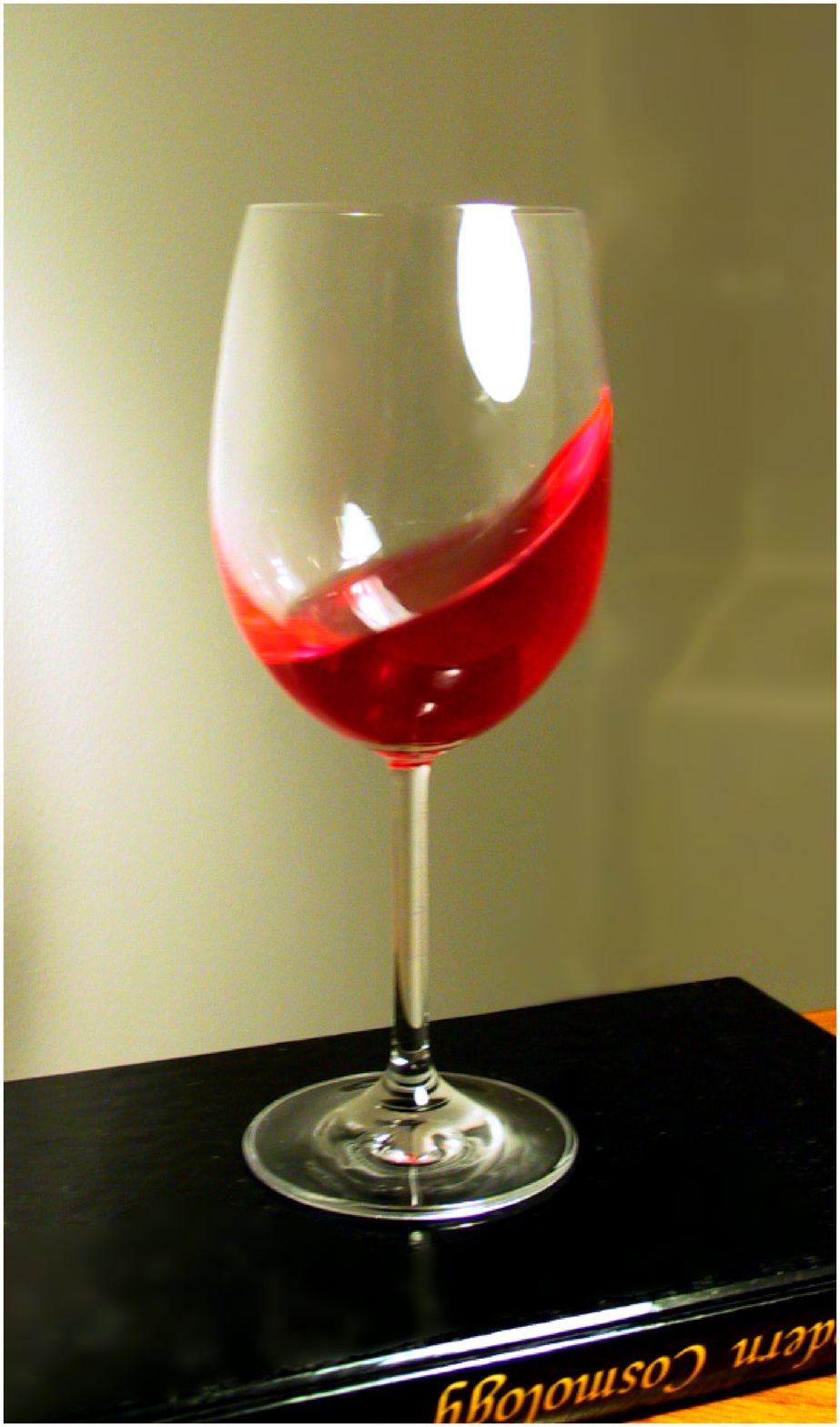}

\caption{The origin of cold fronts in the dense cluster cores.}

\label{glass}
\end{figure}

We studied such a feature in A1795 --- another one of the most-relaxed
nearby clusters (Buote \& Tsai 1996) --- and showed that the gas forming a
cold front is not in hydrostatic equilibrium in the cluster gravitational
potential (M01). Figure \ref{a1795_profs} (reproduced from M01) shows radial
profiles of X-ray brightness and gas temperature in the sector of A1795 that
contains the cold front, along with the best-fit gas density model and the
resulting pressure profile. The brightness edge is barely noticeable (and
would certainly go undetected without the \chandra's arcsecond resolution)
and corresponds to a density jump by only a factor of 1.3 (for comparison,
in the more prominent merger cold fronts discussed in \S\ref{sec:merg}, the
density jumps by factors 2--5). The pressure profile turns out to be almost
exactly continuous, leaving little or no room for a relative gas motion,
since the ram pressure from such a motion would cause the inner thermal
pressure to be higher compared to that on the outside (beyond a small
stagnation region; see \S\ref{sec:vel} below).  Thus the inner and outer
gases appear very nearly at rest and in pressure equilibrium.  Therefore,
one might expect them to be in hydrostatic equilibrium in the cluster
gravitational potential.

For a spherically symmetric cluster in hydrostatic equilibrium, one can
derive the cluster total (mostly dark matter) mass from the above radial
profiles of the gas density and temperature using eq.\ (\ref{eq:mass}).  The
resulting mass profile in the immediate vicinity of the edge in A1795 is
shown in Fig.\ \ref{a1795_profs}{\em e}.  The profile reveals an unphysical
discontinuity by a factor of 2 at the front. For comparison, a similar
analysis was performed using a sector on the opposite side from the center,
which has smooth distributions of gas density and temperature. The total
mass profile derived using that sector is overlaid as a dashed line.  If the
gas around the cluster center were in hydrostatic equilibrium, both sectors
would measure the same enclosed cluster mass.  Indeed, outside the edge
radius, the masses derived from the two opposite sectors agree, strongly
suggesting that the gas immediately outside the edge is indeed near
hydrostatic equilibrium. But the gas inside the edge is not --- even though
there is pressure equilibrium between the two sides of the edge.

Such an unphysical mass discontinuity at the cold front was first reported
by Mazzotta et al.\ (2001) for the cluster RXJ\,1720+26, which is similarly
relaxed on large scales.  Although the statistical accuracy of the available
temperature profile was not sufficient to exclude a significant bulk
velocity of the cool gas, the situation appears similar to A1795.

\subsubsection{Gas sloshing}
\label{sec:a1795}

Given the above evidence, we proposed (M01) that the low-entropy gas in the
A1795 core is ``sloshing'' in the central potential well (Fig.\ 
\ref{glass}).  The observed edge delineates a parcel of cool gas that has
moved from the cluster center and is currently near the point of maximum
displacement, where it has zero velocity but nonzero centripetal
acceleration.  In agreement with this scheme, there is a $30-40\;h^{-1}$ kpc
cool gas filament extending from the cD galaxy in the center of A1795 toward
this cold front (Fabian et al.\ 2001), suggesting that the bulk of the
central gas has indeed been flowing around the cD galaxy (which most
probably sits in the gravitational potential minimum).  Such an oscillating
gas parcel would not be in hydrostatic equilibrium with the potential ---
instead, the gas distribution would reflect the reduced gravity force in the
accelerating reference frame, resulting in the above unphysical mass
underestimate. M01 made an estimate of this acceleration from the apparent
mass jump $\Delta M$, assuming that the gas outside the edge is hydrostatic:
$a\sim G \Delta M r^{-2} \approx 3\times 10^{-8}h$ cm~s$^{-2}$ or
$800h\;{\rm km}\;{\rm s}^{-1}\;(10^8\;{\rm yr})^{-1}$, where $r$ is the
radius of the edge, which is a sensible number for an oscillation on this
linear scale. More recent detailed simulations (A06, see below) have shown
that this picture is somewhat oversimplified, but the physics in it is
correct.

\begin{figure}[t]
\centering
\includegraphics[width=0.85\textwidth,bb=1 11 394 363,clip]%
{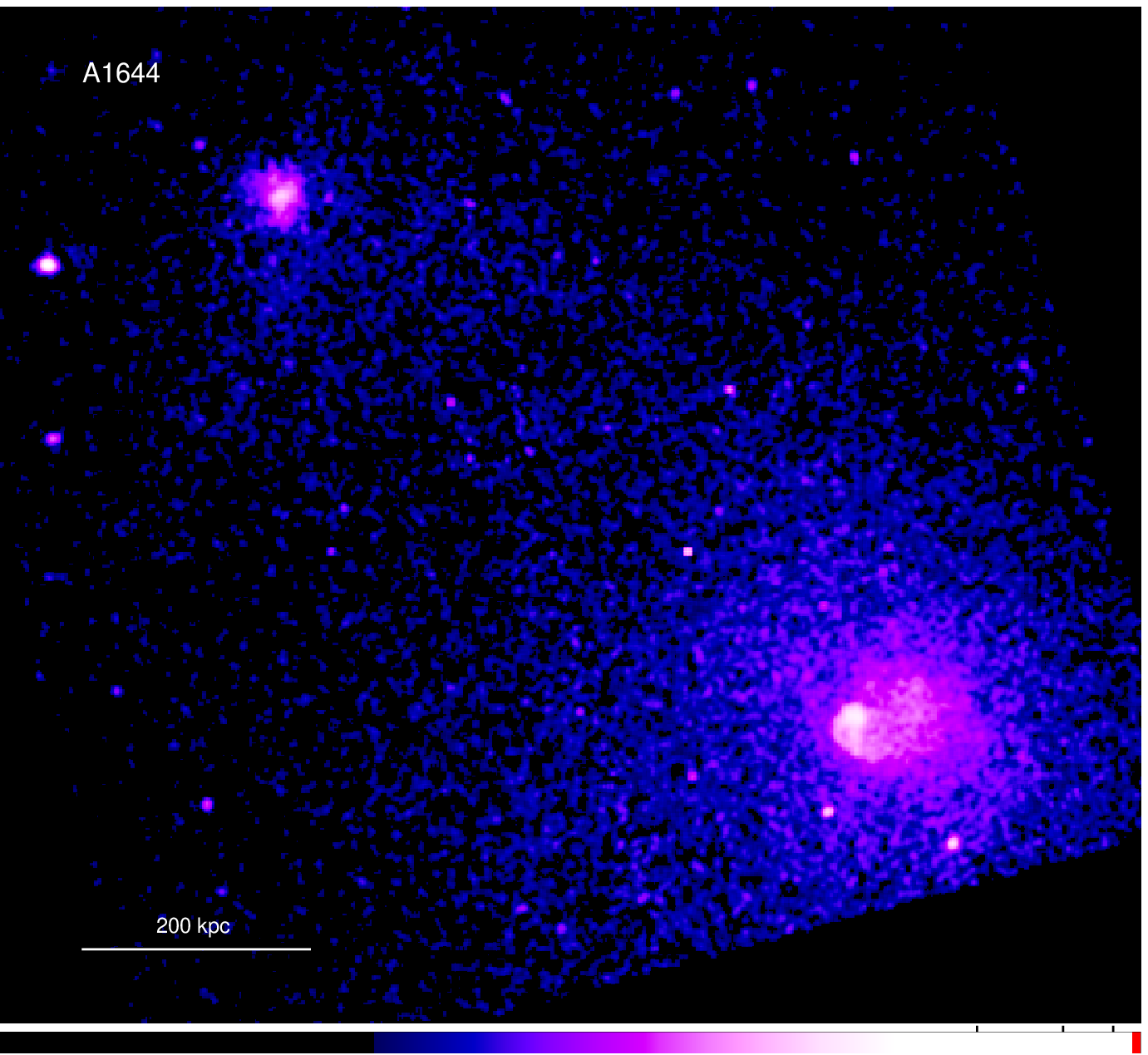}

\caption{\chandra\ X-ray image of the merging system A1644.  An infalling
  subcluster (the northeastern clump) apparently has passed near the center
  of the main cluster, disturbed its mass distribution and set off sloshing
  in its core.}

\label{a1644}
\end{figure}

In M01 we suggested that this subsonic sloshing of the cluster's own cool,
dense central gas in the gravitational potential well may be the result of a
disturbance of the central potential by past subcluster infall.  There are
striking examples suggesting that this is the case at least in some clusters
(Fig.\ \ref{a1644}). Alternatively, one can imagine some off-center
disturbance in the gas from the activity of the central AGN; AGNs blowing
bubbles in the intracluster gas are observed in many cooling flow clusters
(e.g., Fabian et al.\ 2000; Nulsen et al.\ 2005). However, the absence of
any visible merger or AGN disturbance in the X-ray images of two of the most
undisturbed clusters, A2029 and A1795, presented an apparent difficulty,
which has motivated some of the numerical studies reviewed below.

\begin{figure}[t]
\centering
\includegraphics[width=0.32\textwidth]%
{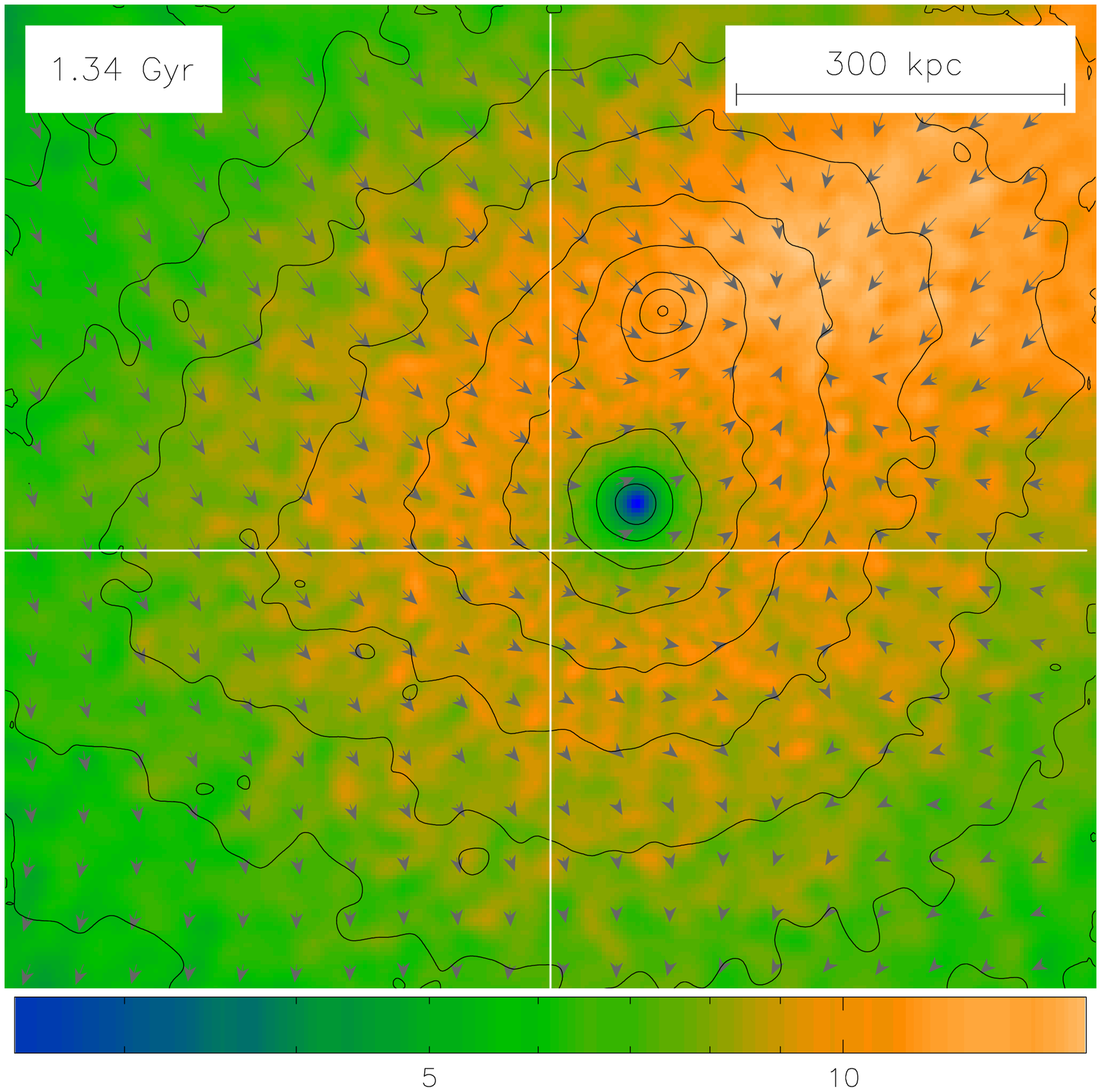}
\includegraphics[width=0.32\textwidth]%
{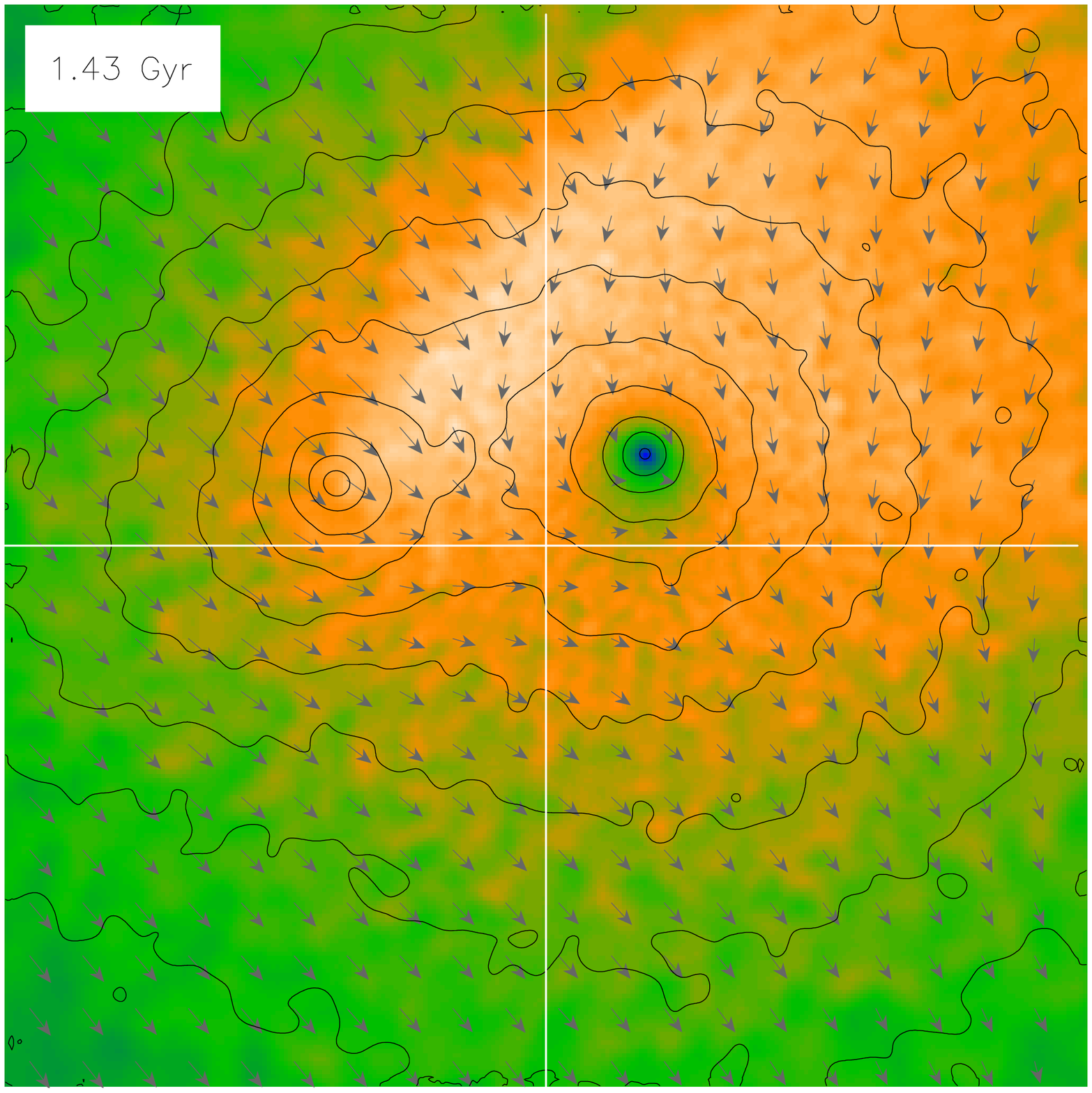}
\includegraphics[width=0.32\textwidth]%
{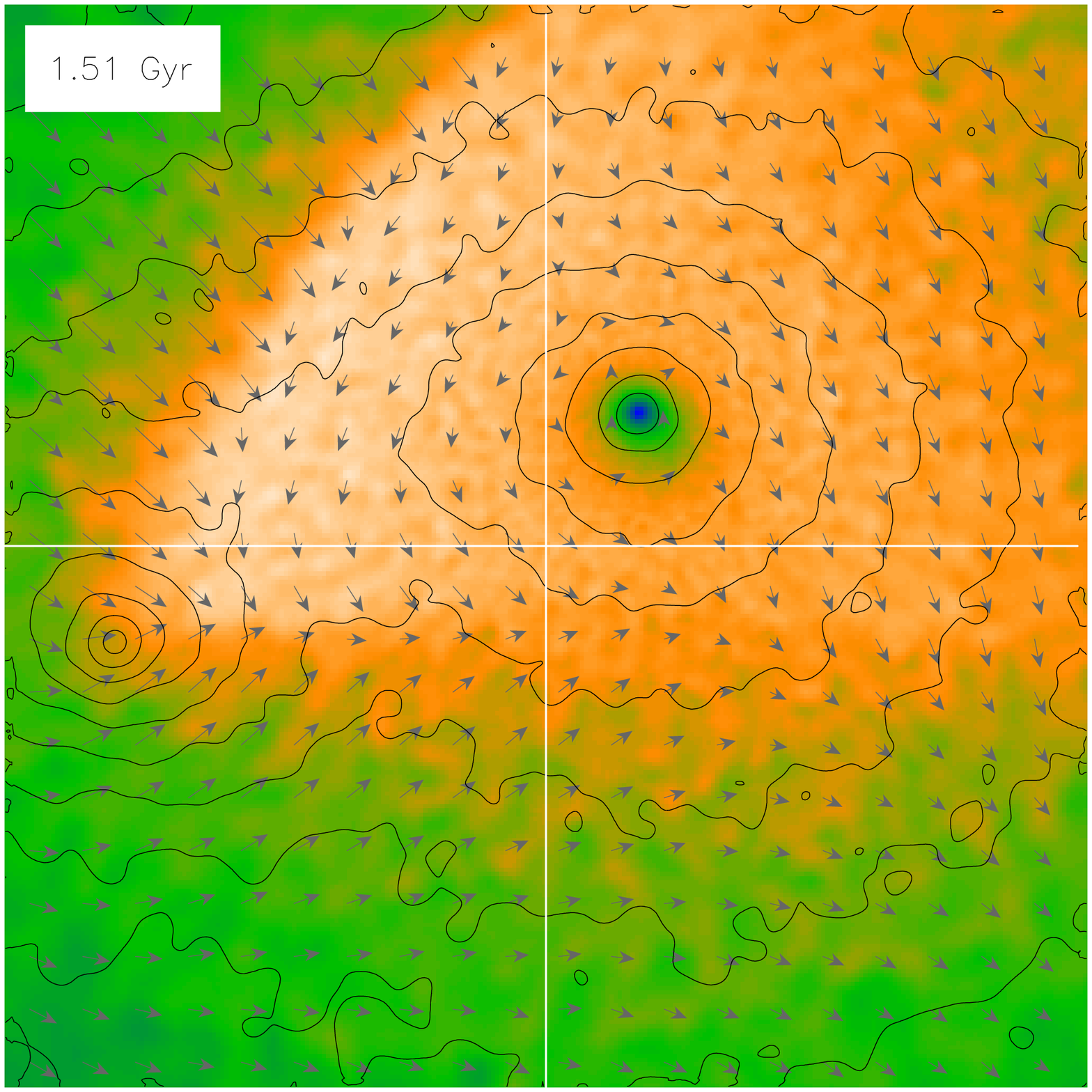}\\[1mm]
\includegraphics[width=0.32\textwidth]%
{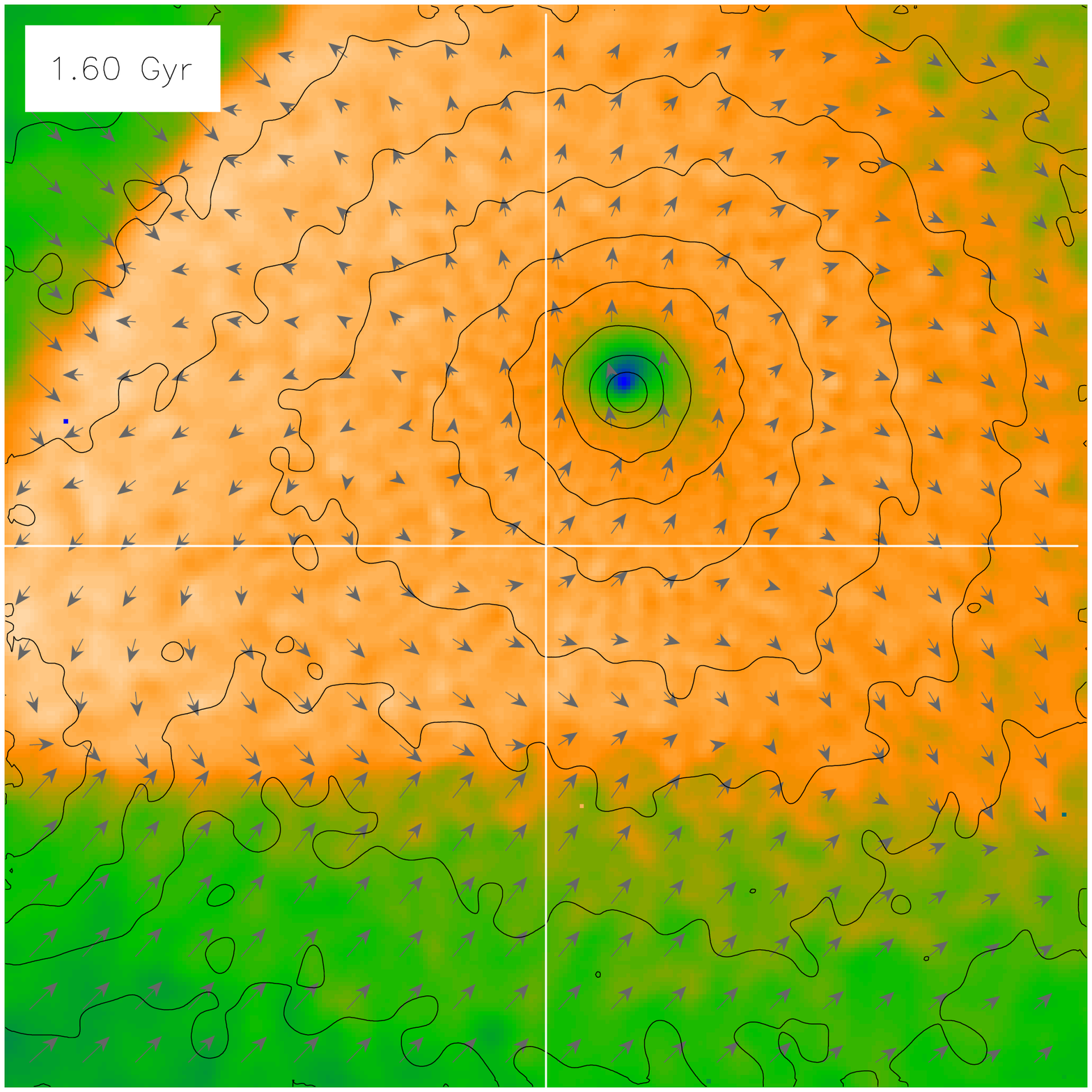}
\includegraphics[width=0.32\textwidth]%
{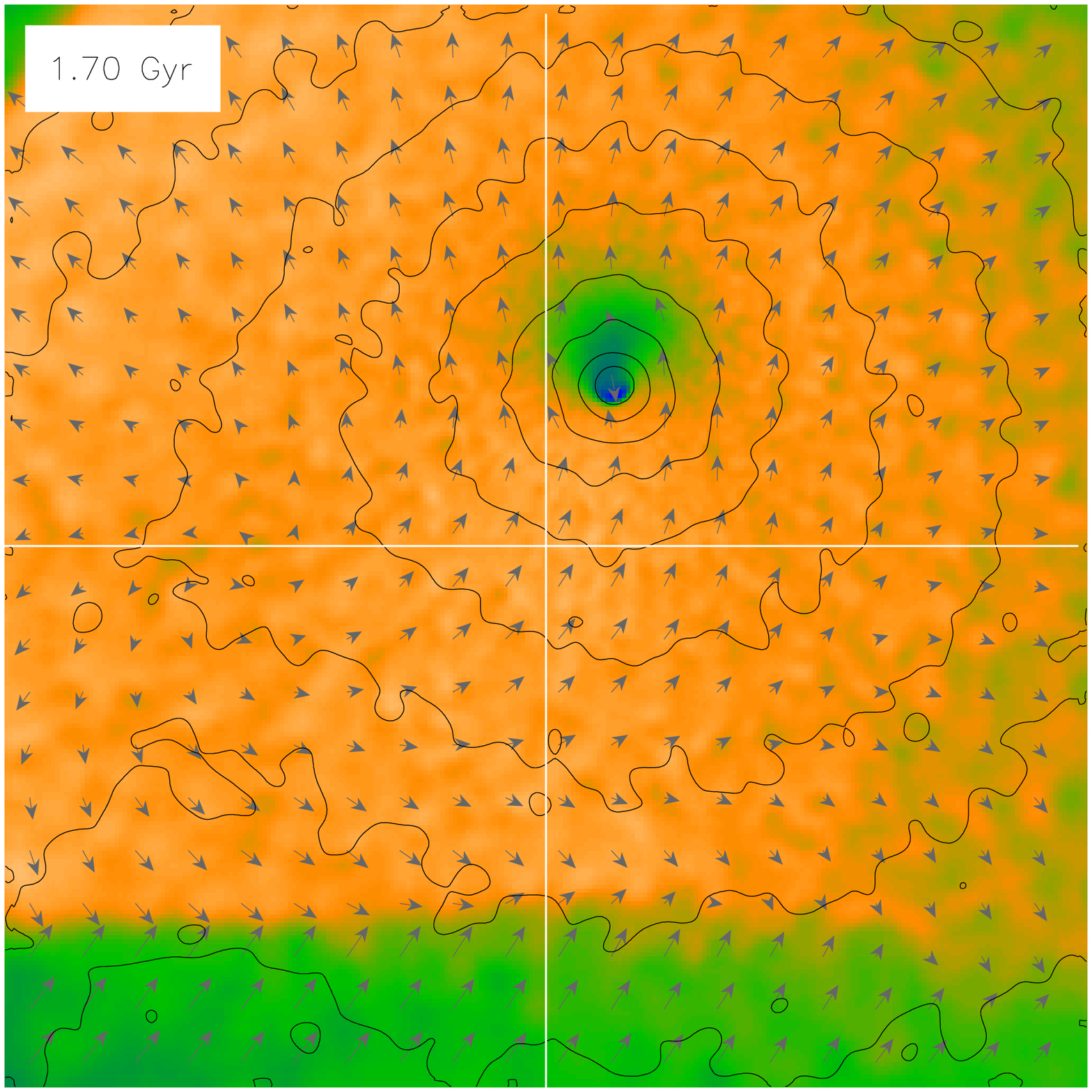}
\includegraphics[width=0.32\textwidth]%
{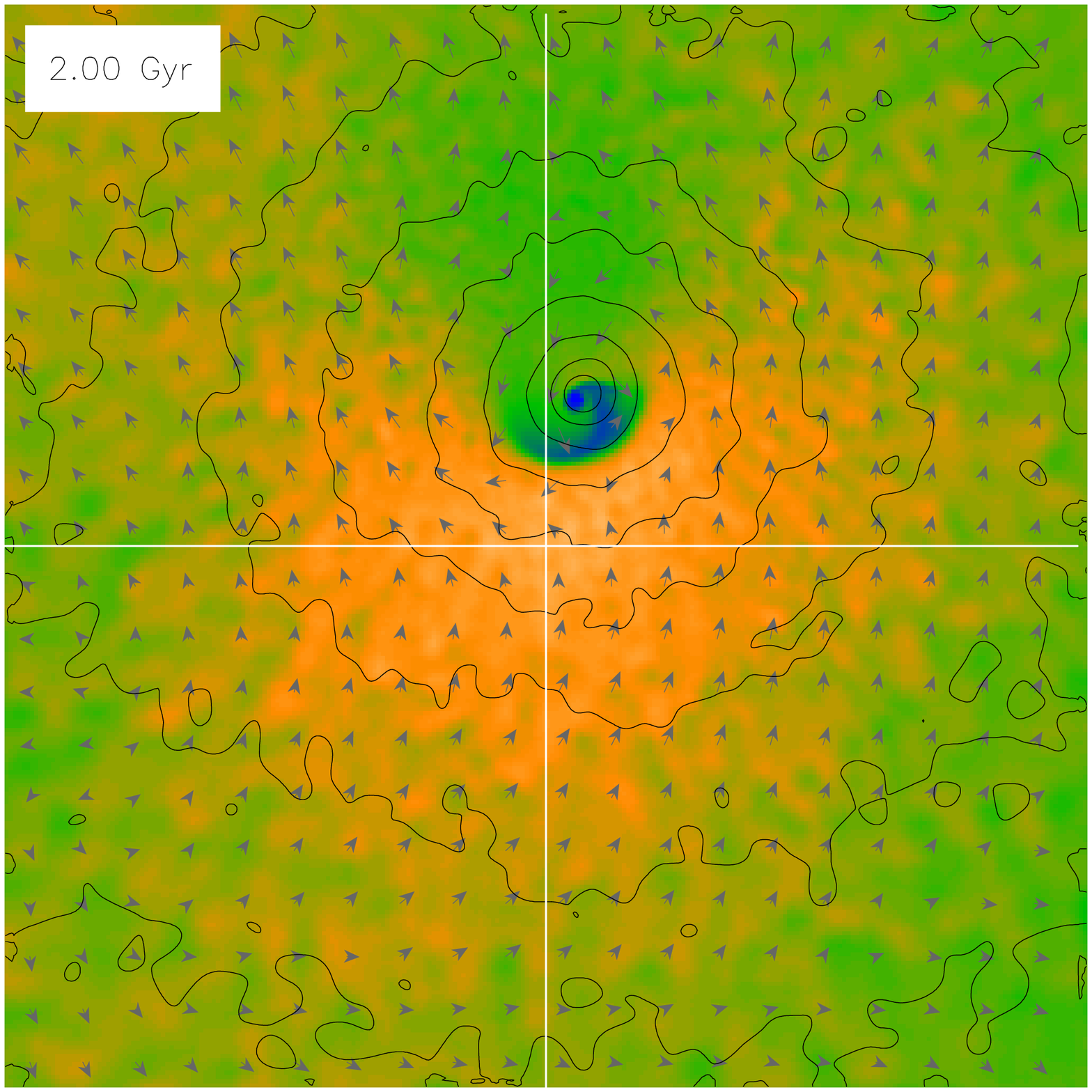}\\[1mm]
\includegraphics[width=0.32\textwidth]%
{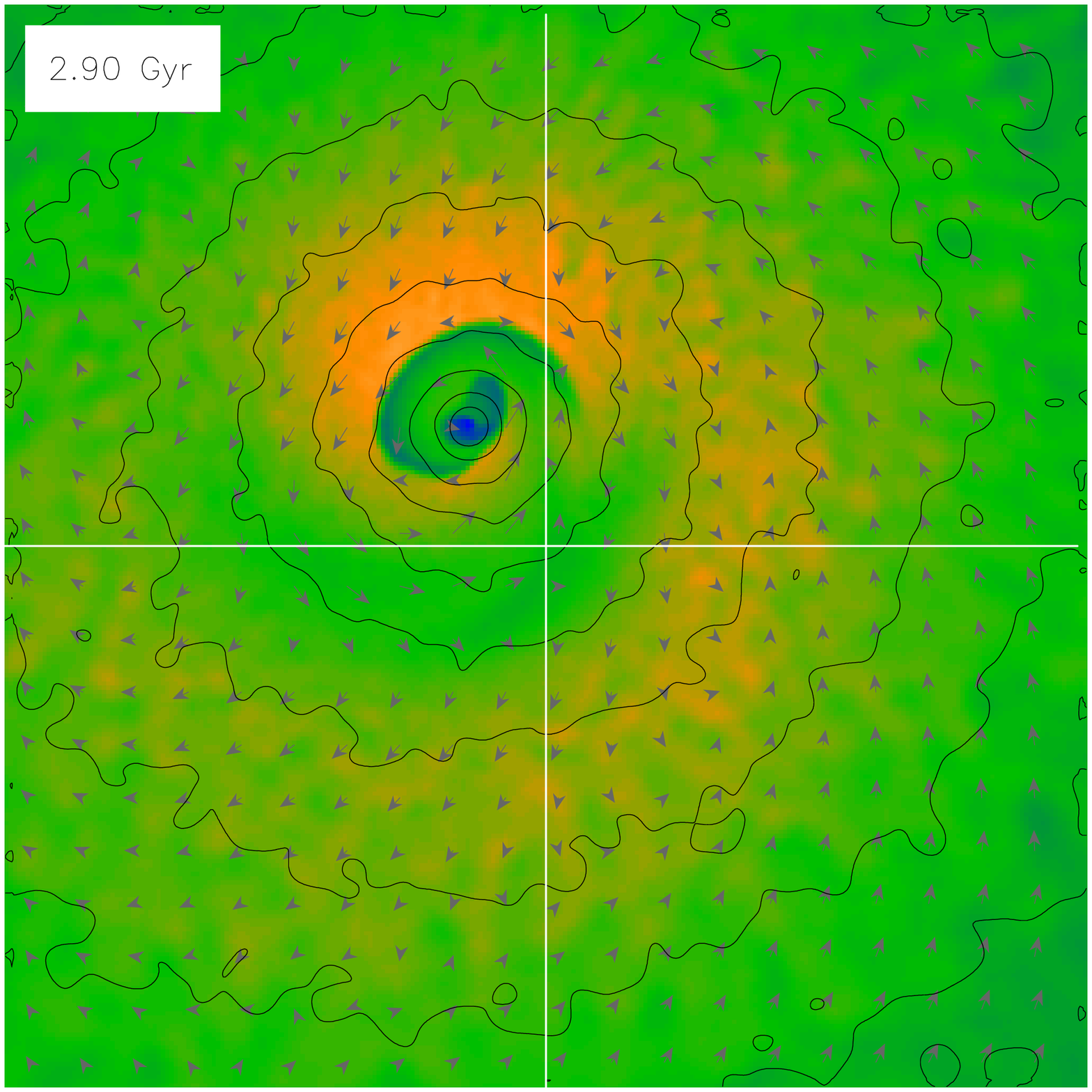}
\includegraphics[width=0.32\textwidth]%
{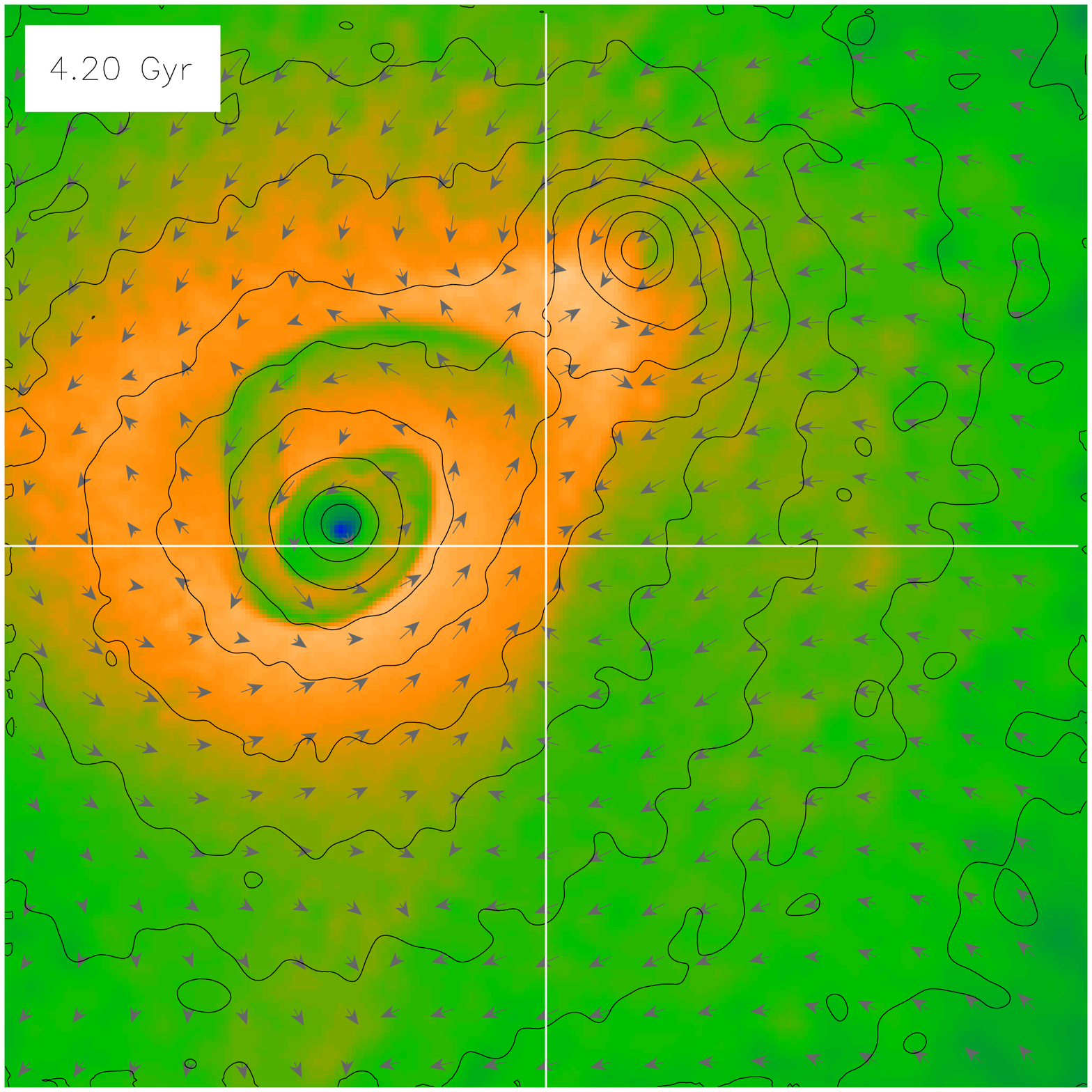}
\includegraphics[width=0.32\textwidth]%
{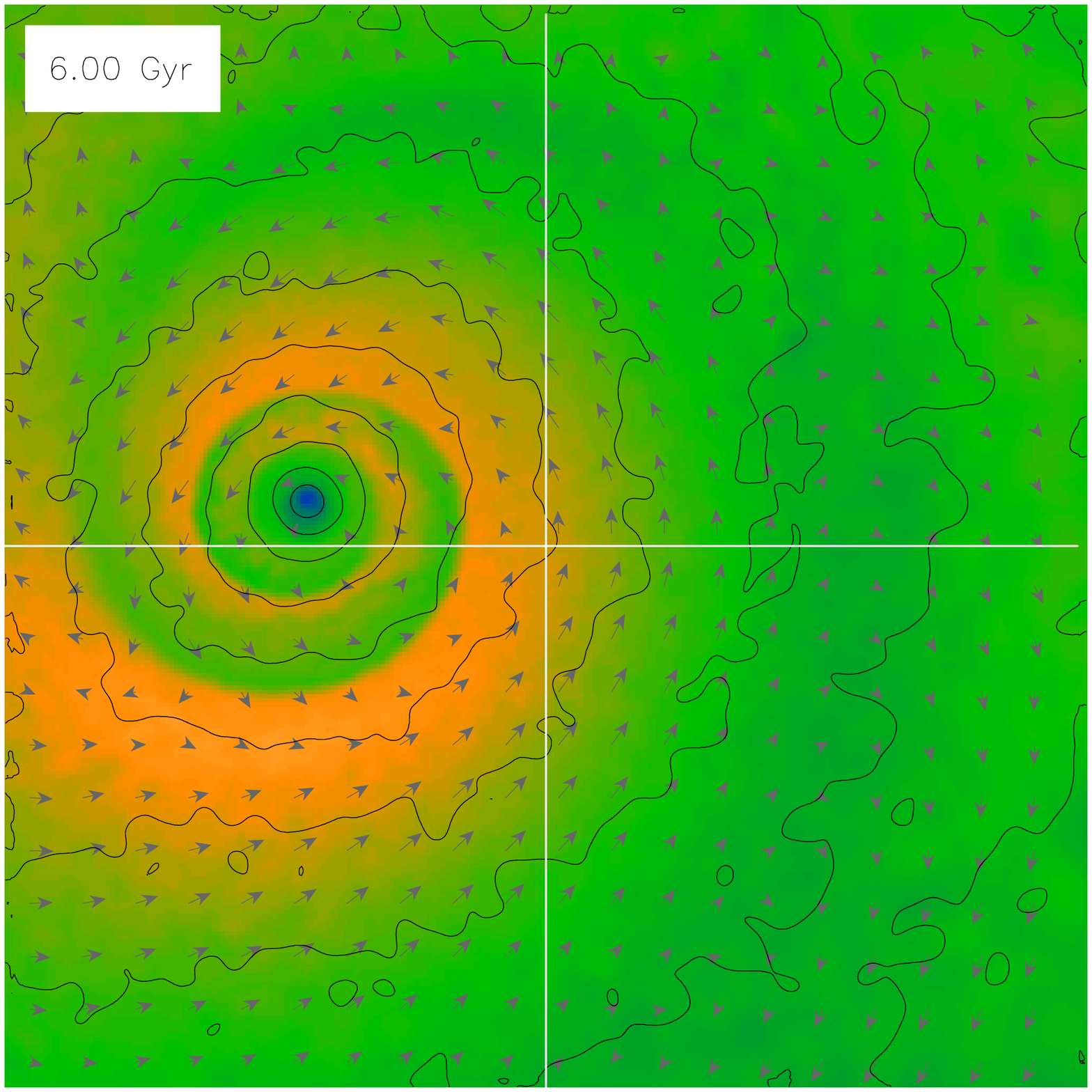}

\caption{
  Time sequence from a simulated infall of a small, purely dark matter
  satellite into a relaxed, cooling flow cluster.  Color shows the gas
  temperature (in keV) in the orbital plane.  Arrows represent the gas
  velocity field w.r.t.\ the main dark matter density peak. Contours show
  the dark matter density. The white cross shows the center of mass of the
  main cluster's particles.  The panel size is 1 Mpc. Such a merger induces
  gas sloshing in the center, but not much disturbance elsewhere, as will be
  seen in the next figure.  (Reproduced from A06.)}

\label{yago_dm}
\end{figure}

\begin{figure}[p]
\centering
\includegraphics[width=0.31\textwidth]%
{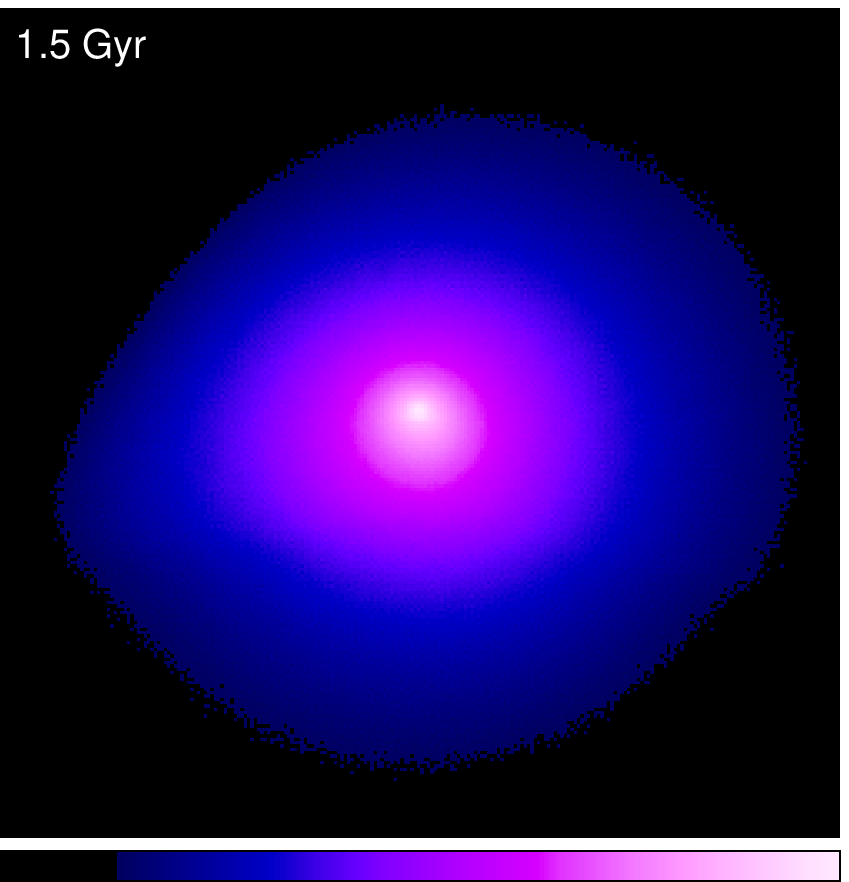}
\includegraphics[width=0.31\textwidth]%
{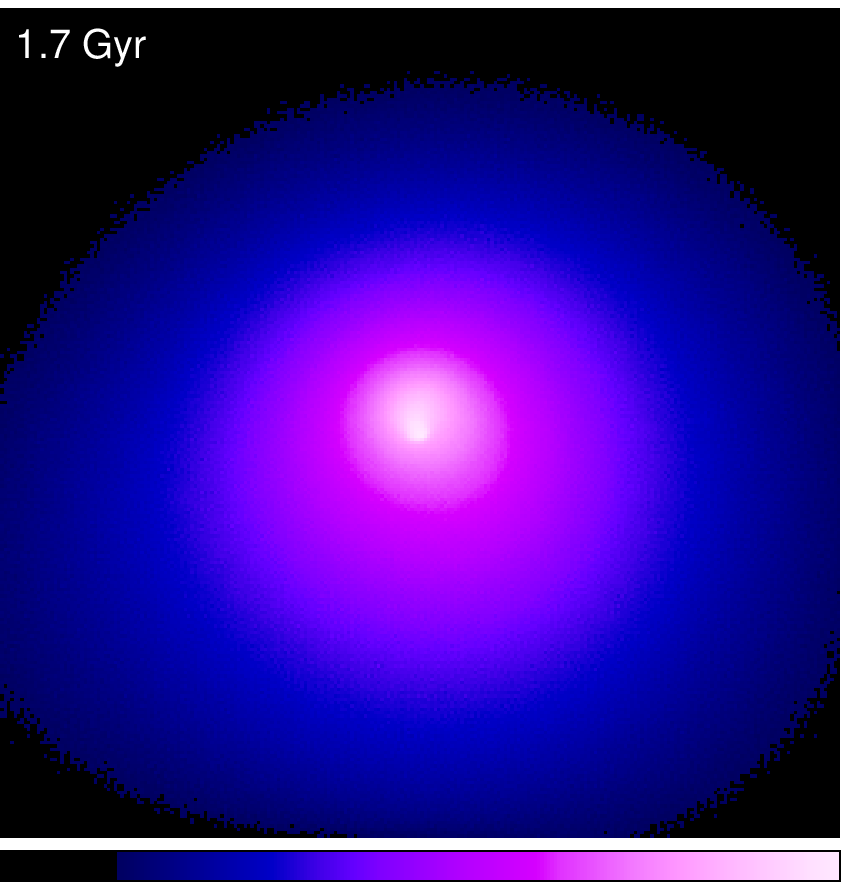}
\includegraphics[width=0.31\textwidth]%
{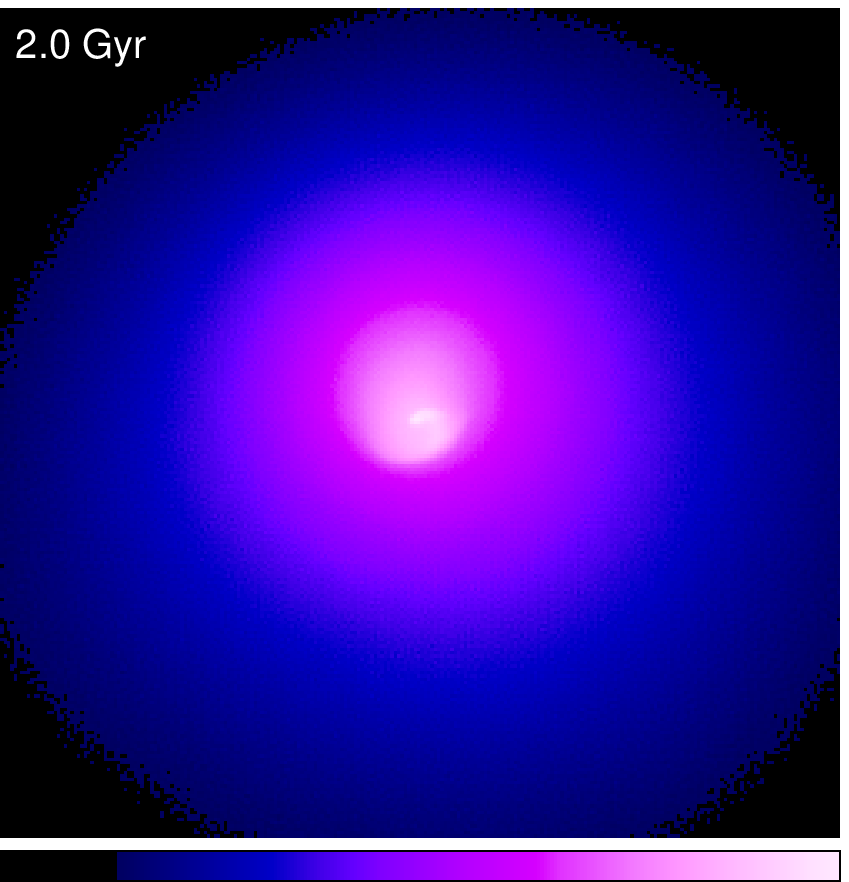}\\
\includegraphics[width=0.31\textwidth]%
{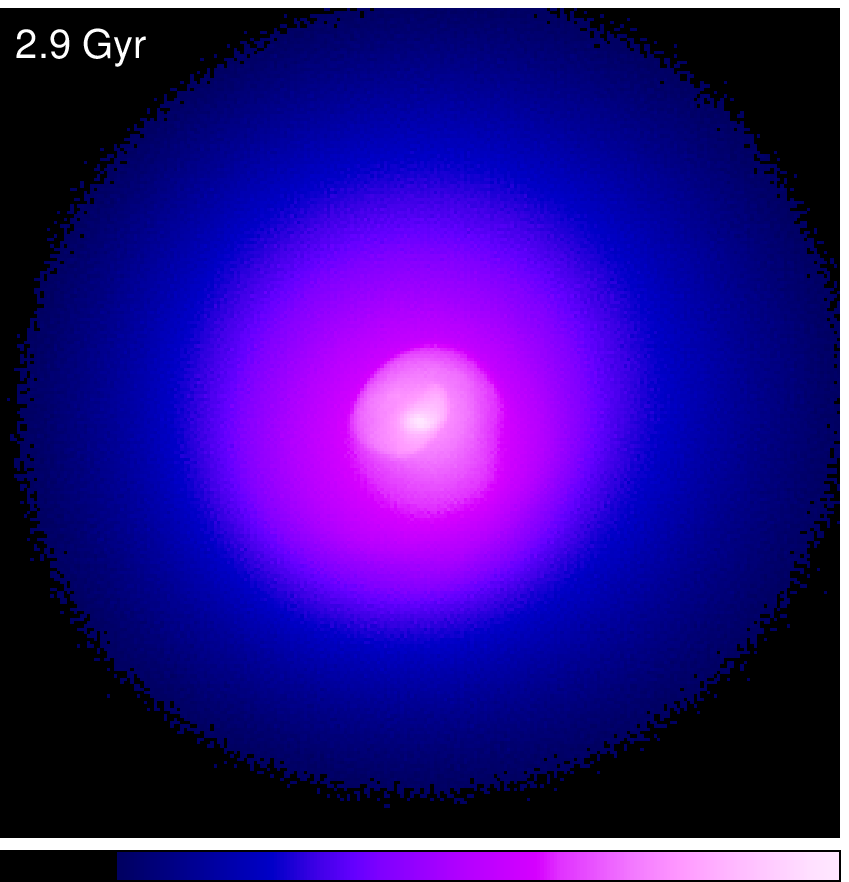}
\includegraphics[width=0.31\textwidth]%
{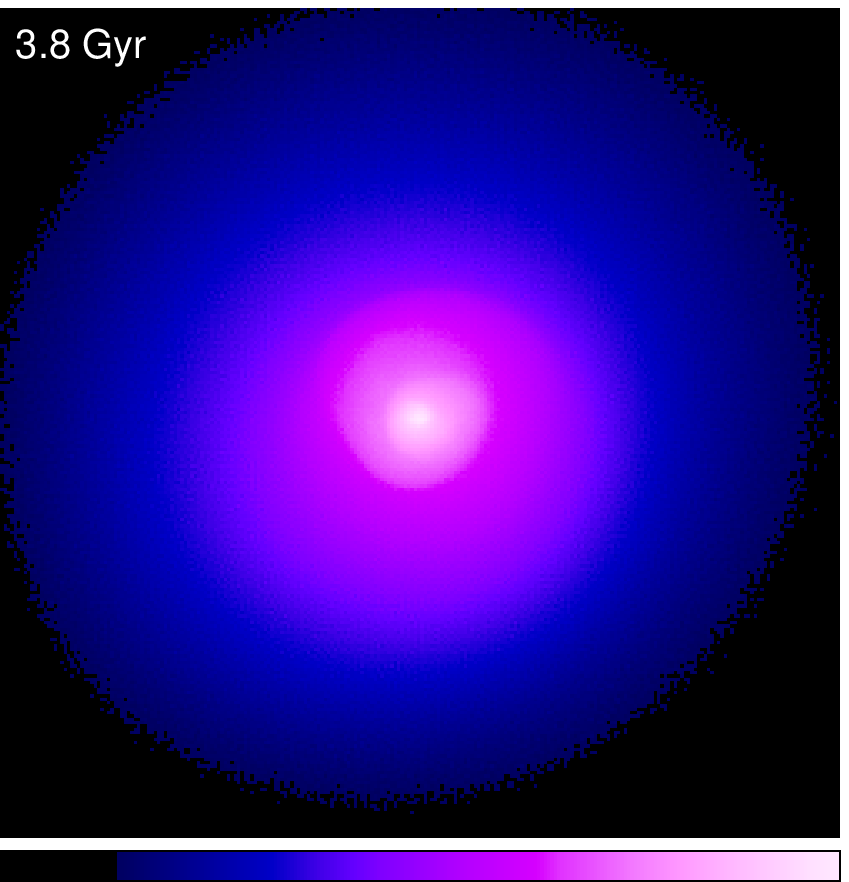}
\includegraphics[width=0.31\textwidth]%
{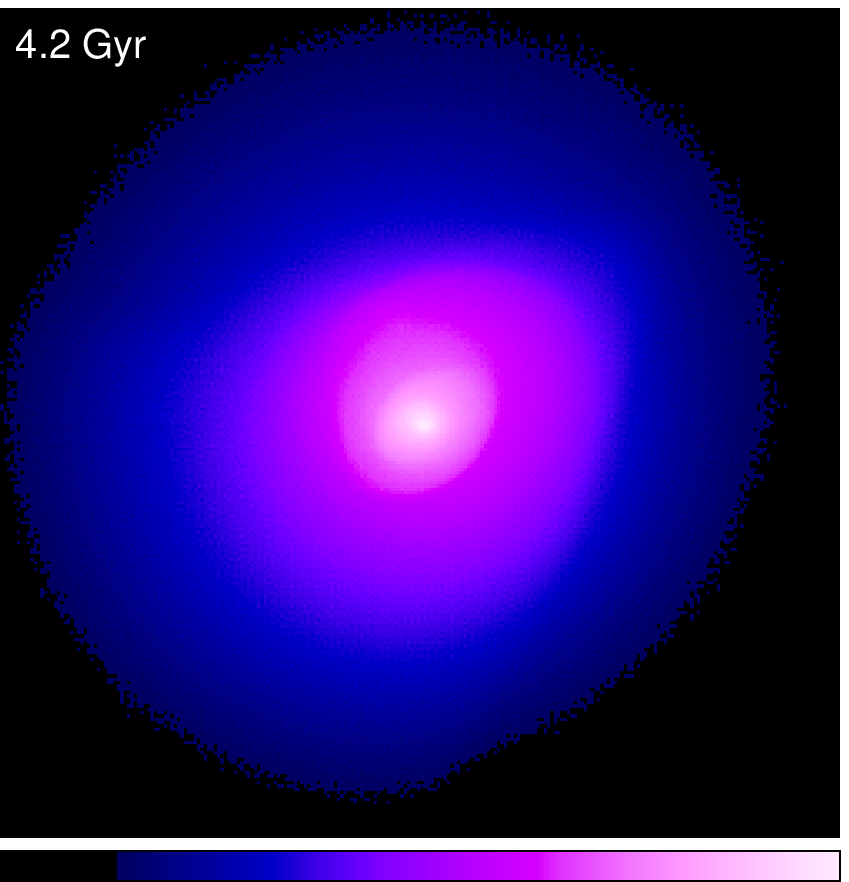}

\caption{
  Projected X-ray brightness for the merger shown in Fig.\ \ref{yago_dm}.
  The panel size is 1 Mpc. With the possible exception of short moments when
  the subcluster flyby generates a conical wake (at 1.5 Gyr and 4.2 Gyr),
  the cluster stays very symmetric on large scales; the only visible
  disturbance is cold fronts in the center.  (Reproduced from A06.)}

\label{yago_lx_dm}
\end{figure}

\begin{figure}[p]
\centering
\includegraphics[width=0.31\textwidth]%
{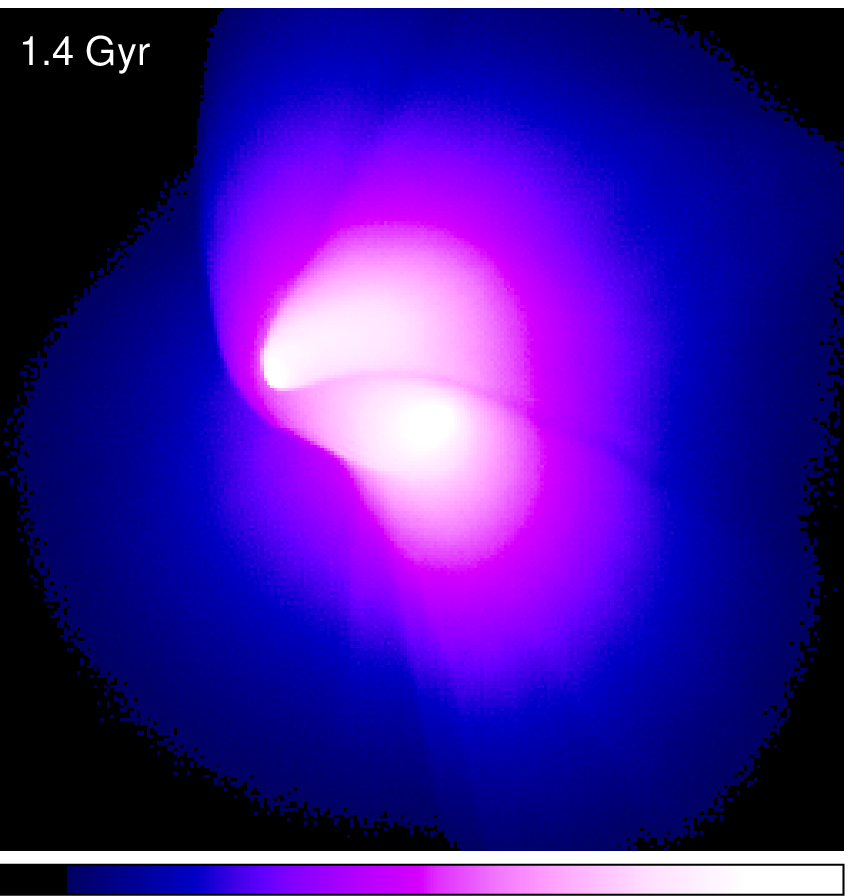}
\includegraphics[width=0.31\textwidth]%
{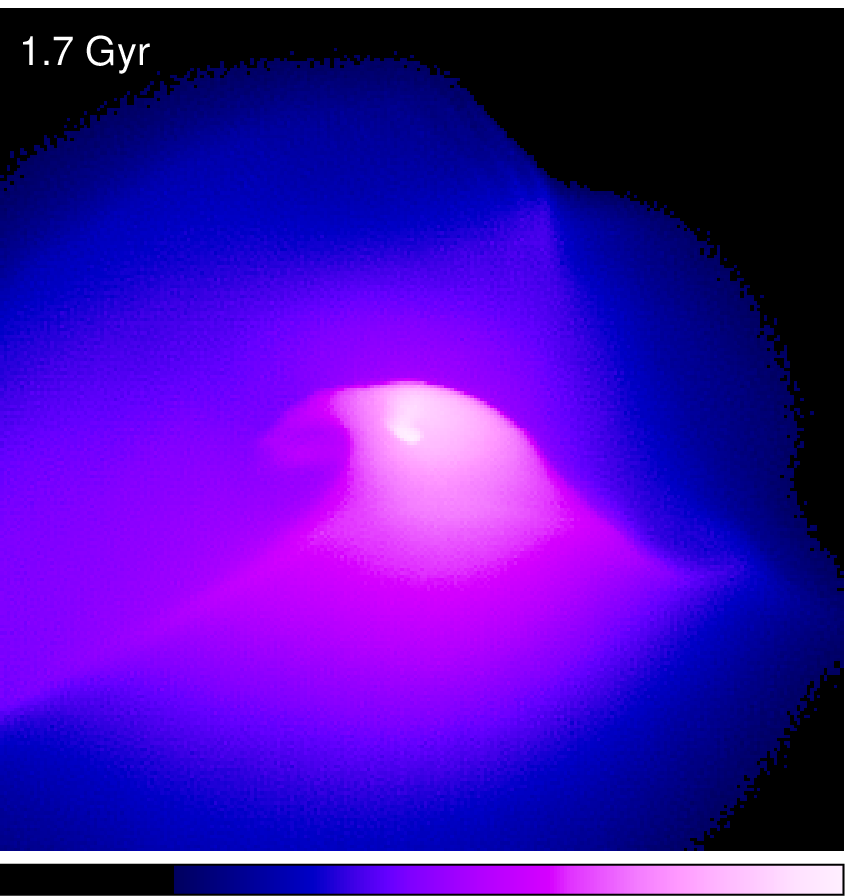}
\includegraphics[width=0.31\textwidth]%
{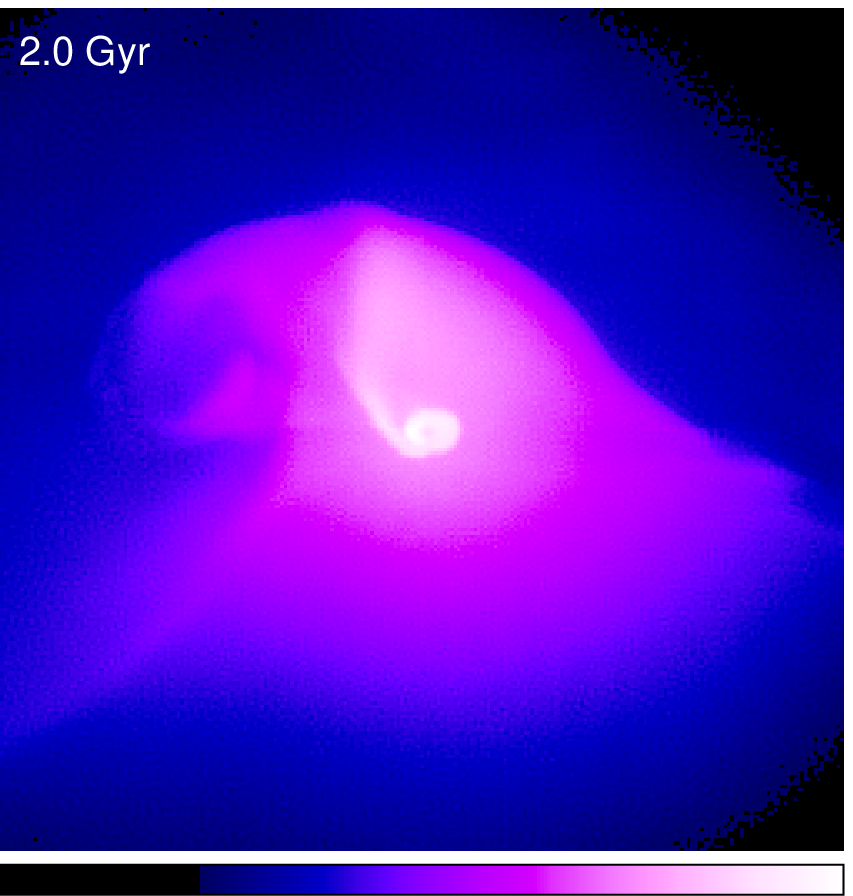}\\
\includegraphics[width=0.31\textwidth]%
{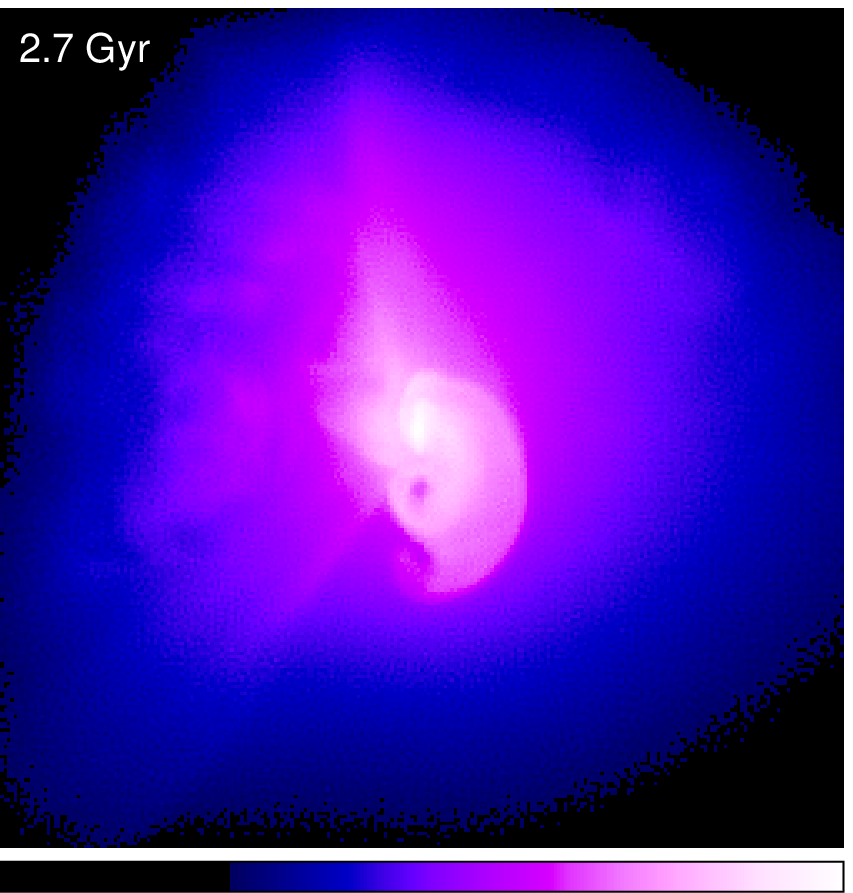}
\includegraphics[width=0.31\textwidth]%
{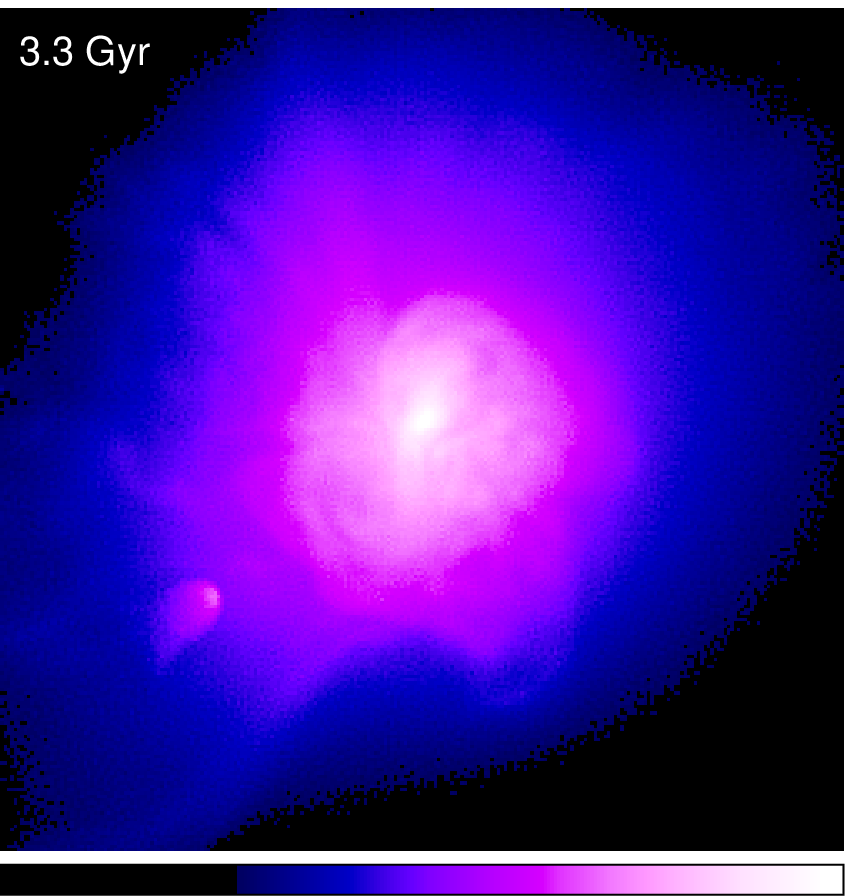}
\includegraphics[width=0.31\textwidth]%
{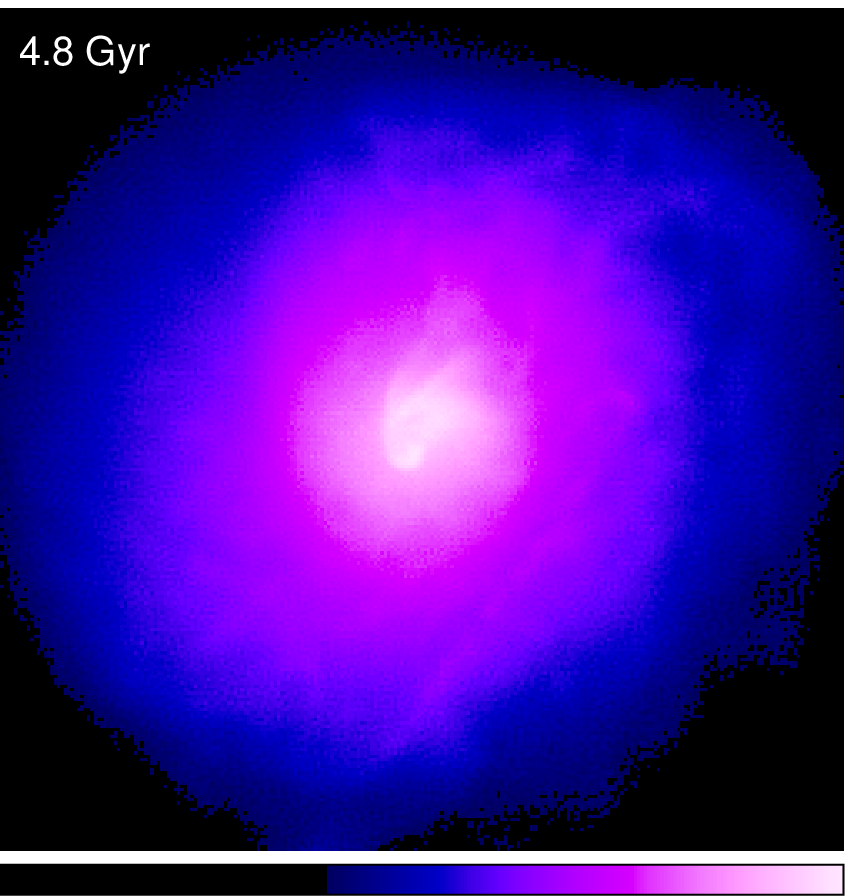}

\caption{
  Projected X-ray brightness for a merger similar to that in Figs.\ 
  \ref{yago_dm} and \ref{yago_lx_dm}, but now the subcluster has gas. Cold
  fronts still form in the center of the main cluster, but the system is
  very disturbed on all scales. It needs 3--4 Gyr after the first core
  passage to regain a symmetric appearance on large scales (last panel).
  (Reproduced from A06.)}

\label{yago_lx_gas}
\end{figure}

\subsubsection{Simulations of gas sloshing}
\label{sec:simslosh}

Several simulation works addressed the possibility that mergers can create
cold fronts in the cluster centers.  Churazov et al.\ (2003) and Fujita,
Matsumoto \& Wada (2004) used 2D simulations to show that a weak shock or
acoustic wave propagating toward the center of a cooling flow cluster can
displace the cool gas from the gravitational potential well and cause gas
sloshing, giving rise to cold fronts. On the other hand, Tittley \&
Henriksen (2005), using mergers extracted from a cosmological simulation,
suggested that cold fronts in the cores arise when the gas peak is dragged
along as the dark matter peak oscillates around the cluster centroid because
of a gravitational disturbance from a merging subcluster.

The most detailed simulation that addressed the specific question of
whether a merger can generate cold fronts in the cores, but no visible
disturbance elsewhere, was presented by A06. They found that sloshing is
indeed easily set off by any minor merger and can persist for gigayears,
producing concentric cold fronts just as those observed. The only necessary
(and obvious) condition for their formation is a steep radial entropy drop
in the gas peak, such as that present in all cooling flows.  Most
interestingly, fronts form even if the infalling subcluster has not had any
gas during core passage. This may occur if the gas was stripped early in the
merger, leaving a clump of only the dark matter and galaxies. It is mergers
with such gasless clumps that can set the central cool dense gas in motion,
while leaving no other long-lived visible traces in X-rays.

Figure \ref{yago_dm} (from A06) shows a time sequence from a simulated
merger with the dark matter-only subcluster whose mass is 1/5 of that of the
main cluster and which falls in with a nonzero impact parameter (or angular
momentum). The figure shows how the subcluster makes two passes, around 1.4
Gyr and 4.2 Gyr from the start of the simulation run.  During the first
pass, the gravitational disturbance created by the subhalo causes the
density peak of the main cluster (DM and gas together) to swing along a
spiral trajectory relative to the center of mass of the main cluster (white
cross in Fig.\ \ref{yago_dm}).  The gas and DM peaks feel the same gravity
force and initially start moving together toward the subcluster.  However,
during the core passage, the direction of this motion quickly changes,
leading to a rapid change of the ram-pressure force exerted on the gas peak.
Mainly as a result of this change, the cool gas core shoots away from the
potential minimum in a ram-pressure slingshot similar to the one described
above in \S\ref{sec:origin}. Subsequently, the densest gas turns around and
starts falling back toward the minimum of the gravitational potential (a
cuspy NWF mass profile has a well-defined sharp potential minimum). It then
starts sloshing around the DM peak and generating long-lived cold fronts, as
will be discussed below.

Mock X-ray images of this simulated merger (Fig.\ \ref{yago_lx_dm}) show
these cold fronts quite clearly, at the same time exhibiting very little
disturbance on the cluster-wide scale. The exception is several brief
periods when the DM satellite crosses the cluster and generates a subtle
conical wake (first and last panels in Fig.\ \ref{yago_lx_dm}), but the
subcluster spends most of the orbiting time in the outskirts.  This
simulation looks very much like the most relaxed cooling-flow clusters in
the real world, such as A2029 and A1795.

For comparison, Fig.\ \ref{yago_lx_gas} shows mock X-ray images from a
simulation of a similar merger, but with the gas subcluster.  Hydrodynamic
effects of the collision of two gas clouds are now overwhelming and the gas
is disturbed on all scales for a long time.  Sloshing and central cold
fronts are generated, too --- in fact, sloshing occurs on a wider scale,
because the initial displacement caused by the subcluster shock and stripped
gas is much stronger than that caused by a swinging motion of the DM peak.
There are many real clusters that look similar to this simulation.

\begin{figure}[p]
\centering
\includegraphics[width=0.45\textwidth]%
{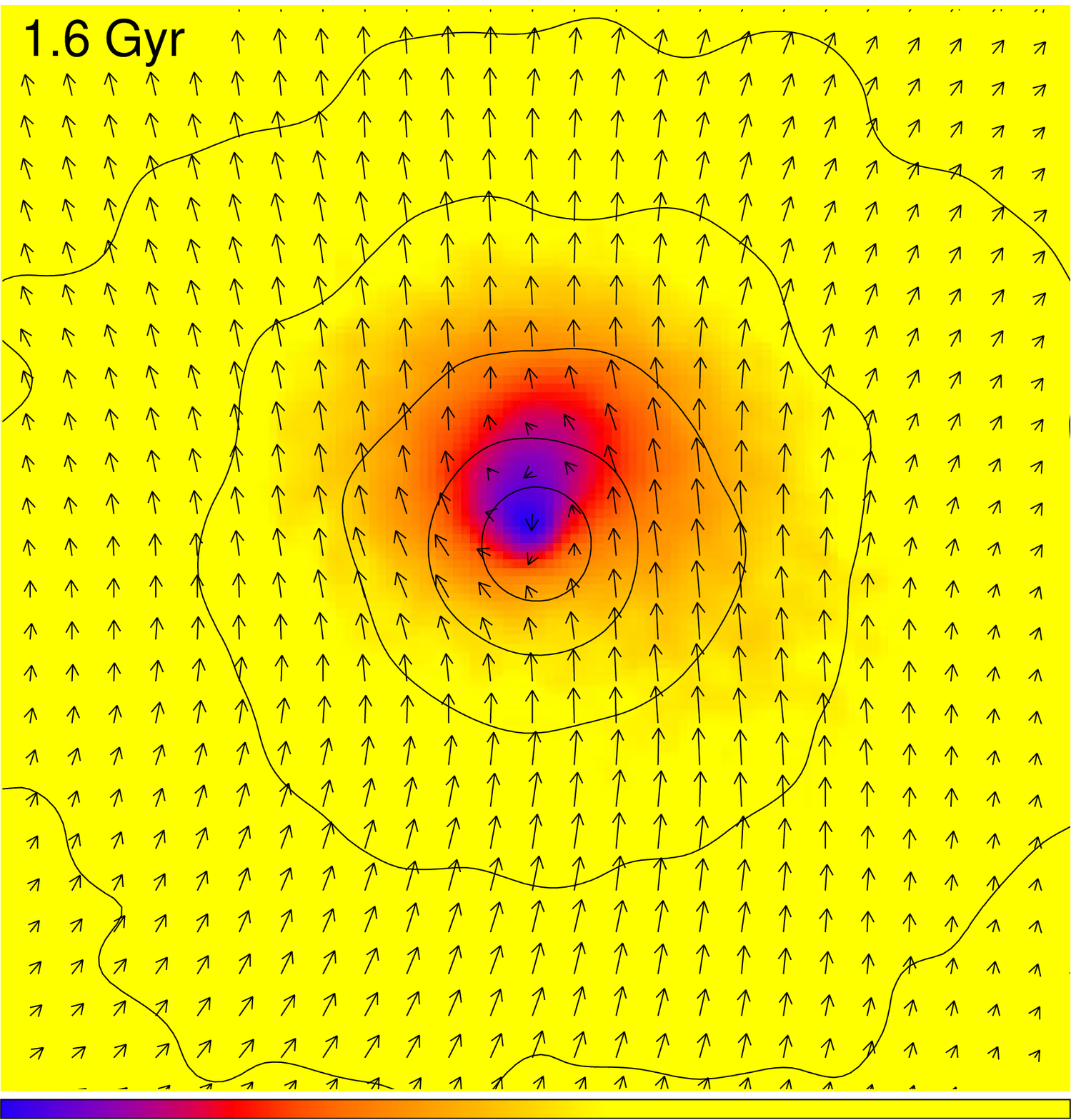}
\includegraphics[width=0.45\textwidth]%
{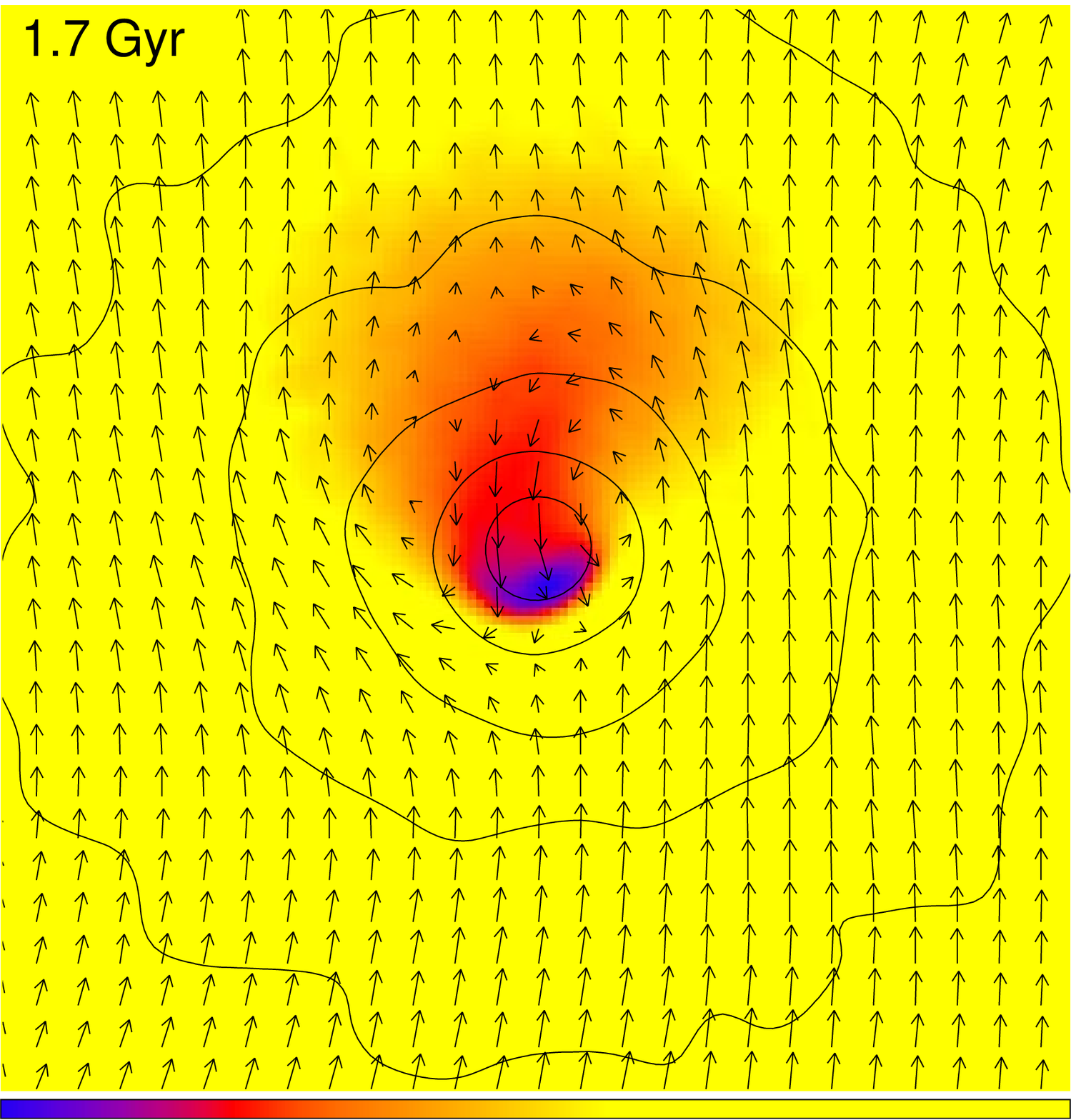}\\[1mm]
\includegraphics[width=0.45\textwidth]%
{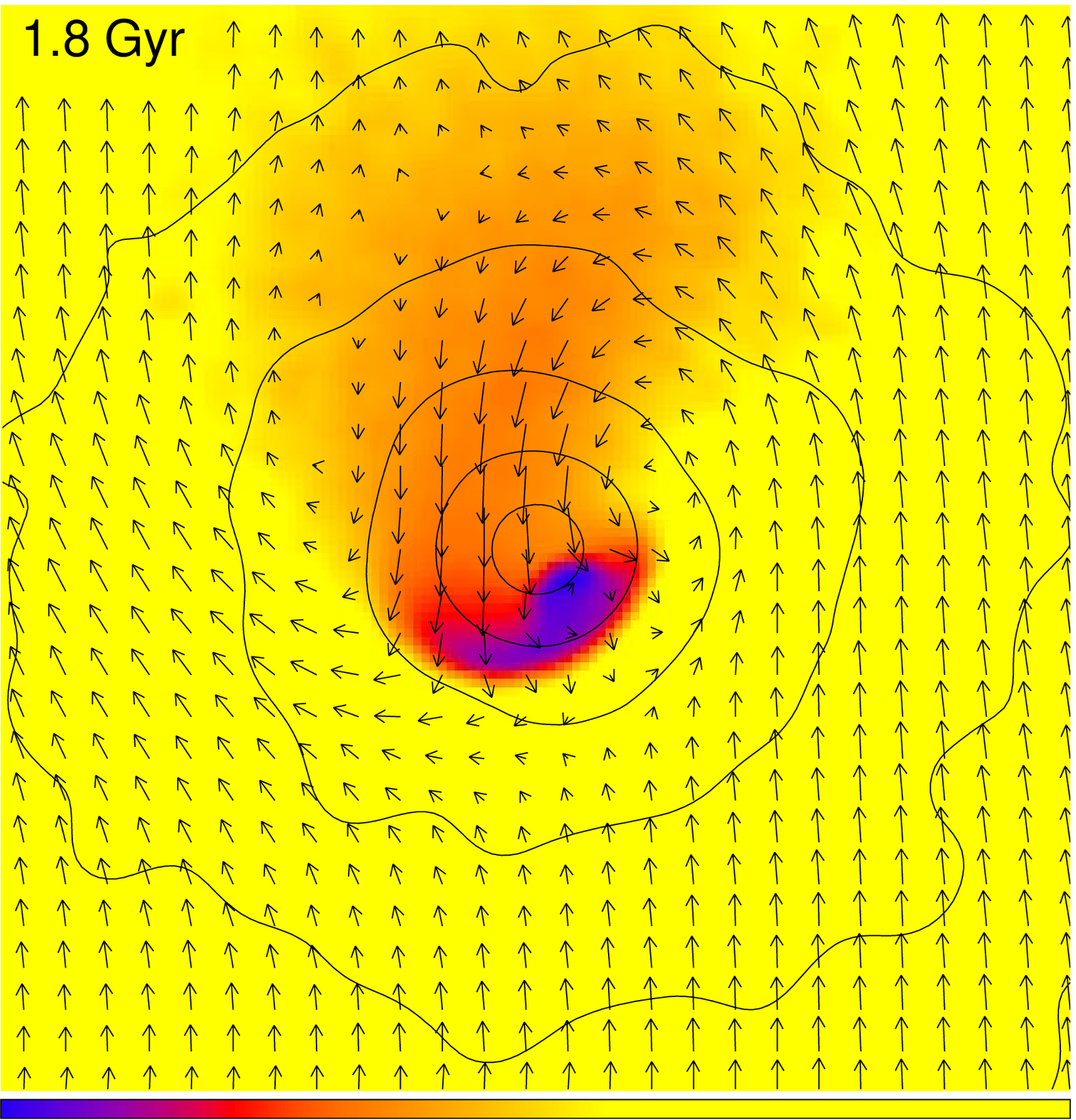}
\includegraphics[width=0.45\textwidth]%
{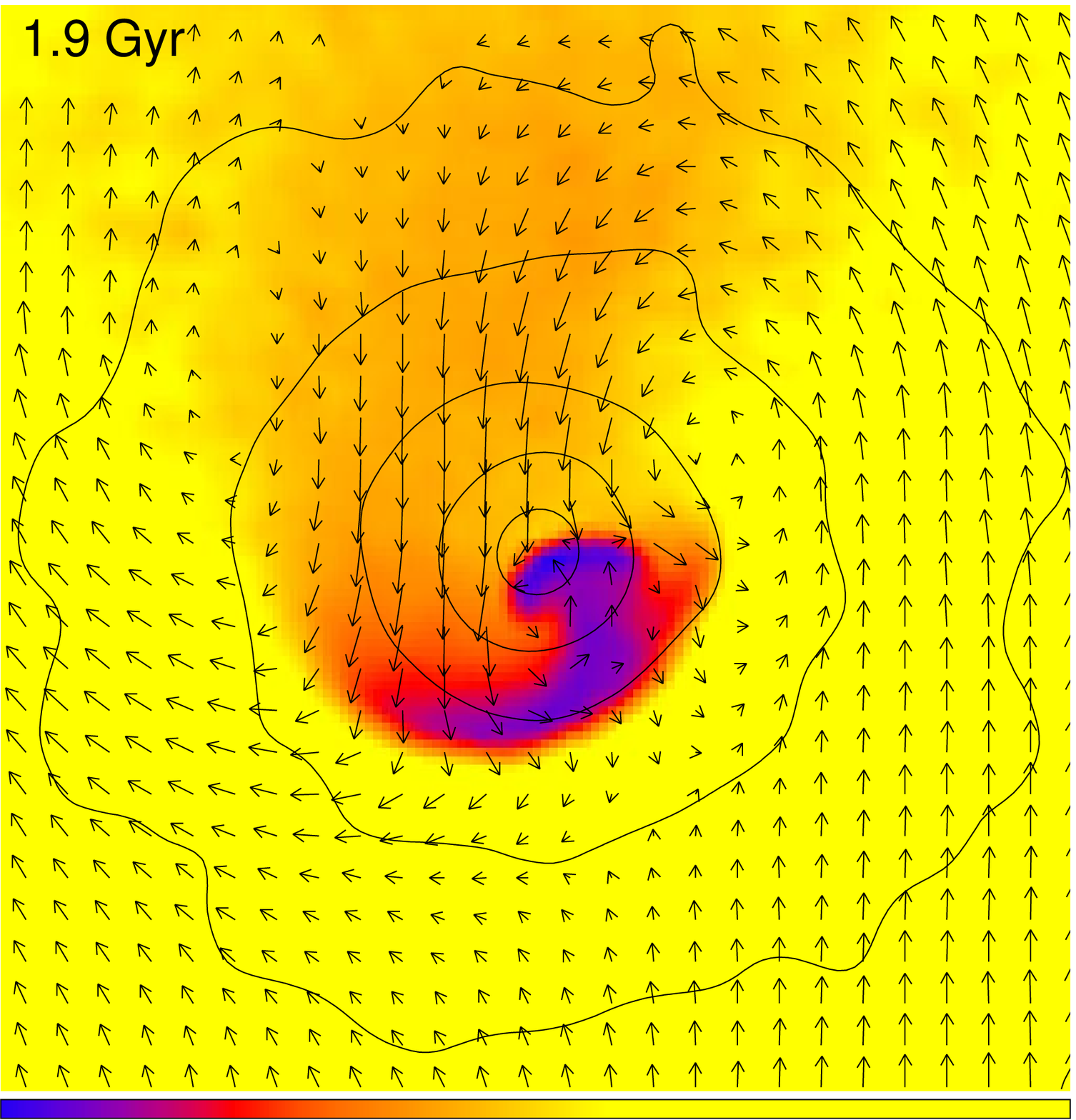}\\[1mm]
\includegraphics[width=0.45\textwidth]%
{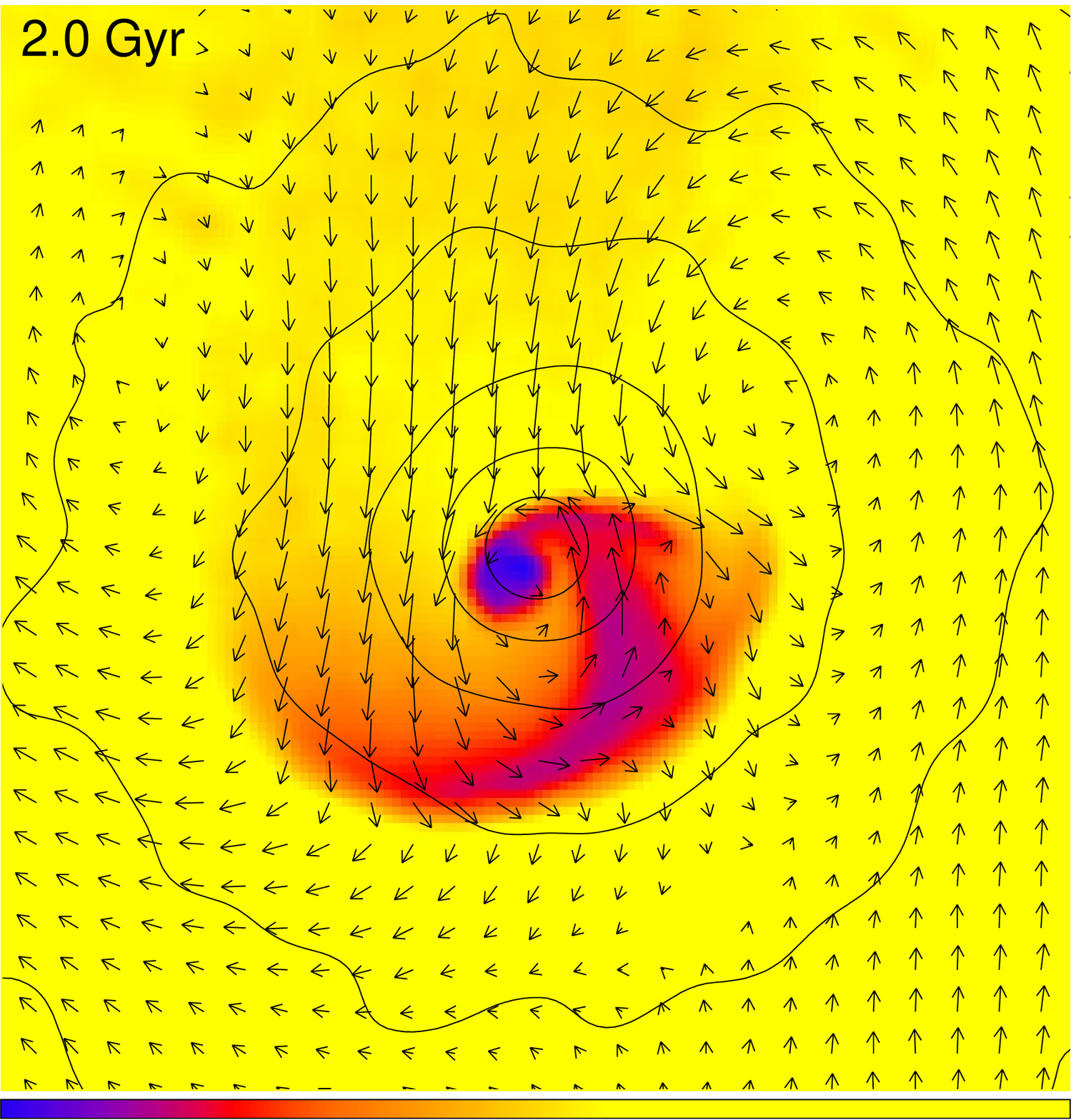}
\includegraphics[width=0.45\textwidth]%
{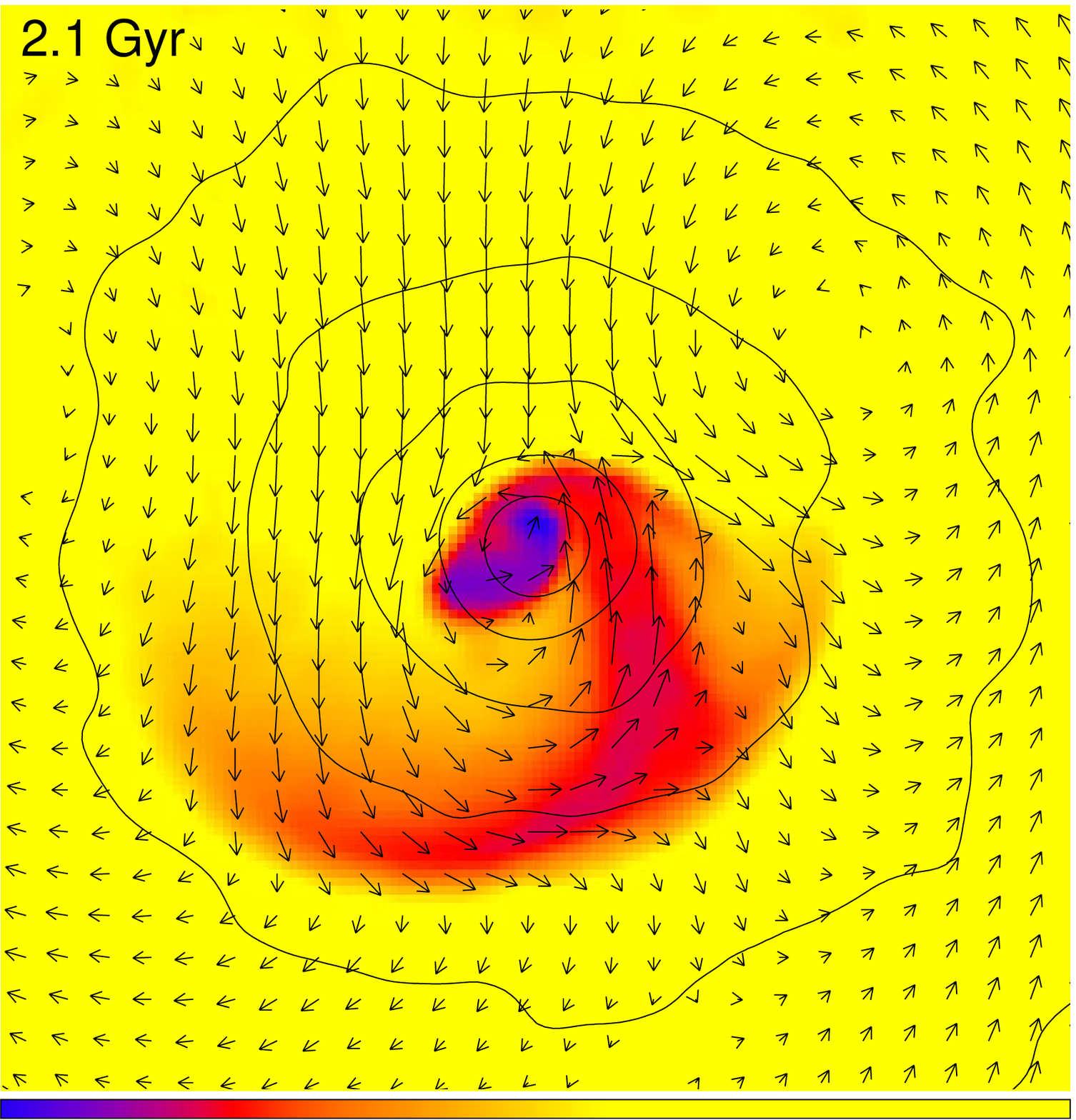}

\caption{
  A zoom-in view of the center of the main cluster in the merger simulation
  shown in Fig.\ \ref{yago_dm} (where the infalling subcluster did not have
  any gas).  The panel size is 250 kpc. Color shows gas temperature (between
  2--10 keV) in the merger plane; contours show dark matter density; arrows
  show gas velocities relative to the dark matter peak. One can see the
  onset of sloshing and the emergence of cold fronts.  (Reproduced from
  A06.)}

\label{yago_dm_zoom}
\end{figure}

\subsubsection*{The emergence of multiple cold fronts}

Let us now examine in details how the sloshing central gas gives rise to
cold fronts.  Figure \ref{yago_dm_zoom} presents a zoomed-in view of the gas
temperature and velocity field from the A06 simulation of a merger with the
gasless subcluster. The dark matter peak has a cuspy NFW density profile,
and the initial gas density and temperature profiles are similar to those
observed in cooling flow clusters, so this is what should happen in clusters
such as A2029.  The figure shows several snapshots following the initial
displacement of the gas density peak from the central potential dip.  The
displaced cool gas expands adiabatically as it is carried further out by the
flow of the surrounding gas (the orange plume above the center in the
1.6--1.7 Gyr panels of Fig.\ \ref{yago_dm_zoom}).  However, in a process
similar to the onset of a Rayleigh-Taylor (RT) instability, the densest,
coolest gas turns around and starts sinking towards the minimum of the
gravitational potential, as seen most clearly in the 1.6 Gyr and again in
the 1.8 Gyr snapshots.  The coolest gas overshoots the center at 1.7 Gyr
and, subjected to ram pressure from the gas on the opposite side still
moving in the opposite direction, eventually spreads into a characteristic
mushroom head with velocity eddies (von Karman vortices) at the sides. This
is a classic structure seen in numeric and real-life experiments involving a
gas jet flowing through a less dense gas (for cluster-related works see,
e.g., Heinz et al.\ 2003, our Fig.\ \ref{heinz}; Takizawa 2005). The
mushroom head forms where the dense gas is slowed by the ambient ram
pressure and spreads sideways into the regions of lower pressure created by
the flow of the ambient gas around a blunt obstacle (Bernoulli's law). The
front edges of these mushroom heads are sharp discontinuities, as confirmed
by a detailed look at these structures (A06).  We will discuss how exactly
they arise in \S\ref{sec:discont} below.

At 1.8 Gyr, one can see how the inner side of the first mushroom head
sprouts a new RT tongue --- the densest, lowest-entropy gas separates and
again starts sinking toward the potential minimum (compare this, e.g., with
a structure in the center of the Ophiuchus cluster in Fig.\ 
\ref{slosh_examples}). Meanwhile, the rest of the gas in the first mushroom
head continues to move outwards, expanding adiabatically as it moves into
the lower-pressure regions of the cluster, and roughly delineating the
equipotential surfaces. The process repeats itself, and these mushroom heads
are generated on progressively smaller linear scales. Sloshing of the
densest gas that is closest to the center occurs with a smaller period and
amplitude than that of the gas initially at greater radii (the reason is
explained in Churazov et al.\ 2003). It is this period difference that
eventually brings into contact the gas phases that initially were at
different off-center distances and had different entropies (recall that a
cooling flow gas profile has a sharp radial entropy gradient).

\begin{figure}[t]
\centering
\includegraphics[width=1\textwidth]%
{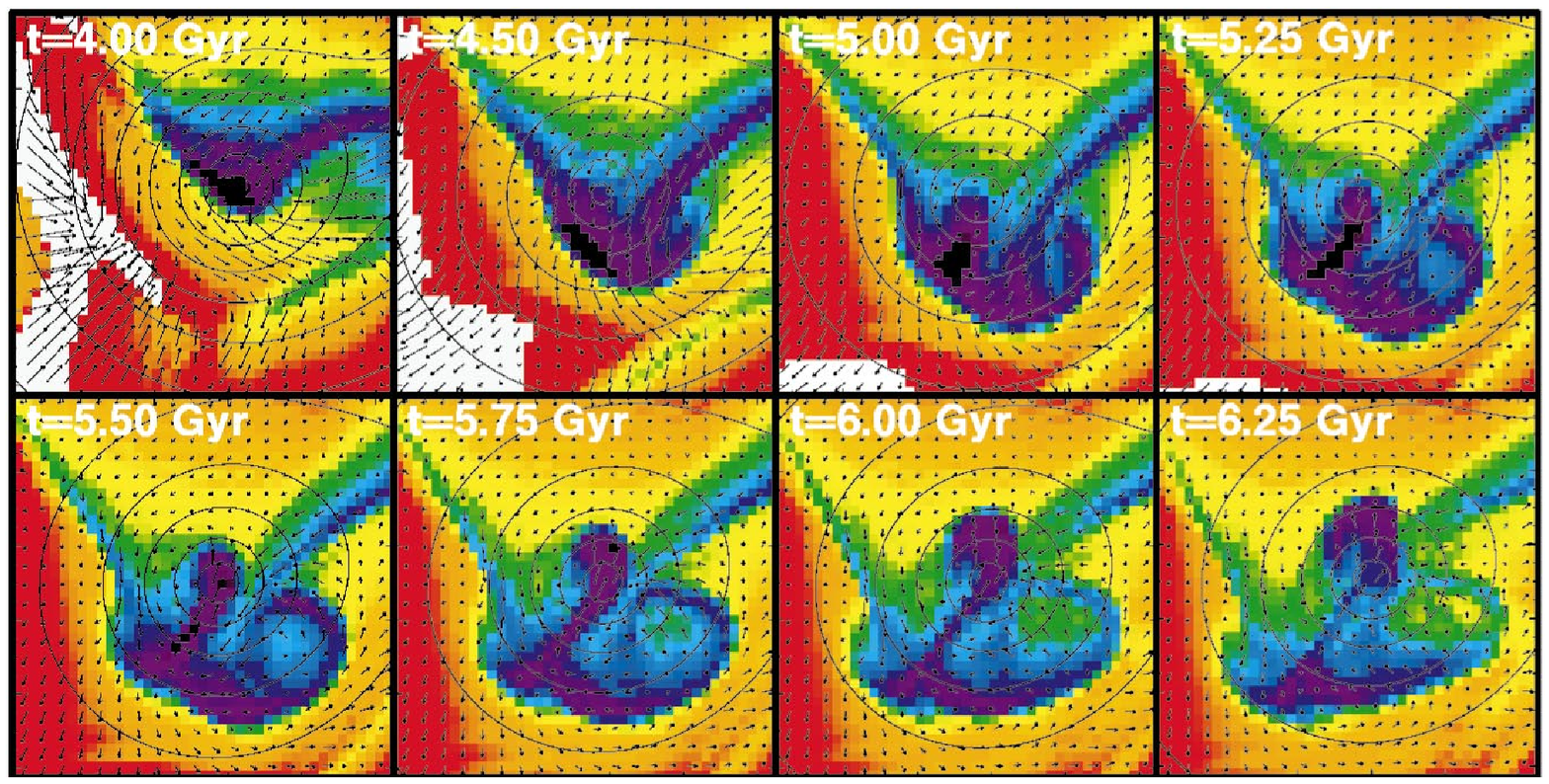}

\caption{Sloshing of the low-entropy central gas set off by an off-axis
  merger in an initially isothermal cluster with a cuspy dark matter
  profile. Color shows the gas specific entropy in the merger plane
  (increasing from black to white), contours show the gravitational
  potential.  Panel size is 1.5 Mpc.  One can see the ram pressure slingshot
  (the low-entropy gas is pushed to the upper-right side and then shoots
  back), followed by the development of a convective plume flowing toward
  the center. The resulting edges in the entropy map are analogs of the cold
  fronts seen in Fig.\ \ref{yago_dm_zoom}.  (Reproduced from Ricker \&
  Sarazin 2001.)}

\label{ricker_slosh}
\end{figure}

The picture is qualitatively similar even if there is no cooling flow-like
temperature drop --- as long as there is entropy gradient.  Ricker \&
Sarazin (2001) simulated a merger in an isothermal cluster with a cuspy
potential, which shows the formation of similar multiple mushroom heads
(Fig.\ \ref{ricker_slosh}). The specific entropy declines toward the center,
but less steeply than in a cooling flow cluster, hence the linear scale of
the resulting sloshing is bigger. (Another reason for that is this merger
involves a gas-containing subcluster, so the initial disturbance was
greater.)

Note that the oscillation of the DM peak caused by the subcluster flyby has
a much longer period, of order 1 Gyr (Fig.\ \ref{yago_dm}), than the $\sim
0.1$ Gyr timescale of gas sloshing. Indeed, as seen in Fig.\ 
\ref{yago_dm_zoom}, the DM distribution in the core stays largely centrally
symmetric, while the gas sloshes back and forth in its potential well. The
former timescale is determined by the subcluster masses and impact
parameter, while the latter is determined by the gas and DM profiles at the
main peak (Churazov et al.\ 2003). Thus, sloshing is mostly a hydrodynamic
effect in a quasi-static central gravitational potential (cf.\ Tittley \&
Henriksen 2005), although in the long term, the DM peak oscillation can feed
additional kinetic energy to the sloshing gas. For this reason, it is
unlikely that looking at the pattern of cold fronts in the center of the
cluster, one will be able to determine, for example, the mass and impact
parameter of the subcluster. It may be possible to get an upper limit on the
time since the disturbance if the velocity of the sloshing gas can be
determined (\S\ref{sec:vel}).

\paragraph*{Long-term evolution of a cold front}

The A06 simulations further showed that, although the lowest-entropy gas
indeed oscillates back and forth in the potential minimum, a cold front,
once formed, always propagates outward from the center, and does not ``turn
around'' with the gas or ``straighten out'' (Figs.\ \ref{yago_dm} and
\ref{yago_dm_zoom}).  This is somewhat counterintuitive, because it has to
be difficult to move the low-entropy central gas out to large radii against
convective stability in the radially increasing entropy profile.  But in
fact, the central gas does not move out to large radii.  In later panels of
Fig.\ \ref{yago_dm_zoom}, one can discern a flow pattern inside the cold
front in which the lowest-entropy gas initially forming the front, turns
around and sinks back towards the center. It is replaced at the front by
higher-entropy gas that arrives later and whose origin traces back to larger
radii. In other words, the cold front as a geometric feature moves out, but
the low-entropy gas stays close to the center of the potential.

\paragraph*{Spiral pattern}

Finally, Fig.\ \ref{yago_dm} reveals a curious spiral pattern that the
central cold fronts develop with time. A similar spiral structure (if not so
well-developed) is seen in the X-ray image of A2029 (Fig.\ 
\ref{slosh_examples}; Clarke et al.\ 2004) and in the temperature map of
Perseus (Fabian et al.\ 2006, discussed in A06).  The simulated merger in
A06 (along with most real-life mergers) has a nonzero impact parameter.  So
when the cool gas is displaced from the center for the first time, it
acquires angular momentum from the gas in the wake and does not fall back
radially.  As a result, the subsequent cold fronts of different radii are
not exactly concentric, but combine into a spiral pattern (Fig.\ 
\ref{yago_dm_zoom}).  Initially, it does not represent any coherent
spiraling motion --- each edge is an independent structure.  However, as the
time goes by and the linear scale of the structure grows, circular motions
that are against the average angular momentum subside, and the ``mushrooms''
become more and more lopsided. On large scales, the spiral does indeed
become a largely coherent spiraling-in of cool gas --- the mushroom stems,
through which the low-entropy gas flows from one mushroom cap toward the
smaller-scale mushroom cap, shift more and more to the edge of the cap.

As a side note, the spiraling-in central gas should have the same direction
of the angular momentum as the infalling subcluster. Thus, looking at the
brightness peak in A1644 (Fig.\ \ref{a1644}), we can immediately say that
the subcluster must have passed it on the eastern side.  Indeed, Reiprich et
al.\ (2004) conclude the same from their analysis of the temperature and
abundance distributions obtained with \xmm.

\begin{figure}[b]
\centering
\includegraphics[width=0.85\textwidth, bb=20 280 590 576, clip]%
{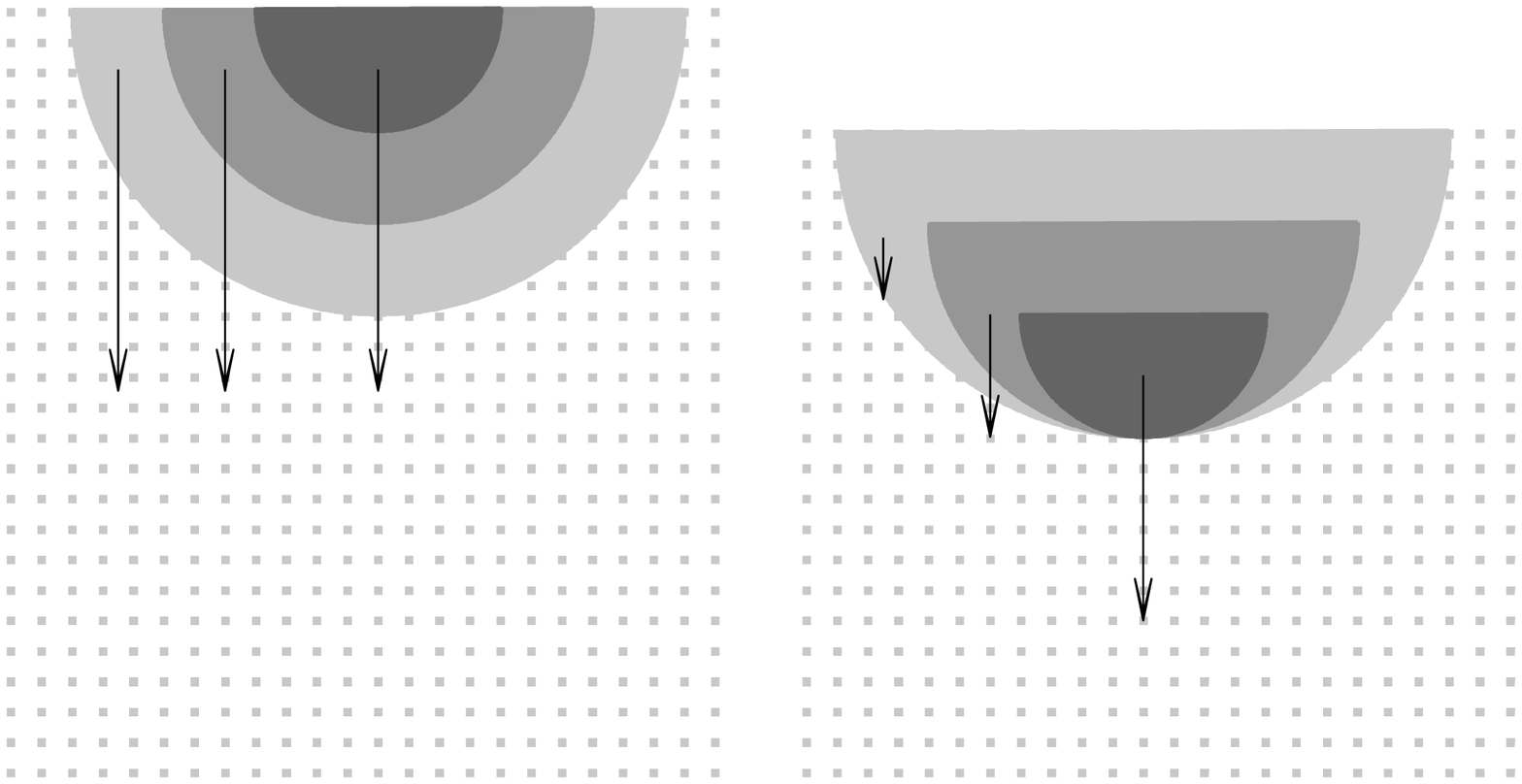}

\caption{The emergence of a density discontinuity from an initially
  continuous density distribution. The area-proportional drag force causes
  different deceleration of gases with different densities, which eventually
  brings the densest gas forward.}

\label{slosh_scheme}
\end{figure}

\subsubsection{Origin of density discontinuity}
\label{sec:discont}

While cold fronts may be caused by different events in the cluster, the
density discontinuities in them form for the same basic reason, which is
worth a clarifying aside. Simulations show that whenever a gas density peak
encounters a flow of ambient gas, a contact discontinuity quickly forms.
This occurs even when the initial gas distribution was perfectly smooth (no
shocks, etc.), as in a merger with a pure dark matter subcluster considered
above.  Stripping by a shear flow is usually quoted as the cause of the
discontinuity (e.g., M00; V01).  Indeed, the gas pressure immediately inside
the cold front in A3667 was found to be equal to the pressure of the outer
gas everywhere along the front, if one models the velocity field around the
spherical front and uses the Bernoulli equation (Vikhlinin \& Markevitch
2002, hereafter V02; see \S\ref{sec:RT} below). This suggests that the
outer, less dense layers of the subcluster's gas are quickly removed until
the radius is reached where the pressure in the cold gas equals the pressure
outside.  It is easy to imagine how a shear flow would strip the
subcluster's gas at the sides of the front. However, at the forward tip of
the front (the stagnation point), there is no shear flow for symmetry
reasons, but the fronts are just as sharp.

A simple reason for the emergence of a discontinuity at the stagnation point
is illustrated in Fig.\ \ref{slosh_scheme}. When an initially smooth
spherical density peak starts moving w.r.t.\ the surrounding gas, it starts
experiencing ram pressure, which creates roughly the same
(area-proportional) net force for each cubic centimeter of the gas in the
core (near the symmetry axis and assuming subsonic motions).  Denser gas
experiences smaller resulting acceleration. This produces a velocity
gradient inside the core along the direction of the force.  The
lower-density, outer layer of the core gas is then squeezed to the sides,
and the ambient gas eventually meets the dense gas for which the
forward-pulling, density-proportional gravity force (as in the bullet
cluster) or inertial force (as in the sloshing central gas) prevails over
the area-proportional ram pressure force. The initial density peak has to be
sufficiently sharp to ensure that the compressed intermediate gas does not
decelerate the denser gas behind it before being squeezed to the sides, a
condition which appears to be easily satisfied in real clusters.  Thus, a
contact discontinuity at the stagnation point forms by ``squeezing out'' the
gas layers not in pressure equilibrium with the flow.  Of course, stripping
by the shear flow does occur away from the axis of symmetry of the cold
front. For an illustration based on simulations, see Fig.\ 23 in A06.

\subsubsection{Effect of sloshing on cluster mass estimates}

Any motion of gas in the cluster core obviously represents a deviation from
hydrostatic equilibrium and thus poses a problem for the derivation of the
total masses based on this assumption (eq.\ \ref{eq:mass}). As we saw above
(A06), in clusters that may be perfectly relaxed outside their cool cores,
the central low-entropy region is easily disturbed and may not subsequently
come to equilibrium for a long time.  Unfortunately, ``relaxed'' clusters
almost always have those easily disturbed cooling flow regions.

How this may affect the hydrostatic mass estimates is illustrated by the
example of A1795 summarized above (Fig.\ \ref{a1795_profs}).  Using the gas
profiles from the sector containing the front, the total mass within the
edge radius was underestimated by a factor of 2.  If one uses the radial
profiles averaged over the full 360\deg\ azimuth, the effect is diluted; on
the other hand, the edge in A1795 is relatively small.  Pending a more
quantitative analysis of this issue (e.g., emulating the hydrostatic mass
estimates for the simulated clusters with sloshing), we can estimate roughly
that masses within the cooling flow regions can be underestimated by up to a
factor of 2. (The average result should always be an underestimate, since
the cool gas is gravitationally bound but has a mechanical component to its
total pressure, which we omit by measuring only the thermal component.)
Recall that even if a cooling flow cluster does not exhibit cold fronts,
statistically, it is likely to have one (or more) hidden by projection.  To
keep this in proper perspective, the radii of the cooling flow regions,
$r\lax 100$ kpc, contain only a few percent of the cluster total mass, so
this systematic mass error is relevant only for a narrow range of studies,
such as the exact shapes of the central dark matter cusps in clusters, or
comparison of X-ray derived masses with those from strong gravitational
lensing.

\subsubsection{Effects of sloshing on cooling flows}

\xmm\ and \chandra\ observations have not found the amounts of cool gas in
the centers of cooling flow clusters predicted by simple models based on the
cluster X-ray brightness profiles (see Peterson \& Fabian 2006 for a
review).  This means that that there has to be a steady energy supply to
compensate for the (directly observed) radiative cooling.  Several
mechanisms have been proposed; the currently favored view is that AGNs,
found in most cD galaxies in the centers of cooling flows, provide the
heating via the interaction of AGN jets with the ICM (e.g., Voit \& Donahue
2005; Fabian et al.\ 2005). A difficulty of this mechanism is that heating
has to be steady and finely tuned (to avoid blowing up the entire cluster
core), whereas AGNs have different powers in different clusters, and some
clusters do not even have a presently active central AGN. In the latter
clusters, other heating mechanisms may be needed for the cooling flow
suppression. One of the possible alternatives is sloshing, which may have
two effects.  First one is obvious --- M01 estimated that the mechanical
energy in the sloshing gas in A1795 is around half of its thermal energy (an
estimate for an analogous edge in Perseus is 10--20\%, Churazov et al.\ 
2003). As the gas sloshes, this energy is converted into heat at a steady
rate.

Another effect is more interesting and possibly more significant. As seen in
Fig.\ \ref{yago_dm_zoom}, sloshing brings hot gas from outside the cool core
into the cluster center, where it comes in close contact with the cool gas
that oscillates with a different period, as discussed in
\S\ref{sec:simslosh}.  Provided that the two phases can mix, this should
result in heat inflow from the large reservoir of thermal energy in the gas
outside of the cool core.  A classic electron heat conduction was proposed
to tap that reservoir, but was shown to be insufficient to balance the
cooling (Voigt \& Fabian 2004 and references therein), mainly because of the
strong temperature dependence of this process. However, a ``heat
conduction'' caused by the above mixing may be an attractive mechanism
(Markevitch \& Ascasibar, in preparation).

\subsubsection{Effect of sloshing on central abundance gradients}

Heavy elements in cooling flow clusters are concentrated toward the center
(e.g., Fukazawa et al.\ 1994; Tamura et al.\ 2004; for ideas why see, e.g.,
B{\"o}hringer et al.\ 2004). Their relative abundance starts to increase
just at the radii where the temperature starts to decrease (e.g., Vikhlinin
et al.\ 2005).  Cold fronts found around these gas density peaks form as a
result of displacement of the central, higher-abundance gas outwards. Thus,
the abundance should be discontinuous across these fronts, as long as
sloshing occurs within the region with the strong gradient.  Such abundance
discontinuities were indeed observed, e.g., in A2204 (Sanders et al.\ 2005)
and Perseus (Fabian et al.\ 2006), although Dupke \& White (2003) did not
see them in A496 (but their measurement uncertainties were relatively
large). In general, sloshing should spread the heavy elements from the
center outwards --- but not too far, because, as we have seen above
(\S\ref{sec:simslosh}), the low-entropy, high-abundance gas eventually flows
back into the center even as a cold front continues to propagate outwards.

\subsection{Zoology of cold fronts}
\label{sec:zoo}

In the sections above, we have discussed cold fronts in merging clusters and
in cooling flows. Since we now know more than two clusters with cold fronts,
we ought to propose a classification scheme, which will also help to
summarize the above observations and simulations. First, in mergers with
cold fronts that are a boundary between gases from two distinct subclusters,
the front can be at the ``stripping'' and the ``slingshot'' stages.  At the
most intuitive ``stripping'' stage, ram pressure of the ambient gas pushes
the cool subcluster gas back from its dark matter host; the examples are
\1e\ (Figs.\ \ref{1e} and \ref{1e_lens}) and NGC\,1404 (Fig.\ \ref{n1404}).
This is likely to occur on the inbound part of the subcluster trajectory or
around the time of core passage, when the ram pressure increases and reaches
its maximum. A less massive subcluster may be completely stripped of gas at
this stage (e.g., right panel in Fig.\ \ref{yago_strip}).  If it does retain
gas, on the outbound leg of the trajectory, the ram pressure drops rapidly
(because both the ambient gas density and the velocity decrease), and the
displaced gas rebounds as in a ``slingshot'', overtaking the subcluster's
mass peak.  An example is A168 (Fig.\ \ref{a168}). In Fig.\ 
\ref{yago_sling}, the subcluster in the left panel and the main cluster in
the right panel exhibit cold fronts at the ``slingshot'' stage, while the
main cluster in the left panel is at the ``stripping'' stage.

A third variety is the ``sloshing'' cold fronts observed in the centers of
clusters that exhibit sharp radial entropy gradients (i.e., cooling flows).
Here, multiple near-concentric discontinuities divide gas parcels from
different radii of the same cluster that came into contact due to sloshing
(\S\ref{sec:simslosh}).  Simulations show (A06) that it is set off easily by
any minor merger and may last for gigayears. This is why this cold front
species is very common, often with multiple fronts in the same cluster. For
comparison, the core passage stage of a merger is very short (of order
$10^8$ yr), and we should also be lucky enough to observe it from the right
angle, which makes ``stripping'' cold fronts the rarest.

\subsection{Non-merger cold fronts and other density edges}

Because the central dense gas in cooling flow clusters is so easily
disturbed, in principle, sloshing can also be induced by bubbles blown by
the central AGNs. This possibility has not yet been addressed with detailed
simulations, although some works suggest that such disturbance is possible
(Quilis, Bower, \& Balogh 2001). A rising bubble can also push the
low-entropy gas in front of it (provided the ensuing instabilities can be
suppressed), which would develop a cold front when it moves into a
lower-density, lower-pressure outer region.

Edge-like features in the X-ray images of clusters and groups may have an
altogether different nature.  The obvious bow shocks caused by mergers will
be discussed later.  Edges in the cores of clusters harboring powerful AGNs
may be weak shocks propagating in front of large AGN-blown bubbles, as
observed, e.g., in the Hydra-A cluster (Nulsen et al.\ 2005).  Such edges
look somewhat different from the ``sloshing'' edges considered above,
spanning a larger sector --- in Hydra-A, it can be traced almost all the way
around the cluster core. In addition, very subtle brightness edges or
``ripples'' observed in the core of the Perseus cluster have been attributed
to sound waves from the central AGN explosions (Fabian et al.\ 2006).
Because such features always have very low brightness contrast and therefore
are strongly affected by line-of-sight projection, and because weak shocks
have inherently low temperature contrast, it is difficult to distinguish
such features from cold fronts by simply looking at their temperature
profiles.

Finally, we mention a more exotic possibility of an ``iron front'', as
reported for the NGC\,507 group (Kraft et al.\ 2004). An X-ray image of this
cool group exhibits an edge, and the spectral analysis shows that most of
the brightness difference is due to a higher abundance of heavy elements on
one side of the edge (which strongly increases the emissivity for a plasma
at $T\lax 1$ keV). Physically, this is still a contact discontinuity similar
to a cold front.

\section{COLD FRONTS AS EXPERIMENTAL TOOL}
\label{sec:tools}

\begin{figure}[t]
\centering
\includegraphics[width=0.6\textwidth]%
{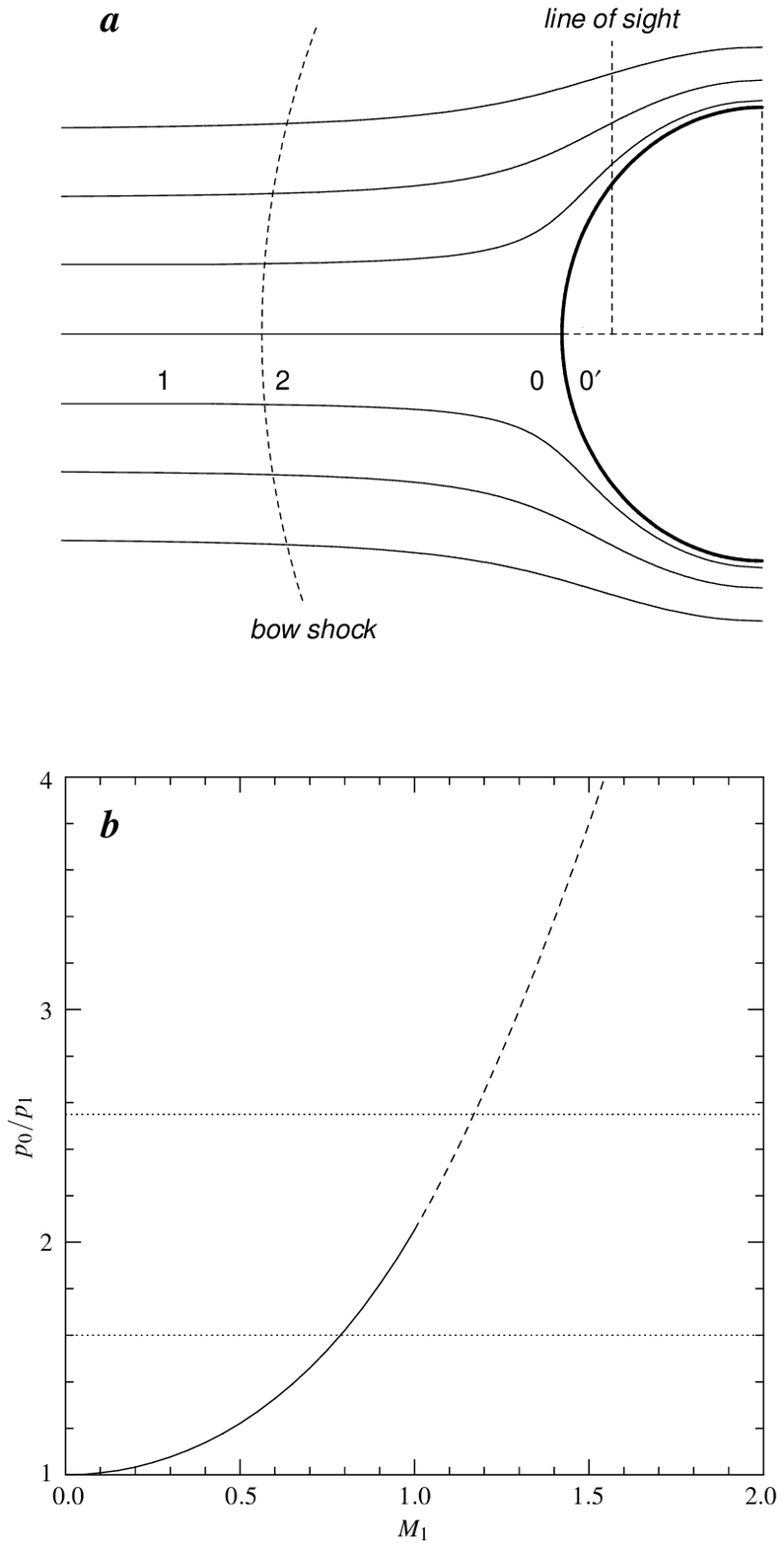}

\caption{({\em a}) Geometry of a flow past a spheroidal cold front. Zones
  0, 1, and 2 are those near the stagnation point, in the undisturbed free
  stream, and past the possible bow shock, respectively.  Zone $0^\prime$ is
  within the body.  ({\em b}) Ratio of pressures at the stagnation point 0
  before the tip of the cold front (which is equal to that just inside the
  front in zone $0^\prime$) and in the free stream 1, as a function of the
  Mach number in the free stream. The solid and dashed line corresponds to
  the sub- and supersonic regimes, respectively (eqs.\ 
  \ref{eq:p:ratio:subsonic} and \ref{eq:p:ratio:supersonic}). The dotted
  lines show the confidence interval for the pressure ratio in A3667.
  (Reproduced from V01.)}

\label{p_M}
\end{figure} 

The origin of cold fronts is certainly interesting, but the most useful
thing about this phenomenon is that it provides a unique tool to study the
cluster physics, including determining the gas bulk velocity (and sometimes
acceleration, as we already saw above), the growth of hydrodynamic
instabilities (or lack thereof), strength and structure of the intracluster
magnetic fields, thermal conductivity, and perhaps viscosity of the ICM.  We
will discuss some of these possibilities below.  From a technical viewpoint,
what makes these studies possible is the high contrast and symmetric shape
of cold fronts in the X-ray images, which enables accurate deprojection of
various three-dimensional quantities near the front.

\subsection{Velocities of gas flows}
\label{sec:vel}

As we mentioned above, in broad terms, thermal pressure of the gas inside
the cold front balances the sum of thermal and ram pressures of the gas
outside. Both components of thermal pressure, the gas density and
temperature, can be measured directly from the X-ray data, but not the
velocity in the plane of the sky. The difference of thermal pressures across
the front gives the ram pressure and thus the velocity of the gas cloud.
This method was first applied by V01 to the cold front in A3667 (Fig.\ 
\ref{2142_3667}).

For a quantitative estimate, we must consider a more exact physical picture,
schematically shown in Fig.\ \ref{p_M} (reproduced from V01).  Panel {\em
  a}\/ shows a uniform flow around a stationary blunt body of dense gas. The
flow forms a stagnation region at the tip of the body (zone 0), where the
velocity component along the axis of symmetry goes to zero.  Note that
thermal pressure increases in the stagnation region as one moves closer to
the front, and is continuous across the front (unlike that across a shock).
The gas is compressed adiabatically, i.e., there will be a density and
temperature increase in the stagnation region compared to the values in the
flow (absent complications such as those discussed in \S\ref{sec:depl}
below).  The ``outer gas pressure'' in the argument above is the pressure in
the free-stream region of the flow (zone 1), at a sufficient distance from
the front beyond the stagnation region ($\sim 0.5$ of the front's radius of
curvature for a transonic flow), or ahead of the shock front if $M>1$. In
practice, the stagnation region is small and difficult to detect because of
the line-of-sight projection, so a typical observed pressure profile derived
in wide radial bins across a moving cold front would exhibit a jump.

The ratio of thermal pressures at the stagnation point, $p_0$, and in the
free stream, $p_1$, is a function of the cloud speed (Landau \& Lifshitz
1959, \S114):
\begin{equation}
\label{eq:p:ratio:subsonic}
   \frac{p_0}{p_1}=\left(1+\frac{\gamma-1}{2}\;M_1^2\right)^{
     \frac{\scriptstyle\gamma}{\scriptstyle\gamma-1}},\quad M_1\le1
\end{equation}
\begin{equation}\label{eq:p:ratio:supersonic}
  \frac{p_0}{p_1}=
        \left(\frac{\gamma+1}{2}\right)^{
           \frac{\scriptstyle\gamma+1}{\scriptstyle\gamma-1}}
        M_1^2
        \left(\gamma-\frac{\gamma-1}{2M_1^2}\right)^{
           -\frac{\scriptstyle 1}{\scriptstyle\gamma-1}},
         \quad M_1>1,
\end{equation}
where $M_1$ is the Mach number of the cloud relative to the sound speed in
the free stream region and $\gamma=5/3$ is the adiabatic index of the gas.
The subsonic equation~(\ref{eq:p:ratio:subsonic}) follows from Bernoulli's
equation.  The supersonic equation~(\ref{eq:p:ratio:supersonic}) accounts
for the gas entropy jump at the bow shock.  Figure \ref{p_M}{\em b}\/ shows
these ratios $p_0/p_1$ as a function of $M_1$.

The gas parameters at the stagnation point usually cannot be measured
directly, because the stagnation region is physically small and its X-ray
emission is strongly affected by projection.  However, as we mentioned,
thermal pressure at the stagnation point equals thermal pressure within the
cloud, which is easily determined.

Because the cluster has a gradient of the gravitational potential, the gas
pressure increases toward the center of a cluster in hydrostatic
equilibrium, which is of course not included in eqs.\ 
(\ref{eq:p:ratio:subsonic}-\ref{eq:p:ratio:supersonic}).  This change may
not be negligible on a distance between zones 1 and 0. Because most clusters
are reasonably centrally symmetric on large scales, one can usually correct
the free-stream pressure for this effect with sufficient accuracy by fitting
a centrally symmetric pressure model in a representative image area that
excludes the front and its disturbed vicinity, and evaluating it at the
radius of the front.

For the cold front in A3667, V01 obtained the ratio of the pressures
$p_0/p_1=2.1\pm0.5$ (horizontal dashed lines in Fig.\ \ref{p_M}), which
corresponds to $M_1=1.0\pm0.2$, i.e.\ the gas cloud moves at the sound speed
of the hotter gas.  Evaluating the sound speed from the X-ray temperature,
the cloud velocity is $1400\pm300$ \kms. In another example, Machacek et
al.\ (2005) performed a similar analysis of the cold front at the boundary
of the galaxy NGC\,1404 and derived $M=0.8-1.0$, which corresponds to the
galaxy's velocity of $530-660$ \kms\ relative to the ambient gas in the
Fornax cluster. Mazzotta et al.\ (2003) obtained $M\approx 0.75\pm 0.2$ for
a prominent cold front in 2A\,0335+096, and O'Hara et al.\ (2004) obtained
$M\approx 1$ for a front in A2319, which are examples of clusters with
sloshing cool cores. In A1795, the pressures at two sides of the cold front
are equal, which corresponds to zero velocity (M01 and \S\ref{sec:slosh}
above).  There are only a few observed mergers with $M>1$.

\subsection{Thermal conduction and diffusion across cold fronts}
\label{sec:cond}

\begin{figure}[t]
\centering
\includegraphics[width=0.65\textwidth, bb=29 183 534 640]%
{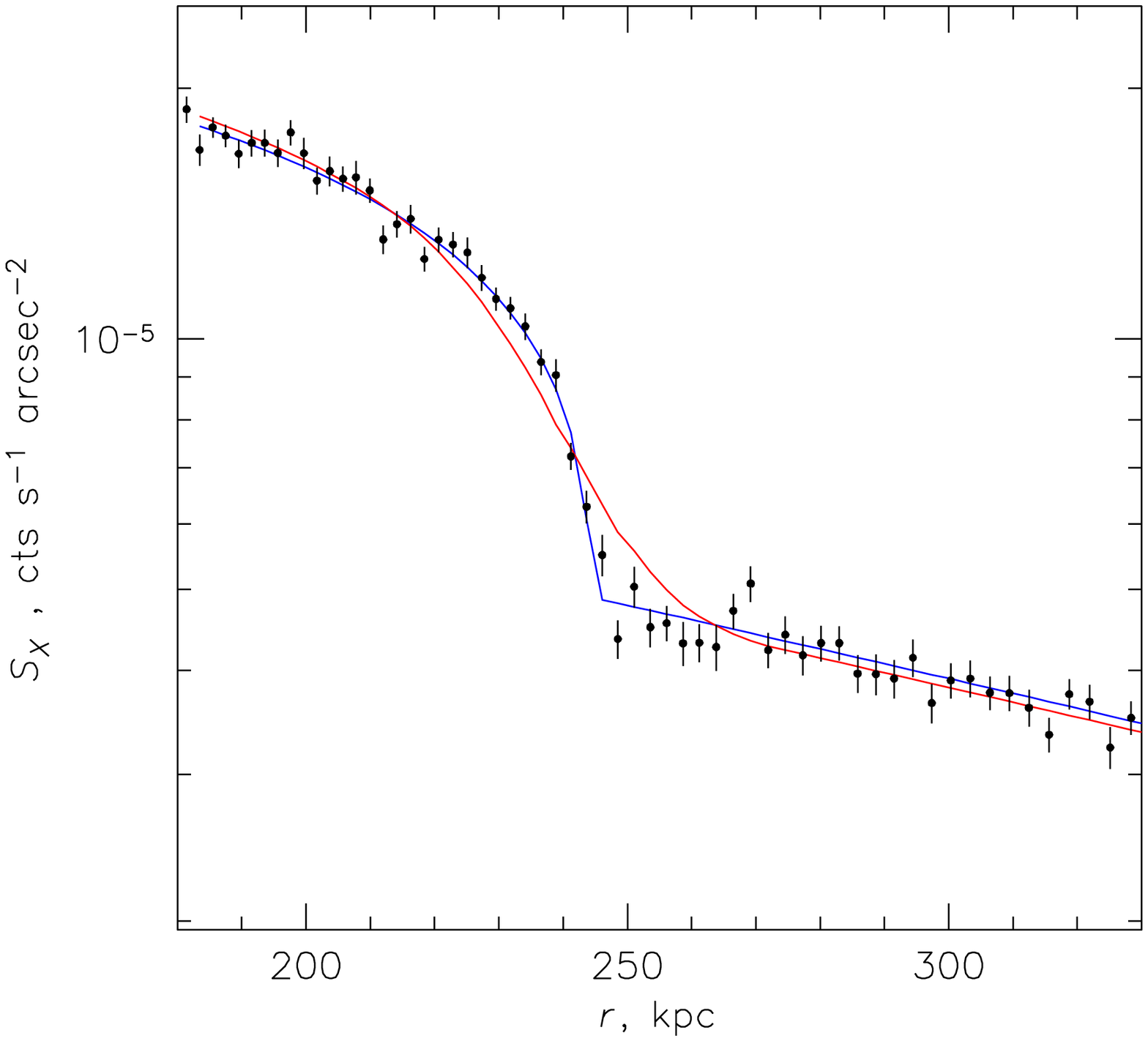}

\caption{X-ray surface brightness profile in a narrow sector near the tip of
  the cold front in A3667 (whose image is shown in Fig.\ \ref{2142_3667}).
  Blue line shows the best-fit projected spherical density discontinuity
  with an infinitely small width, which describes the data well. Red line
  shows a discontinuity smeared with a Gaussian with $\sigma=13$ kpc, which
  corresponds to the collisional m.f.p.\ $\lambda_{\rm in\rightarrow out}$
  (eq.\ \ref{eq:lambda:in:out}). It is ruled out by the data. The $r$\/
  coordinate is from the center of curvature of the front. (This figure is
  an update of a similar plot in V01, using a much longer \chandra\ 
  observation and a narrower sector.)}

\label{a3667_lambda}
\end{figure}

Cold fronts are remarkably sharp, both in terms of the density and the
temperature jumps. Ettori \& Fabian (2000) first pointed out that the
observed temperature jumps in A2142 require thermal conduction across cold
fronts to be suppressed by a factor of order 100 compared to the collisional
Spitzer or saturated values. Furthermore, V01 have found that for A3667, the
gas density discontinuity at the cold front is several times narrower than
the electron mean free path with respect to Coulomb collisions.  Figure
\ref{a3667_lambda} (an update of a similar plot in V01) shows a detailed
X-ray surface brightness profile across the tip of the front.  The X-ray
brightness increases sharply within 2--3 kpc from the front position. We can
compare this width with the Coulomb mean free path of electrons (and
protons, $\lambda_e=\lambda_p$) in the plasma on both sides of the front.
The Coulomb scattering of particles traveling across the front can be
characterized by four different mean free paths: that of thermal particles
in the gas on each side of the front, $\lambda_{\rm in}$ and $\lambda_{\rm
  out}$, and that of particles from one side of the front crossing into the
gas on the other side, $\lambda_{\rm in\rightarrow out}$ and $\lambda_{\rm
  out\rightarrow in}$. From Spitzer (1962), we have for $\lambda_{\rm in}$
or $\lambda_{\rm out}$:
\begin{equation}
  \label{eq:lambda:thermal}
  \lambda = 15~{\rm kpc}\left(\frac{T}{7~{\rm keV}}\right)^2 
  \left(\frac{n_e}{10^{-3}~{\rm cm}^{-3}}\right)^{-1},
\end{equation}
and for $\lambda_{\rm in\rightarrow out}$ and $\lambda_{\rm out\rightarrow
  in}$:
\begin{equation}
  \label{eq:lambda:in:out}
  \lambda_{\rm in\rightarrow out} = \lambda_{\rm out}\,
      \frac{T_{\rm in}}{T_{\rm out}}\,
      \frac{G(1)}{G\left(\sqrt{T_{\rm in}/T_{\rm out}}\right)}
\end{equation}
\begin{equation}
  \label{eq:lambda:out:in}
  \lambda_{\rm out\rightarrow in} = \lambda_{\rm in}\,
      \frac{T_{\rm out}}{T_{\rm in}}\,
      \frac{G(1)}{G\left(\sqrt{T_{\rm out}/T_{\rm in}}\right)},
\end{equation}
where $G(x)=[\Phi(x)-x\Phi'(x)]/2x^2$ and $\Phi(x)$ is the error function.
For the front in A3667, $\lambda_{\rm out}\approx 20-40$~kpc, $\lambda_{\rm
  in}\approx 2$~kpc, $\lambda_{\rm in\rightarrow out}\approx 10-13$~kpc,
$\lambda_{\rm out\rightarrow in}\approx 4$~kpc. The upper bounds in the
above intervals correspond to the expected temperature increase in the
stagnation region (which is difficult to measure due to strong projection
effects), and the lower bounds correspond to no increase from the observed
outer temperature.


The hotter gas in the stagnation region has a low velocity relative to the
cold front. Therefore, diffusion, undisturbed by the gas motions, should
smear any density discontinuity by at least several mean free paths on a
very short time scale.  Diffusion in our case is mostly from the inside of
the front to the outside, because the particle flux through the unit area is
proportional to $nT^{1/2}$. Thus, if Coulomb diffusion is not suppressed,
the front width should be at least several times $\lambda_{\rm in\rightarrow
  out}$. Indeed, the time for $T=4$ keV protons to travel 10 kpc is $10^7$
yr, compared to the age of the structure of at least $R/v \approx 2\times
10^8$ yr, where $R$\/ and $v$\/ are the front radius and velocity.  Such a
smearing is ruled out by the sharp rise in the X-ray brightness at the front
(Fig.\ \ref{a3667_lambda}).  Fitting the observed surface brightness profile
by a projected abrupt density discontinuity smeared with a Gaussian, we
obtain a formal upper limit on the Gaussian $\sigma$ of 4 kpc. One should
remember that because the front is seen along the surface in projection, any
deviations from the ideal spherical shape would smear the edge, and yet the
observed front is sharper than the Coulomb m.f.p. This can be explained only
if the diffusion coefficient is suppressed by at least a factor of 3 with
respect to the Spitzer value. (This is a very conservative upper limit,
simply equal to the ratio of the Spitzer m.f.p.\ and our upper limit on the
front width; in fact, the front should spread by much more than one m.f.p.
during its presumed lifetime.)

The suppression of transport processes in the intergalactic medium is most
naturally explained by the presence of a magnetic field perpendicular to the
density or temperature gradient. Even a very small field is sufficient for
the electron and proton gyro radii to be many orders of magnitude smaller
than the Coulomb m.f.p. in the ICM, so electrons and ions would move mostly
along the field lines. The observation of a sharp density discontinuity in
A3667 is a first direct indication that such a suppression is possible in
the ICM (although this has, of course, been expected, since radio
observations have provided evidence for microgauss-level magnetic fields in
the ICM, see, e.g., Carilli \& Taylor 2002). In many other clusters, e.g.,
A2142, the front width is also unresolved by \chandra, but the data quality
does not allow such accurate constraints.

The suppression of diffusion and collisional thermal conduction is most
effective if the magnetic field lines do not cross the front surface, that
is, the two sides of the front are magnetically isolated. Below we will see
that the cold front in A3667 provides another, indirect indication of just
such a field configuration, with field lines mostly parallel to the front
surface. We will also see why such a configuration should arise naturally in
a cluster merger.

\subsection{Stability of cold fronts}

Cold fronts are remarkably smooth in shape, considering that they form in a
violent merger environment. This property contains information on their
underlying dark matter distribution, as well as conditions in the ICM,
possibly including its viscosity, the prevalence of turbulence, and strength
and structure of the magnetic fields.  These ICM properties can
significantly impact such diverse problems as the energy balance in the
cluster cooling flows and estimates of the cluster total masses.

\begin{figure}[t]
\centering
\includegraphics[width=0.5\textwidth]%
{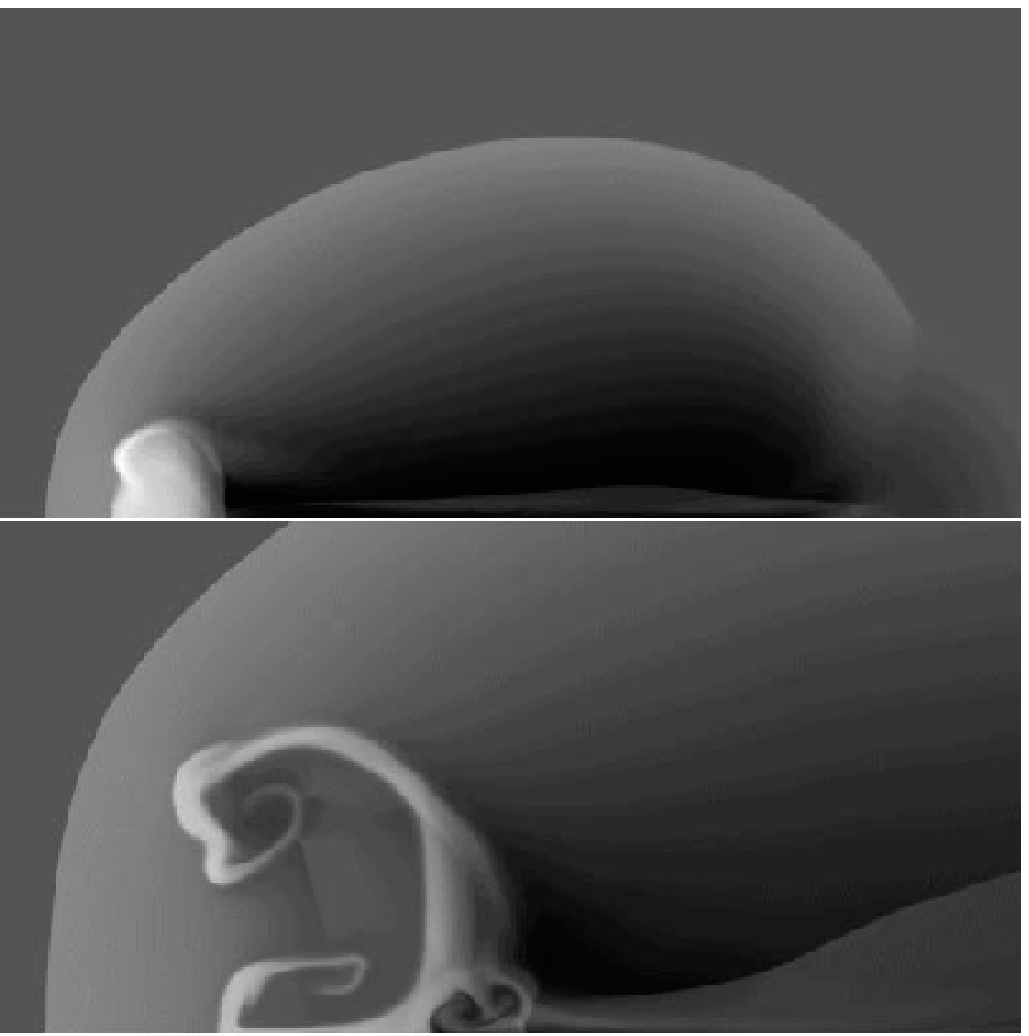}

\caption{A simulation of the gas flow around an initially round dense gas
  bullet without gravity or significant magnetic field support. Density
  increases from dark to light gray; time in the two snapshots increases
  from top to bottom. The bullet is a white blob in the top panel, preceded
  by a shock front. The RT instabilities quickly develop and destroy the
  bullet.  (Reproduced from Jones, Ryu, \& Tregillis 1996.)}

\label{jones}
\end{figure} 

\subsubsection{Rayleigh-Taylor instability and underlying mass}
\label{sec:RT}

When a dense gas cloud moves through a more rarefied medium, ram pressure of
the ambient flow slows it down. In the decelerating reference frame of the
cloud, there is an inertial force directed from the dense phase to the
less-dense phase, which makes the front interface of the cloud
Rayleigh-Taylor unstable.  As a result, the cloud quickly disintegrates, as
observed in laboratory and numerical experiments, see Fig.\ \ref{jones}.  If
a gas cloud is bound by gravity, it can prevent the onset of the RT
instability. It is interesting to see if, for example, the cold front in
A3667 is stable in this respect (V02). The drag force on the cloud is
$F_{\rm d}=C \rho_{\rm out} v^2 A/2$, where $\rho_{\rm out}$ and $v$\/ are
the gas density and velocity of the ambient flow, $A$\/ is the cloud
cross-section area, and $C$\/ is the drag coefficient.  For the particular
geometry of the front in A3667 (a cylinder with a rounded head), $C\approx
0.4$.  From the measured density inside the cloud, $\rho_{\rm in}$, we can
roughly estimate the mass of the cloud as a sphere with radius $r$,
$M\approx (4/3)\pi r^3 \rho_{\rm in}$. Then the drag acceleration is:
\begin{equation}
g_{\rm d}=\frac{F_{\rm d}}{M}
        \approx 0.15\,\frac{\rho_{\rm out}}{\rho_{\rm in}}\,\frac{v^2}{r}.
\end{equation}
For the measured quantities in A3667, $g_{\rm d}\approx 8\times 10^{-10}$
cm~s$^{-2}$.  One can also estimate the gravitational acceleration at the
front surface created by the gas mass inside the cloud, which points in the
opposite direction. It turns out to be $\sim 2$ times smaller, which means
that gravity of the gas itself is insufficient to suppress the RT
instability. Because the front is apparently stable, this means that there
has to be a massive underlying dark matter subcluster, centered inside the
front, that holds the gas cloud together. This is, of course, what we expect
in a merger.  If the total mass of the underlying dark matter halo is the
usual factor of $\sim 10$ higher than the gas mass, its gravity at the front
surface would be more than sufficient to compensate for the drag force,
thereby removing the RT instability condition.

\begin{figure}[t]
\centering
\includegraphics[width=1\textwidth]%
{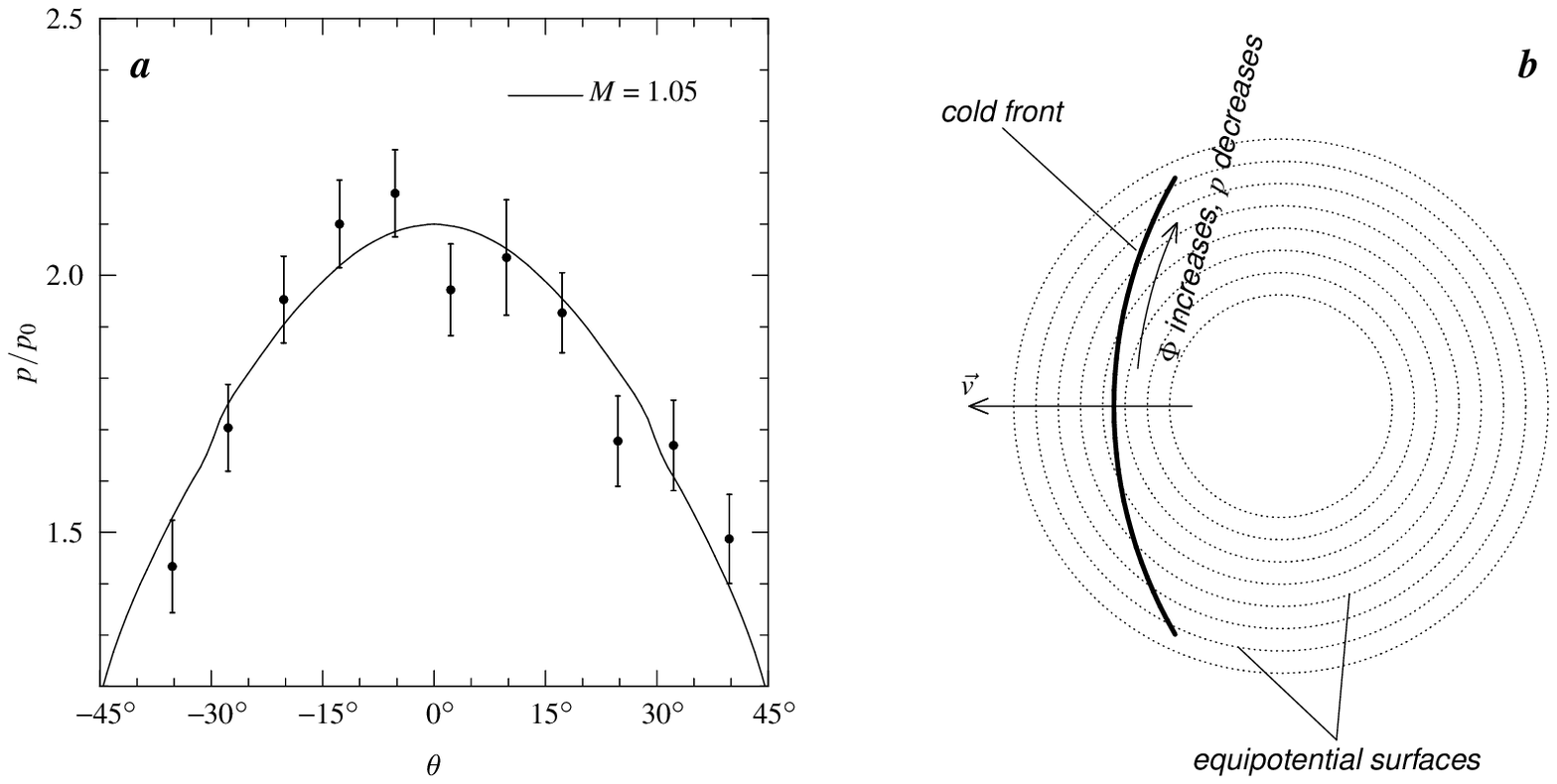}

\caption{({\em a}) Gas pressure under the surface of the cold front in A3667
  as a function of angle from the symmetry axis. The line is a prediction
  for a $M=1.05$ flow. ({\em b}) Schematic depiction of the front surface
  and the gravitational potential $\Phi$\/ of the underlying dark matter
  subcluster.  The presence of an underlying massive subcluster is necessary
  to prevent the development of Rayleigh-Taylor instability (Fig.\ 
  \ref{jones}). The mass center should be offset from the center of
  curvature of the front in order for the front to be in pressure
  equilibrium with the ambient flow. (Reproduced from V02.)}

\label{a3667_p_azim}
\end{figure} 

\paragraph*{Pressure along the front and the underlying dark matter halo}

The A3667 cold front also provides another indication of the existence of a
massive dark matter halo binding the gas.  Following the Bernoulli law, the
pressure of the ambient flow at the surface of the front should have a
maximum at the stagnation point and decline as one moves along the front
surface away from the symmetry axis, as the shear velocity increases.
Figure \ref{a3667_p_azim}{\em a}\/ (from V02) shows the measured thermal
pressure just inside the surface of the front as a function of the angle
from the axis of symmetry.%
\footnote{The data used by V02 were consistent with an isothermal gas
  inside the front, so the pressure in Fig.\ \ref{a3667_p_azim}{\em a}\/ was
  derived assuming a constant temperature (i.e., what changes in the plot is
  the gas density).  A more recent \xmm\ observation uncovered spatial
  temperature variations there (Briel et al.\ 2004), in particular, a cooler
  spot at the tip of the front. This would reduce the peak pressure
  somewhat; however, qualitatively and methodologically, the V02 result
  still holds.}
It behaves as expected if the front surface was in pressure equilibrium with
the ambient flow. The expected ambient pressure for a $M\approx 1$ flow
(from a simple simulation) is also shown for comparison; it describes the
measured profile very well.  This indicates that the front is stationary
(i.e., it is not an expanding shell, for example).

Furthermore, because the density is not constant inside the cool gas along
the front, there has to be an underlying mass concentration that supports
the resulting pressure gradient.  Indeed, under the simplifying assumption
that the gas temperature is constant, the hydrostatic equilibrium equation
(\ref{eq:mass}) can be written as $\rho=\rho_0 \exp (-\mu m_p \Phi /kT)$,
where $\rho$\/ is the gas density and $\Phi$\/ is the gravitational
potential.  The declining gas pressure along the front requires a
corresponding rise of the potential --- Fig.\ \ref{a3667_p_azim}{\em b}\/
schematically shows the required configuration.

In V02, we presented an estimate of what kind of mass concentration is
needed for the gas inside the cold front to be in hydrostatic equilibrium
and exhibit such a pressure profile at the front.  For a spherical halo with
a King radial mass profile, the center of the halo should be located $\sim
70$ kpc under the front surface (compare this to the front radius of
curvature of 250 kpc), and a total mass within $r=70$ kpc should be
approximately $3\times 10^{12}$\msun. This is about 20 times higher that the
gas mass in the corresponding region. Thus, the dark halo and the gas inside
the cold front together look like a typical merging subcluster.  The overall
picture of A3667 is quite similar to the Roettiger et al.\ (1998) simulated
merger shown in Fig.\ \ref{roettiger}, and to that in the cosmological
simulations by Nagai \& Kravtsov (2003).

\begin{figure}[t]
\centering
\includegraphics[width=1\textwidth, bb=50 488 489 703]%
{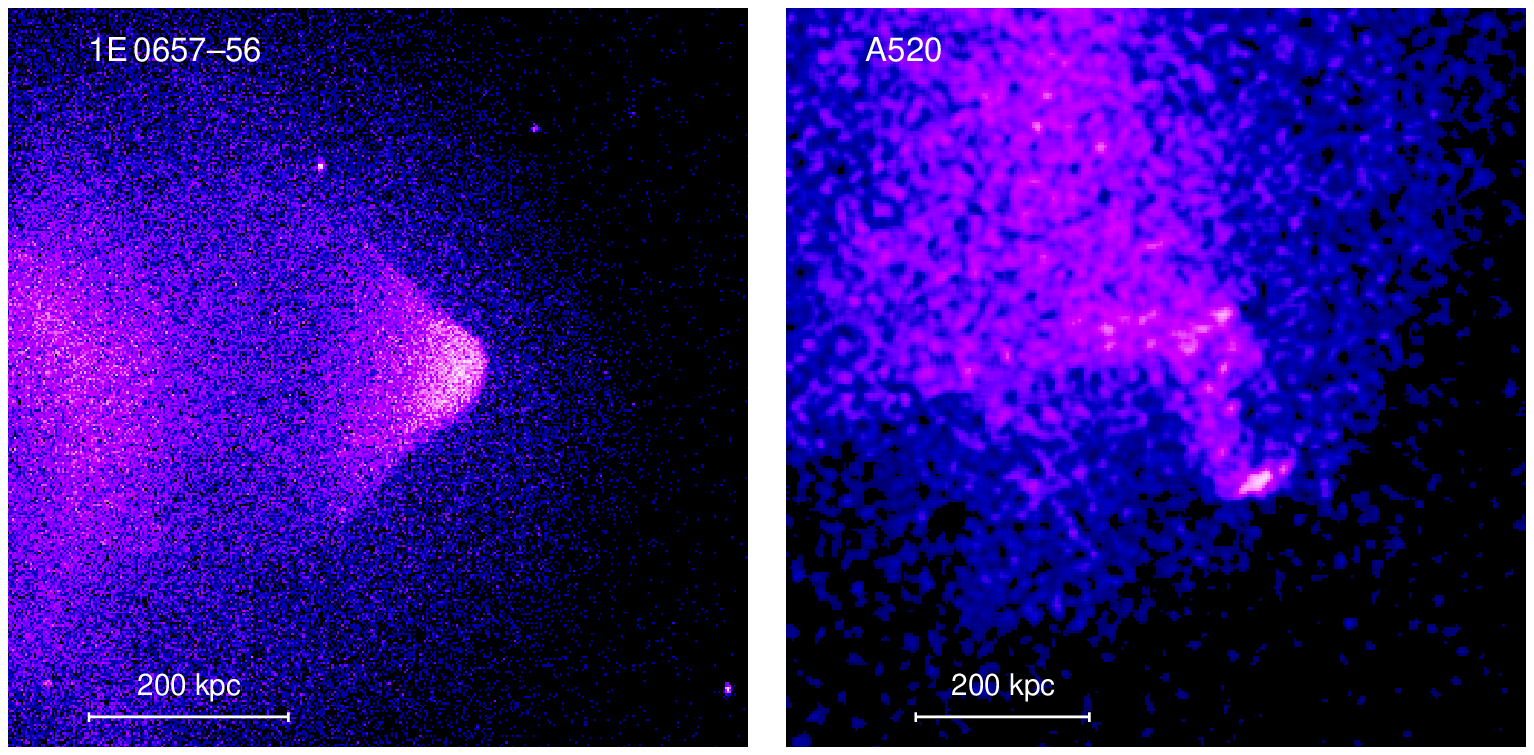}

\caption{X-ray images of cool subclusters in \1e\ and A520 (their
  larger-scale images can be found in Figs.\ \ref{1e}{\em c}\/ and
  \ref{a520_profs}{\em a}, respectively). The A520 image is smoothed;
  small-scale fluctuations are photon noise. Both are experiencing mergers
  in the plane of the sky, and both have cool gas subclusters stripped by
  ram pressure from their associated mass halos, making them Rayleigh-Taylor
  unstable.  In \1e, the gas bullet has a sharp cold front not yet disrupted
  by instabilities, while in A520, the cool subcluster has already been
  broken up.}

\label{1e_a520}
\end{figure}

Let us now take another look at the X-ray and lensing maps for \1e\ (Figs.\ 
\ref{1e}{\em bc}\/ or \ref{1e_lens}).  The ram pressure has just pushed the
gas bullet out of its host dark matter halo. The halo's gravity can no
longer prevent the RT instability, and the gas bullet is expected to fall
apart very quickly, although it has not happened yet (Fig.\ 
\ref{1e_a520}{\em a}).  Indeed, another cluster, A520, exhibits a cool
subcluster at a slightly later stage, see Fig.\ \ref{1e_a520}{\em b}.
Similarly to \1e, A520 is going through a supersonic merger in the plane of
the sky. Its mass map also reveals a small dark matter halo located ahead of
what remains of the gas subcluster (Okabe \& Umetsu 2007; Clowe et al.\ in
preparation).  Apparently, A520 has already proceeded to a stage when the
instabilities have broken up the cool subcluster into several pieces
(Markevitch et al.\ in preparation).

\subsubsection{Kelvin-Helmholtz instability and magnetic field}
\label{sec:KH}

Vikhlinin, Markevitch, \& Murray (2001a) pointed out that
the shape of the cold front in A3667 provides an independent constraint of
the magnetic field at the location of the front. Their argument is that the
cold front in A3667 (Fig.\ \ref{2142_3667}) is sharp and has a smooth shape
within a certain sector ($\pm 30^\circ$) around the symmetry axis. At the
same time, given the measured subcluster velocity, the front should be
quickly disturbed by the Kelvin-Helmholtz (KH) instability. It would be
observed in projection as smearing of the sharp edge on a width scale
comparable to the wavelength of the mode that has reached the nonlinear
growth stage. Thus, within that sector, the KH instability appears to be
suppressed, at least for perturbations with $\lambda$ greater than the
observed upper limit on the width of the front there.

\begin{figure}[t]
\centering
\includegraphics[width=0.5\textwidth, bb=411 421 653 663,clip]%
{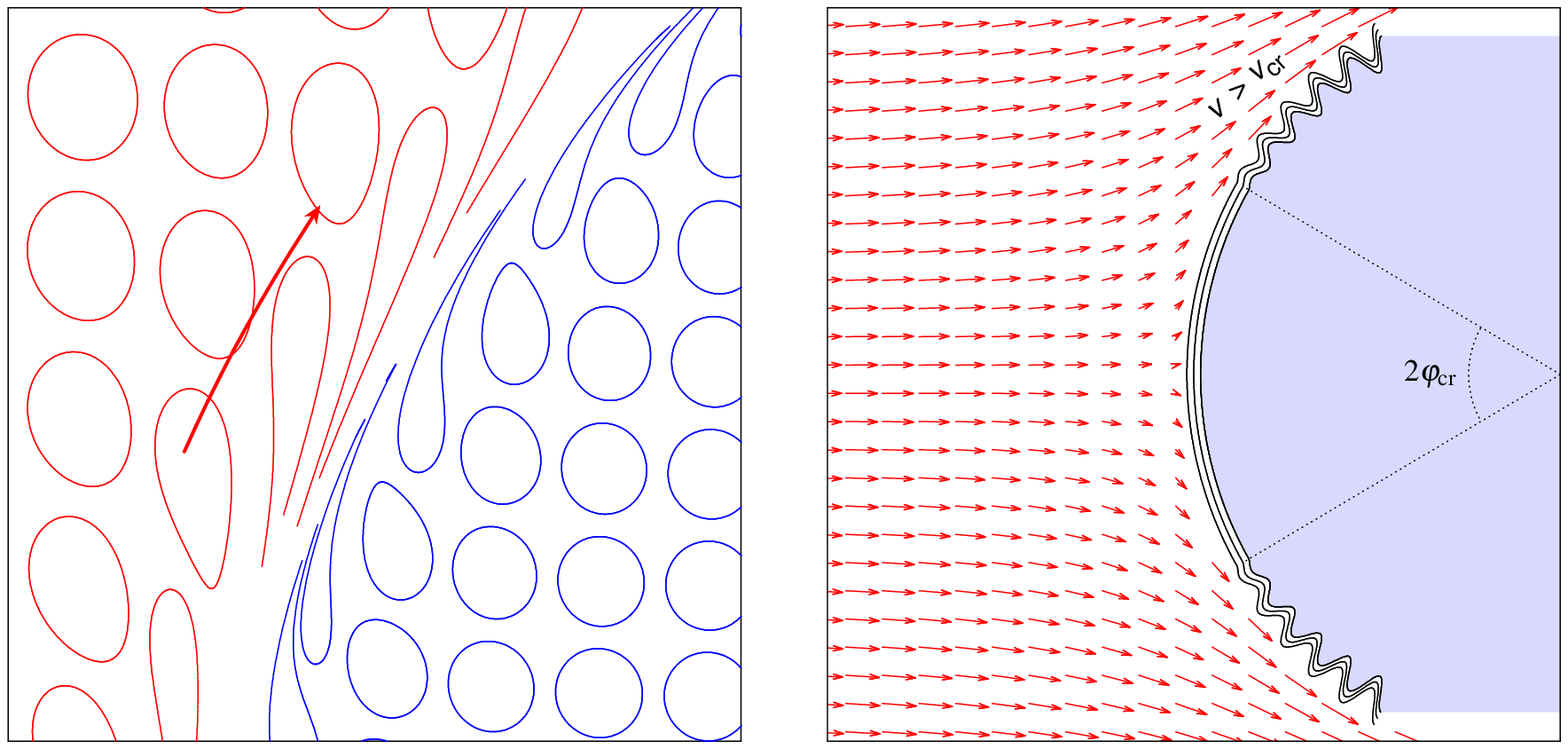}

\caption{A schematic illustration of the suppression of Kelvin-Helmholtz
  instability at the surface of the A3667 cold front. The magnetic layer
  (shown by parallel curves along the front) can provide surface tension
  that suppresses the growth of perturbations in the region where the
  tangential velocity is smaller than some critical value $V_{\rm cr}$. The
  velocity field (arrows) corresponds to the flow of incompressible
  fluid around a sphere.  (Reproduced from Vikhlinin et al.\ 2001a.)}

\label{a3667_magn}
\end{figure}

The gravity of the subcluster can in principle suppress the KH instability,
just as it does the RT instability (\S\ref{sec:RT}). However, an estimate
shows that it is far too small (V02). Assuming the density discontinuity is
sharp (to which we will return below), the next most natural stabilizing
mechanism is the formation of a layer of magnetic field parallel to the
front. Such a layer would provide surface tension and make it difficult for
any deformations of the surface to grow (Fig.\ \ref{a3667_magn}). As we will
discuss below, a layer of increased magnetic field parallel to the surface
is indeed expected to emerge as a result of ``magnetic draping'', i.e.,
stretching of the field in the ambient ICM as it flows around the cold front
(e.g., Asai et al.\ 2005; Lyutikov 2006).

Because the shear velocity of the flow increases as one moves along the
front away from the stagnation point, the surface tension of a magnetic
layer at a certain angle ($\varphi_{\rm cr}$ in Fig.\ \ref{a3667_magn}) may
become insufficient, and the KH instability starts to grow.  Thus, the
extent of the undisturbed sector of a cold front $\varphi_{\rm cr}$ may be
used to derive a lower limit on the strength of the stabilizing magnetic
field.  Detailed calculations for a front in A3667 can be found in Vikhlinin
et al.\ (2001a).%
\footnote{A minor algebraic error in that paper was pointed out by P.
  Mazzotta and corrected in V02, which did not change the result.}
Denoting the magnetic field strengths on the hot and cold sides of the front
as $B_h$ and $B_c$, the respective gas temperatures as $T_h$ and $T_c$, the
gas pressure at the front as $p_{\rm gas}$, and the Mach number of the local
shear flow as $M$, the KH instability is suppressed when
\begin{equation}
\label{eq:kh:mag:P:condition}
  \frac{B_h^2}{8\pi}+\frac{B_c^2}{8\pi}>\frac{1}{2}\,\frac{\gamma
    M^2}{1+T_c/T_h}\, p_{\text{gas}}.
\end{equation}
For the observed temperatures and flow velocities, the observed stability of
the front within the sector $\varphi<30^\circ$ (where $M\le0.55$), and
taking into account the uncertainties, this gives a lower limit on the sum
of magnetic pressures in the two gas phases:
\begin{equation}
\label{eq:pmag/pgas}
\frac{B_h^2}{8\pi}+\frac{B_c^2}{8\pi} > (0.1-0.2) p_{\text{gas}}.
\end{equation}
This gives $B> (7-16)\,\mu G$ for the maximum of the two quantities $B_h$
and $B_c$. If the apparent smearing of the front beyond the stable sector is
interpreted as the onset of the KH instability, from a lower limit this
becomes an estimate of the magnetic field. However, there may be other
mechanisms disturbing the front at large $\varphi$, so it is best to
consider it a lower limit.

Looking at eq.\ (\ref{eq:pmag/pgas}), it is interesting to realize that even
though a magnetic field is far from being ``dynamically important'' in the
usual sense --- it does not dominate the total pressure in the ICM (and thus
will not significantly affect the cluster total mass estimates, for
example), it can still be strong enough to qualitatively alter the evolution
of cold fronts, mixing of gases, and perhaps the development of turbulence.

A more rigorous analysis of the growth of KH instability in the absence of a
magnetic layer was presented by Churazov \& Inogamov (2004).  Their growth
factor estimate for a sharp discontinuity differed from a simplified
estimate in Vikhlinin et al.\ (2001a), but it did not change the conclusion
that such a front would be unstable. More interestingly, they pointed out
that the KH instability will not develop if the density discontinuity had a
finite intrinsic width, for example, because of diffusion. This would
completely suppress the growth of perturbations with wavelengths shorter
than a certain fraction of the front radius, dependent on the smearing
width. An upper limit on the width of the front from the current data
(\S\ref{sec:cond}) does not rule out the Churazov \& Inogamov scenario
(although it comes close).  We note, however, that the observed limit on the
front width is already smaller by a factor of several than the collisional
m.f.p., so the natural candidate to create such a stabilizing layer,
collisional diffusion, does not work. A different physical process would
have to widen the front within a factor of 2--3 of the current upper limit,
which would require a coincidence.

While the above analyses assume a spherical shape for the front, a recent,
much deeper \chandra\ exposure of A3667 (Fig.\ \ref{2142_3667}) shows that
it is not exactly spherical. This would modify the flow pattern, and so the
above instability suppression calculations are, of course, only qualitative.

\begin{figure}[t]
\centering
\includegraphics[width=\textwidth]%
{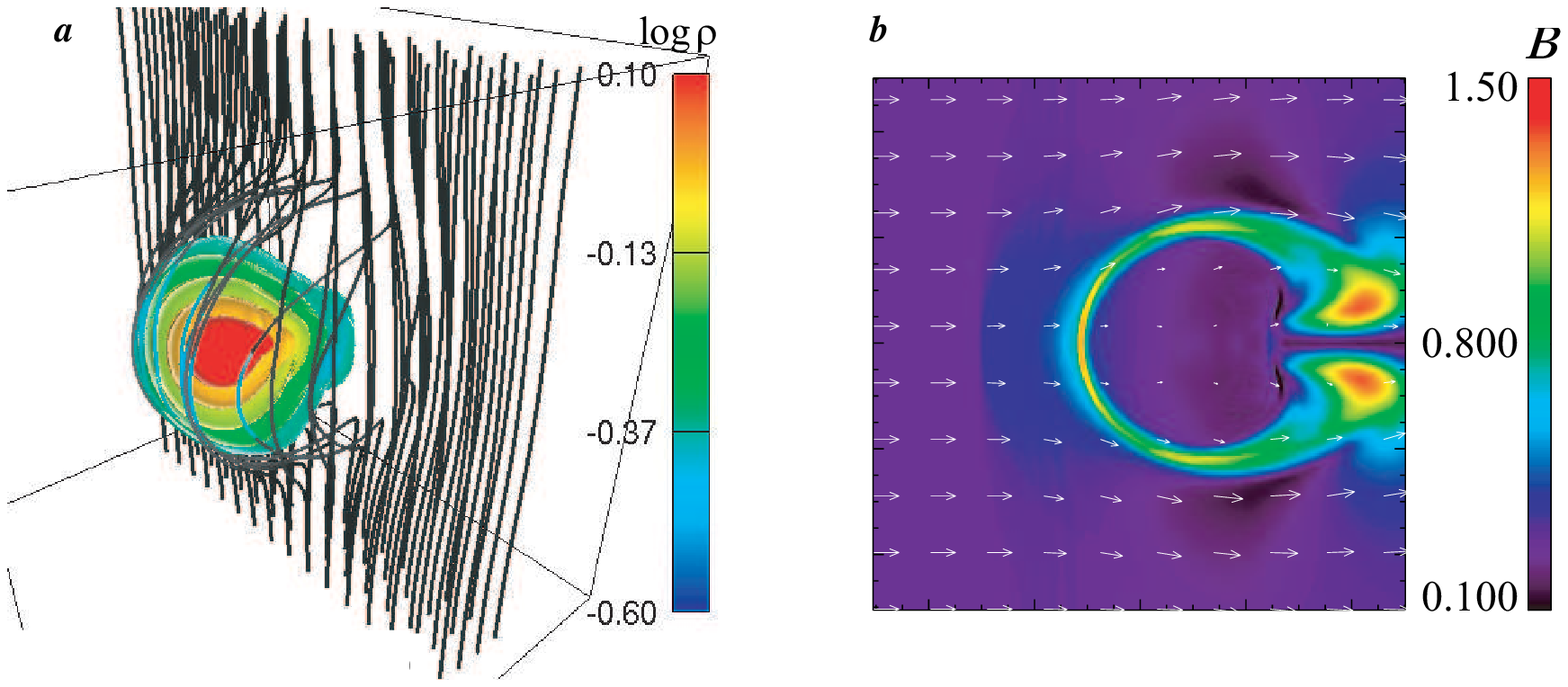}

\caption{Simulation of the magnetic field draping around a cold front in the
  course of a gas cloud's motion through an ICM with a uniform magnetic
  field.  ({\em a}) The subcluster gas density (color) and the magnetic
  field lines.  ({\em b}) A horizontal cross-section through the subcluster:
  color shows magnetic field strength, arrows show gas velocities.
  Compression and shear of the field in the incoming flow creates a narrow
  layer around the front, in which the magnetic field is strongly amplified.
  (Reproduced from Asai et al.\ 2005.)}

\label{asai}
\end{figure}

\subsubsection{Origin of magnetic layer}
\label{sec:drap}

Assuming the KH instability is indeed suppressed by the surface tension of a
magnetic layer, the lower limit on the field in this layer obtained above is
significantly higher than the $B\sim 1\,\mu G$ estimates typically given for
the cluster regions outside cooling flows (e.g., Carilli \& Taylor 2002).
This is not unexpected, because a moving cold front is a special place,
where the field should align with the surface and significantly strengthen
via ``draping'' around an obstacle, a phenomenon originally proposed by
Alfv\'en (1957) to explain the comet tails. It is illustrated in Fig.\ 
\ref{asai} (from Asai et al.\ 2005).  Panel ({\em a}) shows a simulation of
the magnetic field lines draping around a gas cloud as the cloud moves
through an ICM with a frozen-in magnetic field.  These simulations used an
ordered field; however, even if the field is tangled on scales less than the
size of the cold front (as expected in clusters), the loops will stretch and
the field at the front will consist of large two-dimensional patches
(perhaps disconnected), which will have the same qualitative effect on the
instabilities.  Panel ({\em b}) shows the increase of the field strength at
the front as a result of this shear and compression. As pointed out by
Lyutikov (2006), the field can reach equipartition with the gas thermal
pressure, i.e., the ratio of thermal to magnetic pressures $\beta\approx 1$,
in a narrow layer along the surface of the front (recall that in most of the
ICM, $\beta\sim 10^2-10^3$).  This effect has long been known to space
physicists --- it is observed by space probes at locations where the solar
wind flows around the Earth magnetosphere (at magnetopause) and around the
atmospheres of Mars and Venus.  Similarly to the ICM, outside of such
special locations, solar wind has $\beta>1$, so the analogy is physically
meaningful.  Such a magnetic layer would be more than sufficient to suppress
KH instabilities and completely suppress thermal conduction and diffusion
across any moving cold front.

\begin{figure}[t]
\centering

\includegraphics[width=0.8\textwidth]%
{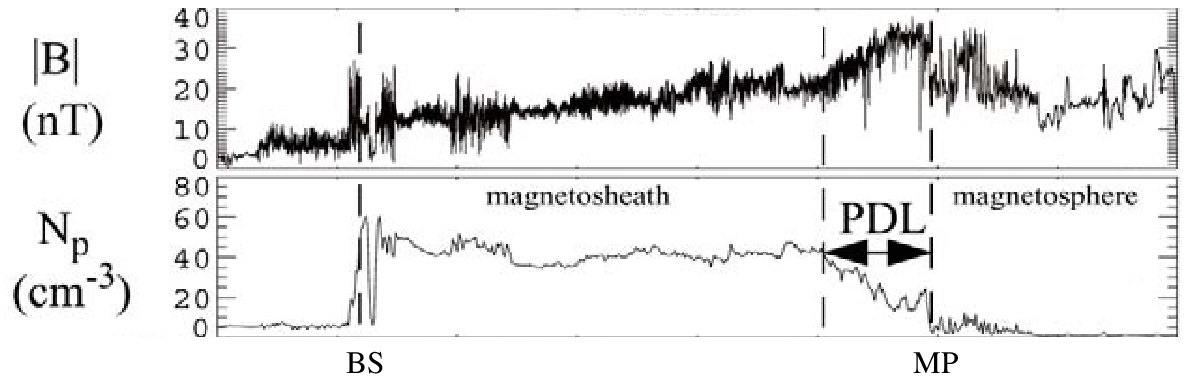}

\caption{Magnetic draping observed in the solar wind as it flows around  the
  Earth's magnetosphere.  Profiles of the field strength and plasma density
  between the Earth's bow shock (vertical dashed line marked BS) and
  magnetopause (MP; an analog of a cluster cold front) as measured by a
  satellite.  Draping results in an increase of the field strength and
  formation of a ``plasma depletion layer'' (PDL) near the magnetopause.
  (Reproduced from {\O}ieroset et al.\ 2004.)}

\label{oieroset}
\end{figure}

\subsection{Possible future measurements using cold fronts}
\label{sec:futurecf}

\subsubsection{Plasma depletion layer and magnetic field}
\label{sec:depl}

There are several interesting tests that are yet to be done using cold
fronts.  First, apart from stabilizing the fronts, magnetic draping
mentioned in the previous section should have another directly observable
effect. When magnetic pressure near the front reaches equipartition, the
total (thermal plus magnetic) pressure should still satisfy the constraints
imposed by the hydrodynamics of the flow. The result is that some of the
plasma is squeezed out of the narrow layer near the front. The width of such
a layer, $\Delta r$, is estimated to be
\begin{equation}
\frac{\Delta r}{R}\approx  \frac{1}{M} \frac{p_{\text{gas}}}{p_B},
\label{eq:drap}
\end{equation}
where $R$\/ is the radius of the front, $M$\/ is the Mach number of the
cloud, and $p_{\text{gas}}$ and $p_B$ are thermal and magnetic pressures in
the undisturbed gas ahead of the front (Lyutikov 2006).  For cluster-sized
transonic cold fronts and $1\;\mu G$ fields, $\Delta r\sim 1$ kpc.  Such
``plasma depletion layers'' have indeed been observed from the satellites in
the corresponding locations of the solar wind (e.g., Zwan \& Wolf 1976;
{\O}ieroset et al.\ 2004; see Fig.\ \ref{oieroset} reproduced from the
latter work). If such a dip is found in the X-ray brightness profile of a
cluster cold front, its width could be used for an independent estimate of
the magnetic field strength in the {\em undisturbed}\/ cluster gas.  

We note that this effect will compete with the density increase in the
stagnation region of the flow (\S\ref{sec:vel}).  There is also a
possibility of a trivial geometric ``depletion'' such as that seen, e.g., in
the X-ray brightness profile of NGC\,1404.  The gas temperature inside that
cold front drops well below 1 keV (Fig.\ \ref{n1404}; Machacek et al.\ 
2005).  Even though the dense gas is present, it does not emit much at $E>1$
keV. Combined with projection, this creates a gap along the front in the
\chandra\ wide-band X-ray image (curious readers can extract it from the
\chandra\ data archive).  Thus, detecting a plasma depletion layer will
require accurate modeling.

\subsubsection{Effective viscosity of intracluster plasma}

Another interesting physical quantity that may be constrained using cluster
cold fronts is the effective viscosity of the ICM. For example, this unknown
property of the ICM has recently attracted the attention of those studying
the cluster cooling flows as a way to transfer the mechanical energy of AGN
explosions into heat (e.g., Fabian et al.\ 2005). As we have already
mentioned, the observed cold fronts appear very smooth and undisturbed by
turbulence. A particularly interesting example is the bullet subcluster in
\1e\ (Fig.\ \ref{1e_a520}), which exhibits long and straight ``wisps'' at
its sides, where one would expect strong turbulence.  There are several
mechanisms that would prevent turbulence from developing; collisional
viscosity is one. Even if it is suppressed (e.g., for similar reasons why
thermal conduction is suppressed), there may be other kinds of viscosity
specific to a magnetized plasma.  Another plausible mechanism is a
stabilizing magnetic layer at the cold front interface, such as the one
discussed above. Such magnetic structures should not be rare; in fact, the
plasma may be filled with them, because any flow in the ICM would create
velocity shear and stretch the magnetic field lines around each moving gas
parcel.  Their net damping effect on gas mixing and turbulence may be
similar to an effective viscosity.  Hydrodynamic simulations that explicitly
include viscosity (e.g., Sijacki \& Springel 2006) would be required to
derive any quantitative constraints.

For an order of magnitude estimate of what one can expect, we can look at
\1e.  To prevent the development of turbulence around the gas bullet, the
Reynolds number of the gas flow should be of order 10 or lower (e.g., Landau
\& Lifshitz 1959, \S26).  One can use a gasdynamic expression $\text{Re}\sim
ML/\lambda$, where $M$\/ is the Mach number of the flow of gas around the
bullet, $L$\/ is the size the bullet, and $\lambda$ is the m.f.p.\ of the
gas particles that determines viscosity. The condition $\text{Re}\lax 10$
gives $\lambda$ in the several kpc range. Interestingly, this is comparable
to the Spitzer m.f.p.\ in the plasma around the bullet.  Thus, the effective
viscosity, whatever its physical nature, may be of the order of the
collisional Spitzer viscosity. Note that it may also be possible to derive
upper limits on viscosity, for example, from the observed RT instability in
A520 (\S\ref{sec:RT}), and perpahs from the presence of turbulence necessary
to explain the cluster radio halos (\S\ref{sec:halos}).

\section{SHOCK FRONTS AS EXPERIMENTAL TOOL}

We now turn to the ICM density discontinuities of another physical nature
which we always expected to find in clusters --- shock fronts.  There are
three types of phenomena that create shocks in the ICM.  As we already
mentioned, in the cluster central regions, powerful AGNs often blow bubbles
in the ICM, which may generate shocks within the central few hundred kpc
regions (e.g., Jones et al.\ 2002; McNamara et al.\ 2005; Nulsen et al.\ 
2005; Fabian et al.\ 2006; Forman et al.\ 2006). These shocks have $M\sim
1$. It is difficult to derive accurate temperature and density profiles for
such low-contrast shocks because of large corrections for the projected
emission, so they are poorly suited for our purpose of using shocks as a
diagnostic tool.

At very large off-center distances (several Mpc), cosmological simulations
predict that intergalactic medium (IGM) should continue to accrete onto the
clusters though a system of shocks that separate the IGM from the hot,
mostly virialized inner regions. The IGM is much cooler than the ICM, so
these accretion (or infall) shocks should be strong, with $M\sim 10-100$
(e.g., Miniati et al.\ 2000; Ryu et al.\ 2003). As such, they are likely to
be the sites of effective cosmic ray acceleration, with consequences for the
cluster energy budget and the cosmic $\gamma$-ray background (see, e.g.,
Blasi, Gabici, \& Brunetti 2007 for a review). However, these shocks have
never been observed in X-rays or at any other wavelengths, and may not be in
the foreseeable future, because they are located in regions with very low
X-ray surface brightness.

If an infalling subcluster has a deep enough gravitational potential to
retain at least some of its gas when it enters the dense, X-ray bright
region of the cluster into which it is falling, we may observe a spectacular
merger shock, such as the one in the X-ray image of \1e\ (Fig.\ \ref{1e}{\em
  c}). Such shocks can be used as tools for a number of interesting studies,
which will be the subject of the rest of this review.

\subsection{Cluster merger shocks}
\label{sec:mergshocks}

Merger shocks have been predicted in the earliest hydrodynamic simulations
of cluster mergers (e.g., Schindler \& M\"uller 1993; Roettiger, Burns, \&
Loken 1993; Burns 1998).  They have relatively low Mach numbers, $M\lax 3$
(e.g., Gabici \& Blasi 2003; Ryu et al.\ 2003), simply because the sound
speed in the gas of the main (bigger) cluster and the velocity of the
infalling subcluster both reflect the same gravitational potential of the
main cluster. Indeed, the central depth of the King potential is
$\Phi_0=-9\sigma_r^2$, where $\sigma_r$ is the radial velocity dispersion of
galaxies in this potential, which in turn is close to the average velocity
of particles in the intracluster gas in hydrostatic equilibrium (e.g.,
Sarazin 1988).  So an infalling test particle, or a small subcluster, should
acquire $M\sim 3$ at the center.  In practice, for a merger of clusters with
comparable masses (what is usually called a ``major merger''), it is
unlikely to observe a shock with $M$\/ much greater than 1. Such a merger
would generate multiple successive shocks, each one significantly preheating
the gas of the whole combined system, thereby reducing the Mach numbed for
any subsequent shock.

Simulations also indicate that it is unlikely to find a shock front during
the inbound leg of the infalling subcluster's trajectory, because the front
must climb up the steep gas density gradient of the main cluster.  At the
merger stage shown in top panel of Fig.\ \ref{a2142_scheme}, shock fronts
may slow down and disappear while ascending the density peaks, if the peaks
are high enough. At the same time, regions of the front away from the
symmetry axis will continue to propagate past the density peak, and may
eventually form a continuous surface behind it.

\begin{figure}
\centering
\includegraphics[width=0.63\textwidth]%
{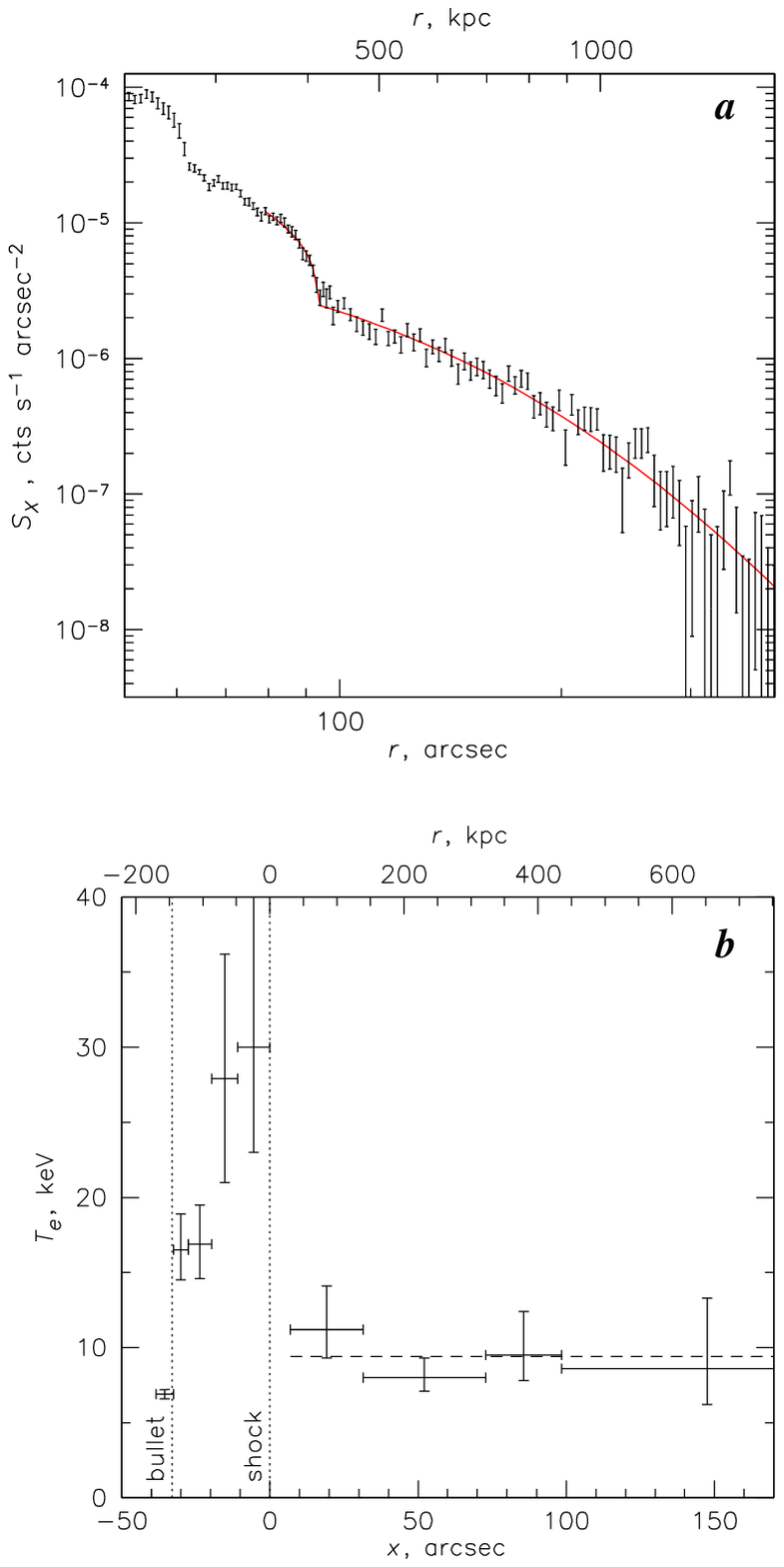}

\caption{Radial profiles of the \1e\ X-ray brightness ({\em a}) and projected
  temperature ({\em b}) in a narrow sector crossing the tip of the bullet
  (the first big drop of the brightness) and the shock front (the second big
  drop).  The $r$\/ coordinate in panel {\em a}\/ is measured from the
  shock's center of curvature; the $x$\/ coordinate in panel {\em b} is
  measured from the shock surface.  Red line in panel {\em a}\/ shows the
  best-fit model for the shock jump (a projected sharp spherical density
  discontinuity by a factor of 3).  Vertical dotted lines in panel {\em b}\/
  show the boundaries of the cool bullet and the shock; dashed line shows
  the average pre-shock temperature.  There is a subtle additional edge
  between the bullet and the shock; the gas temperature inside it is lower
  (the lower two crosses between the vertical lines).  That region is not
  used for any shock models. Error bars are 68\%.  (Reproduced from M06.)}

\label{1e_profs}
\end{figure}

\begin{figure}[t]
\centering
\includegraphics[width=\linewidth,bb=70 274 507 733]%
{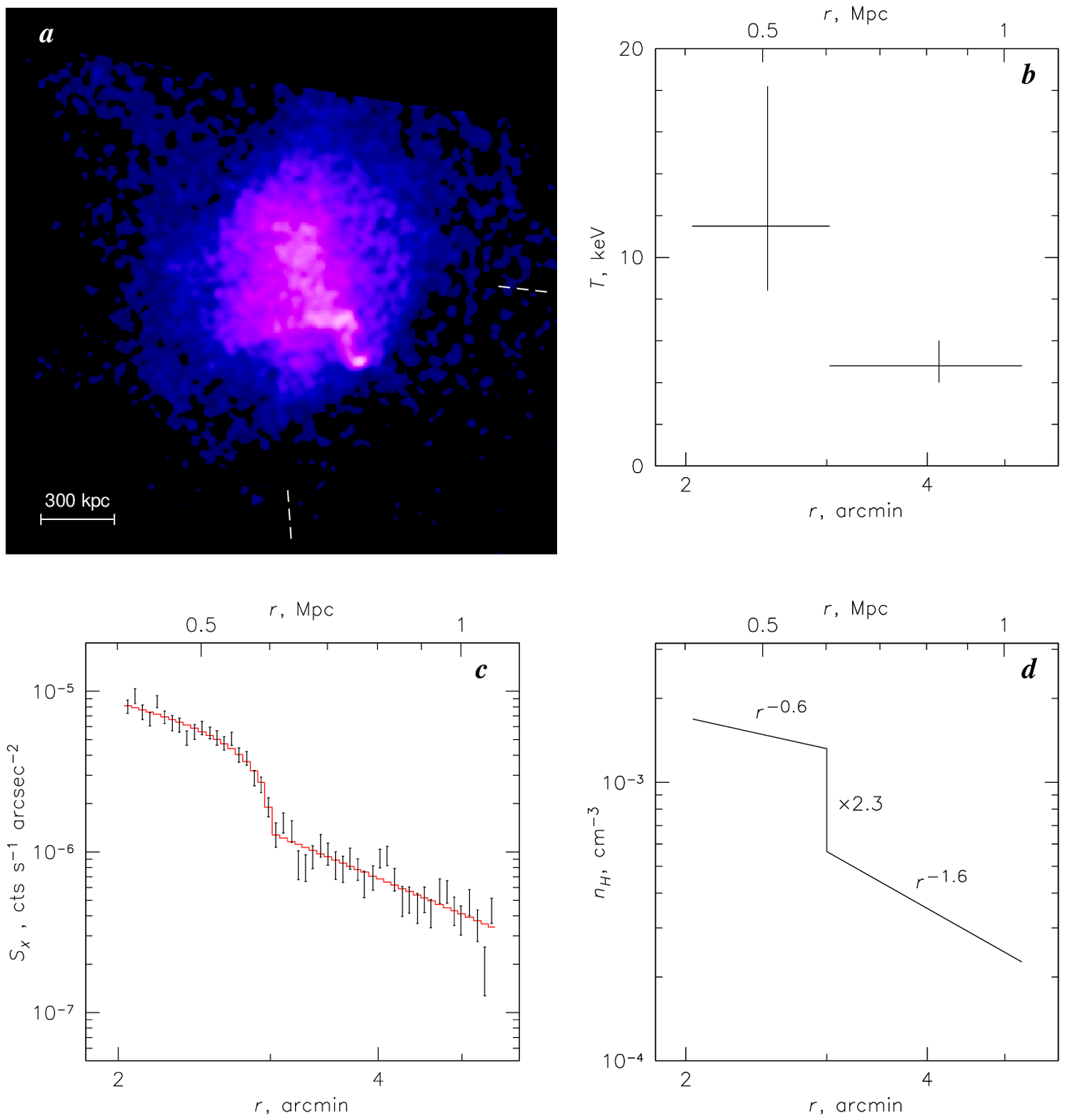}

\caption{Shock front in A520. ({\em a}) A \chandra\ image
  (slightly smoothed) with point sources removed.  The bow shock is a faint
  blue edge southwest of the bright irregular remnant of a dense core. White
  dashed lines mark a sector used for radial X-ray brightness and projected
  temperature profiles across the shock (panels {\em b,c}). The profiles are
  extracted excluding the core remnant. ({\em d}) A three-dimensional model
  fit to the brightness profile; its projection is shown as a red line in
  panel ({\em c}). (Reproduced from M05.)}

\label{a520_profs}
\end{figure}

Regions of high-entropy gas in clusters have been observed with
lower-resolution X-ray telescopes such as \rosat, \asca\ and \xmm\ and
interpreted as the result of shock heating (e.g., Henry \& Briel 1995, 1996;
Markevitch, Sarazin, \& Irwin 1996b; Markevitch et al.\ 1999a; Belsole et
al.\ 2003, 2004).  Shock-heated regions are also routinely found by
\chandra\ (e.g., Markevitch \& Vikhlinin 2001; Markevitch et al.\ 2003a;
Kempner \& David 2004; G04). However, as of this writing, only two shock
{\em fronts}, exhibiting both a sharp gas density edge and an unambiguous
temperature jump, were found by \chandra, those in \1e\ (Fig.\ 
\ref{1e_profs}; M02; Markevitch 2006, hereafter M06) and A520 (Fig.\ 
\ref{a520_profs}; Markevitch et al.\ 2005, hereafter M05).  Such finds are
so rare because one has to catch a merger when the shock has not yet moved
to the low-brightness outskirts, and is propagating nearly in the plane of
the sky, to give us a clear view of the shock discontinuity. In addition,
the shocks in A520 and \1e\ have $M=2-3$, which provides a big enough gas
density jump to enable accurate deprojection.  Merger shock fronts may have
been found in a couple of other clusters (e.g., in A3667, V01; A754,
Krivonos et al.\ 2003; Henry, Finoguenov, \& Briel 2004), but temperature
data for them either do not exist or are uncertain. Below, we will discuss
what can be learned from the fronts in \1e\ and A520.

\subsection{Mach number determination}
\label{sec:1e_M}

Let us now look at the profiles of the X-ray brightness and projected gas
temperature derived in a narrow sector crossing the bullet nose and the
shock front in \1e. They are shown in Fig.\ \ref{1e_profs} (M06). From these
data, we can derive a Mach number and velocity of the shock, which should
also give the approximate velocity of the subcluster.  In Fig.\ 
\ref{1e_densprof}, we showed an approximate gas density model with two
abrupt jumps, which in projection describes this brightness profile. (This
fit corresponds to an early subset of the data shown in Fig.\ 
\ref{1e_profs}. Note also the different reference points for the $r$\/
coordinate used in Figs.\ \ref{1e_densprof} and \ref{1e_profs}.) The inner
brightness edge in Fig.\ \ref{1e_profs}{\em a}\/ is the bullet; its boundary
at $r=65''$ is a cold front, as seen from the temperature jump in panel
({\em b}) at $x=-30''$.  The edge at $r=90''$ is the shock front, as
confirmed by the temperature jump of the right sign.  There is another
subtle brightness edge between these main features (Fig.\ \ref{1e_profs}).
It is unrelated to the shock, so we will exclude it when deprojecting the
shock temperatures and densities.

We can determine the density discontinuity across the shock just as we did
for the cold fronts, by fitting the brightness profile with a model with an
abrupt spherical density jump (\S\ref{sec:merg}).  The best-fit model (red
line in Fig.\ \ref{1e_profs}{\em a}) has a density jump by a factor of 3,
which includes a small correction for the observed gas temperature change
across the front.

The Rankine-Hugoniot jump conditions (e.g., Landau \& Lifshitz 1959, \S89)
relate the density jump at the shock, $r$, and the Mach number of the shock,
$M\equiv v/c_s$, where $c_s$ is the velocity of sound in the pre-shock gas
and $v$\/ is the velocity of that gas w.r.t.\ the shock surface:
\begin{equation}
M=\left[\frac{2r}{\gamma+1-r(\gamma-1)}\right]^{1/2}.
\label{eq:M}
\end{equation}
Using this equation, the above density jump gives a Mach number $M=3.0\pm
0.4$. 

Note that $M$\/ can be derived {\em independently}\/ using the temperature
jump (using eq.\ \ref{eq:r} and the above equation), provided that the
effective adiabatic index $\gamma$ is known (i.e., no dominant relativistic
particle component, significant energy leakage into particle acceleration,
etc.) and that there is temperature equilibrium between electrons and ions
(the latter assumption is needed only for the estimate based on the
temperature jump). We will see in the next section that the observed
temperature jump in \1e\ is in good agreement with the value of $M$\/ from
the density jump.  A similar analysis can be performed for the shock in A520
(Fig.\ \ref{a520_profs}), where a density jump by a factor of 2.3 gives
$M=2.1^{+0.4}_{-0.3}$, again consistent with the temperature jump (M05). For
relatively weak shocks such as these, the accuracy with which we can
determine $M$\/ from the X-ray derived density discontinuity is better than
what we can do from the temperatures.  However, as $M$\/ increases, the
density jump asymptotically goes to a finite value ($r=4$ for $\gamma=5/3$),
while the temperature jump continues to grow, so the situation reverses.

Sound speed in the pre-shock gas can be determined from its observed
temperature. The above Mach numbers then give the shock velocities of 4700
\kms\ and 2300 \kms\ for \1e\ and A520, respectively.  From the continuity
condition, the respective velocities of the post-shock gas w.r.t.\ the shock
surface are $r$\/ times lower, 1600 \kms\ and 1000 \kms. For \1e, the
distance between the centers of merging subclusters is 720 kpc (see the weak
lensing map in Fig.\ \ref{1e_lens}). Assuming that the bullet subcluster's
velocity is close to the shock velocity (although it can be lower, Springel
\& Farrar 2007; Milosavljevic et al.\ 2007), the two subclusters have passed
through each other just 0.15 Gyr ago.

\subsection{Front width}
\label{sec:shockwidth}

The X-ray brightness profile in Fig.\ \ref{1e_profs} does not exclude a
shock front of a finite width; in fact, it marginally prefers a density jump
widened by a Gaussian with $\sigma\approx 8''$ (35 kpc). Curiously, this is
of the order of the collisional m.f.p.\ in the gas around the shock --- but
it may simply be a coincidence, since, e.g., any deformations of the front
shape seen in projection would smear the edge.  The best-fit amplitude of
the jump that was used above to derive the Mach numbers does not depend
noticeably on this width.

In this regard, it is interesting to look at simulations by Heinz \&
Churazov (2005), who studied propagation of a shock in an ICM filled with
bubbles of relativistic plasma. Such a mixture in pressure equilibrium may
form as a result of AGN activity over the cluster lifetime; cluster radio
relics may be examples of such regions containing fossil relativistic plasma
(e.g., En{\ss}lin \& Gopal-Krishna 2001).  Because the sound speed inside
such a bubble is very high, a shock front surface will be deformed on a
linear scale of these bubbles.  Heinz \& Churazov proposed such smearing as
an explanation for the lack of visible strong AGN jet-driven shocks in the
cluster centers.  The fact that the fronts in \1e\ and A520 can be seen at
all in the X-ray images indicates that, at least in these clusters, the bulk
of the ICM is not filled with fossil bubbles greater than several tens of
kpc in size.  On the other hand, there is an example of A665, in which the
overall X-ray morphology and the temperature map suggest that there should
be a shock front, and there is a gas density excess, but no density
discontinuity is seen (Markevitch \& Vikhlinin 2001; G04). It might be that
an observable sharp discontinuity did not form there because the shock has
to propagate in such a mixture.

\subsection{Mach cone and reverse shock?}

Before proceeding to some of the interesting shock-based measurements, we
would like to address two questions that people familiar with shocks in
other astrophysical contexts frequently have: the Mach cone%
\footnote{The Mach cone discussion here differs from the version published
  in {\em Physics Reports}\/ --- it is updated with results from simulations
  by Springel \& Farrar (2007).}
and the reverse shock.

A small body moving supersonically creates a Mach cone with an opening angle
$\sin\varphi = M^{-1}$. For $M=3$, the shock in \1e\ should have an
asymptotic angle of 20\deg\ from the symmetry axis. However, the image
(Fig.\ \ref{1e}{\em c}) shows a more widely open ``Mach cone''. A possible
reason is easy to understand if we consider that the bullet is not a solid
body, but a gas cloud whose outer, less dense gas is being continuously
stripped by the flow of the shocked gas.  The tip of the bullet is becoming
smaller with time, as shown schematically in Fig.\ \ref{slosh_scheme} (see
\S\ref{sec:discont}).  The off-axis parts of the shock front that we see at
present are not driven by the bullet that we see at present, but by a bigger
bullet that has existed just a short while ago. Thus, the shape of the front
(at least of its bright region which we can follow in the image) does not
correspond to a Mach cone, but rather reflects the change of the bullet size
and velocity with time. Another significant effect is the inflow of the
pre-shock gas caused by the gravity of the subcluster (Springel \& Farrar
2007). This inflow is faster at the nose of the shock (which currently
coincides with the subcluster's mass peak, see Figs.\ \ref{1e}{\em bc}),
which flattens the front shape.

We can test this by deriving an X-ray brightness profile across an off-axis
stretch of the shock front, e.g., in a narrow sector pointing 30\deg\ 
clockwise from the bullet velocity direction. For a stationary oblique shock
with such an angle to the uniform upstream flow, the density jump should be
reduced from 3.0 at the nose to 2.8 (e.g., Landau \& Lifshitz 1959, \S92).
However, the observed jump is a factor of 2.1, which corresponds to a
smaller $M$. This is consistent with a higher inflow velocity of the
pre-shock gas at the nose of the shock.

In addition, the main cluster's radially declining density profile, in which
the shock propagates, and deceleration of the bullet by gravity and ram
pressure should also affect the shape of the front (M02).

Those working with supernova remnants (SNR) may also ask where is the
reverse shock in \1e\ --- a front that propagates inwards from the outer
surface of an SNR. If we look at the scheme in Fig.\ \ref{a2142_scheme}, it
is clear that the reverse shock for shock 1 is shock 2, and vice versa.  The
current stage of the merger in \1e\ corresponds to the bottom panel in Fig.\ 
\ref{a2142_scheme}. By this time, the reverse shock has attempted to climb
the sharp density peak of the bullet subcluster, but failed to penetrate
inside the radius where we now have a cold front.  As mentioned in
\S\ref{sec:mergshocks}, the outer regions of that shock must have moved past
the bullet and away from the picture; from the symmetry, it should be
somewhere on the opposite side from the cluster center.

\subsection{Test of electron-ion equilibrium}
\label{sec:tei}

The post-shock temperature that enters the Rankine-Hugoniot jump conditions
is the temperature that all plasma particle species acquire when they reach
equilibrium after the shock passage. In a collisional plasma, protons, whose
thermal velocity is lower than the shock velocity, are heated dissipatively
at the shock layer that has a width of order the collisional m.f.p. The
faster-moving electrons do not feel the shock (for $M\ll
(m_p/m_e)^{1/2}=43$) and are compressed adiabatically. Subsequently,
electrons and protons equilibrate via Coulomb collisions on a timescale
(e.g., Zeldovich \& Raizer 1967)
\begin{equation}
\tau_{\rm c}= 2\times 10^8\;{\rm yr}
\left(\frac{n_e}{10^{-3}\;{\rm cm}^{-3}}\right)^{-1}
\left(\frac{T_e}{10^8\,{\rm K}}\right)^{3/2}.
\label{eq:tei}
\end{equation}
For a shock in a magnetized plasma such as the ICM, the final post-shock
temperature should be the same (provided that the kinetic energy does not
leak into cosmic rays, etc.), but the shock structure can be very different.
Indeed, studies of solar wind shocks in-situ showed that the electron and
proton temperature jump occurs on a linear scale of order several proton
gyroradii, many orders of magnitude smaller than their collisional m.f.p.
(Montgomery, Asbridge, \& Bame 1970 and later works), which is why these
shocks are called collisionless. Therefore, it would not be surprising to
find a different rate of electron heating at the shock, and a shorter
electron-proton equilibration timescale. (To find a longer timescale would
be surprising, because Coulomb collisions do occur even in the
``collisionless'' plasma.)

The bow shock in \1e\ offers a unique experimental setup to determine how
long it takes for post-shock electrons to come to thermal equilibrium with
protons in a magnetized plasma (M06). Given the likely turbulent widening of
the heavy ion emission lines (and, of course, for lack of the required
energy resolution at present), we cannot directly measure $T_i$ in X-rays,
only $T_e$.  However, we can use the accurately measured gas density jump at
the front (\S\ref{sec:1e_M}) and the pre-shock electron temperature to
predict the post-shock adiabatic and instant-equilibration electron
temperatures, using the adiabatic and the Rankine-Hugoniot jump conditions,
respectively, and compare them with the data.  Furthermore, we also know the
downstream velocity of the shocked gas flowing away from the shock (1600
\kms, \S\ref{sec:1e_M}).  This flow effectively spreads out the time
dependence of the electron temperature along the spatial coordinate in the
plane of the sky.  The Mach number of the \1e\ shock is conveniently high,
such that the adiabatic and shock electron temperatures are sufficiently
different (for $M\lax 2$, they become close and difficult to distinguish,
given the temperature uncertainties).  It is also not a strong shock, for
which the density jump would just be a factor of 4 and would not let us
directly determine $M$.  Furthermore, the distance traveled by the
post-shock gas during the time given by eq.\ (\ref{eq:tei}), $\Delta x\simeq
230$ kpc $=50''$, is well-resolved by \chandra.

\begin{figure}[t]
\centering

\includegraphics[width=0.65\linewidth,bb=42 186 548 685,clip]%
{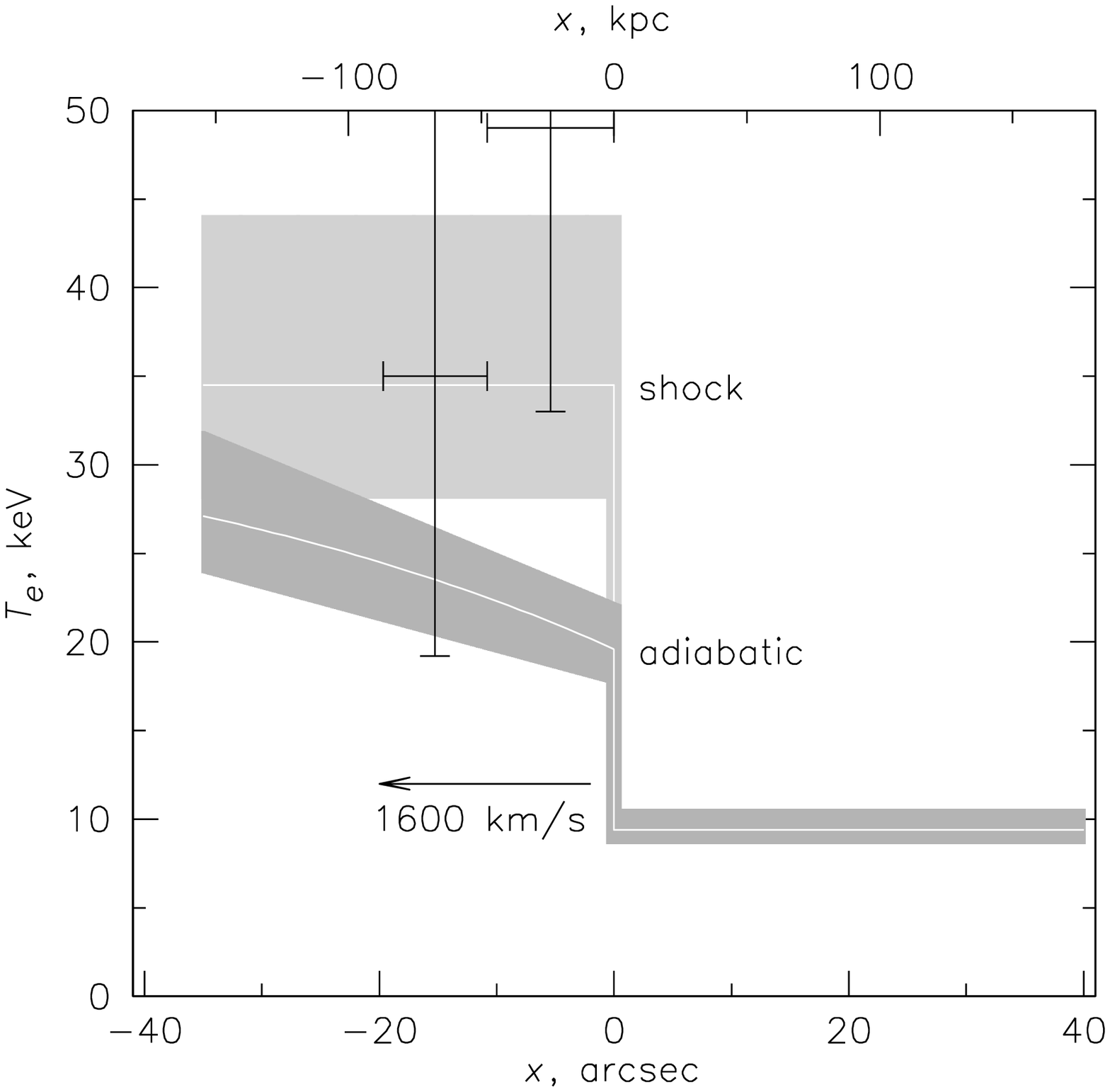}

\caption{Electron-ion equilibrium at shock in \1e.
  Deprojected electron temperatures for the two outer post-shock bins of the
  temperature profile from Fig.\ \ref{1e_profs}, overlaid on the model
  predictions (with error bands) for instant equilibration (labeled
  ``shock'', light gray) and adiabatic compression followed by collisional
  equilibration (dark gray). The velocity shown is for the post-shock gas
  relative to the shock.  Error bars are 68\%.  (Reproduced from M06.)}

\label{1e_tei}
\end{figure} 

As we already mentioned, there is a subtle secondary brightness edge between
the bullet and the shock, behind which the temperature decreases (Fig.\ 
\ref{1e_profs}). It is most likely caused by residual cool gas from the
subcluster in one form or another, and is unrelated to the shock.
Therefore, we can use only the temperature profile between the shock and this
edge, where we have two bins.  Using the gas density profile in front of the
shock, we can subtract the contributions of the cooler pre-shock gas
projected onto these post-shock bins, assuming spherical symmetry in this
small segment of the cluster. The shock brightness contrast is high, which
makes this subtraction sufficiently accurate. In Fig.\ \ref{1e_tei}, the
deprojected temperatures are overlaid on the two models: instant
equilibration and adiabatic compression with subsequent equilibration on a
timescale given by eq.\ (\ref{eq:tei}). (The plot assumes a constant
post-shock gas velocity, which of course is not correct, but we are only
interested in the immediate shock vicinity.)  The deprojected post-shock gas
temperatures are so high compared with the \chandra\ energy band that only
their lower limits are meaningful. The temperatures are consistent with
instant heating; equilibration on the collisional timescale is excluded,
although with a relatively low 95\% confidence. The equilibration timescale
should be at least 5 times shorter than $\tau_{\rm c}$.

A few sanity checks have been performed in M06. In particular, the
possibility of a non-thermal contamination of the spectra was considered.
\1e\ has a radio halo (Fig.\ \ref{1e}{\em d}; Liang et al.\ 2000), which has
an edge right at the shock front (M02; see below). Therefore, there may be
an IC contribution from relativistic electrons accelerated at the shock.
However, in the \chandra\ energy band, the power-law spectrum of such
emission for any $M$\/ would be {\em softer}\/ than thermal (which for these
temperatures has a flat effective photon index of $-1.4$), so it cannot bias
our temperature measurements high.  It is unfortunate that a cluster with
such a perfect geometric setup and $M$\/ is so hot that \chandra\ can barely
measure the post-shock temperatures; however, there are not many of them to
choose from. The available A520 data do not have the accuracy for such a
measurement.

\begin{figure}[t]
\centering

\includegraphics[width=0.8\textwidth,bb=48 157 559 456]%
{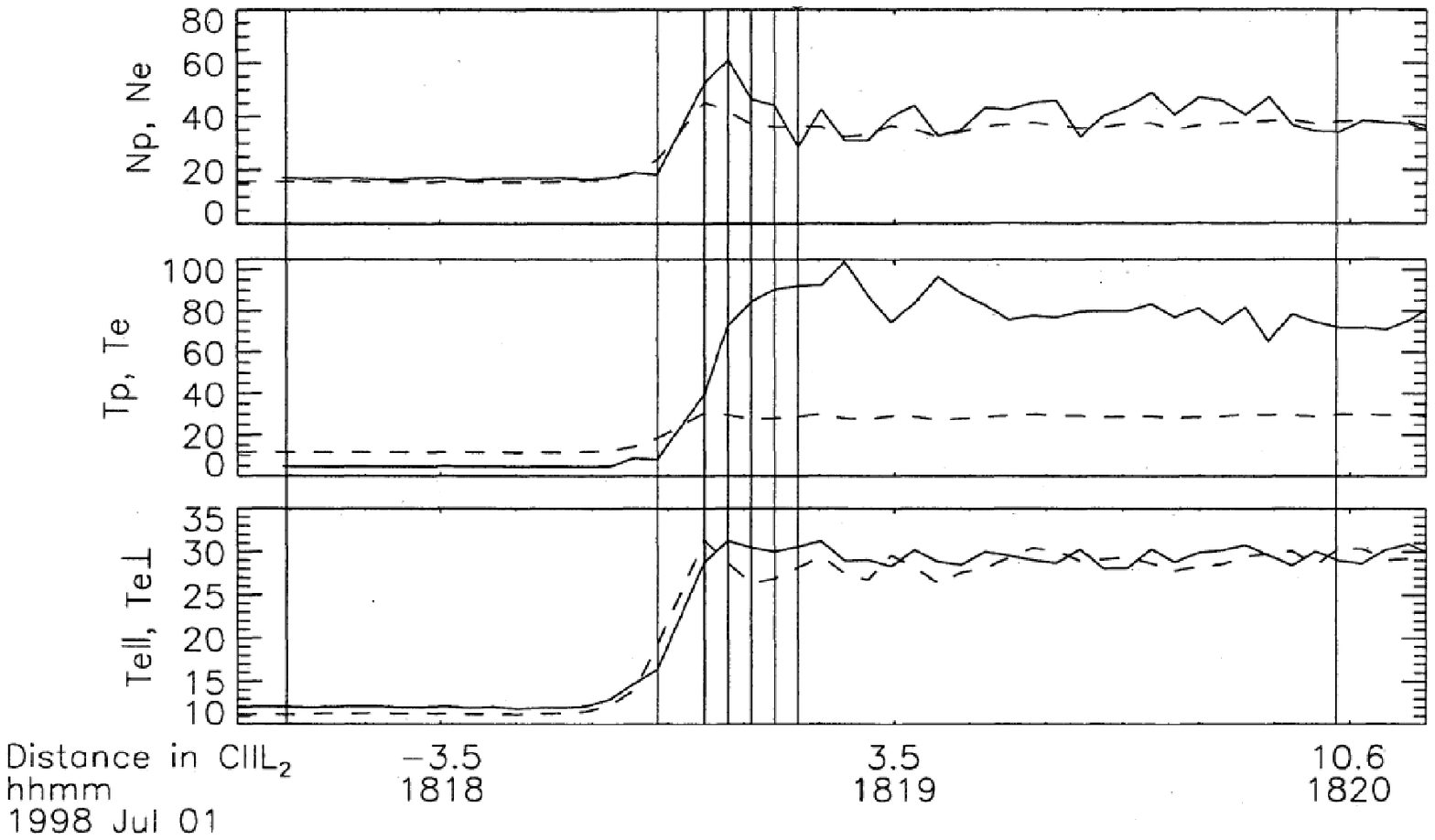}

\caption{Structure of the Earth's bow shock observed in-situ from a 
  satellite.  Top two panels show proton (solid lines) and electron (dashed
  lines) density (in units of cm$^{-3}$) and temperature ($10^4$ K)
  profiles.  Lower panel shows temperature components parallel (solid) and
  perpendicular (dashed) to the magnetic field. The linear scale unit
  CIIL$_2$ (the downstream proton cyclotron frequency times the flow
  velocity) corresponds to about 35 km. This shock propagates
  quasi-perpendicularly to the upstream magnetic field and has $M\approx 3$,
  $M_A\approx 4.7$ and electron $\beta=1.5$.  Electrons are not heated at
  the shock as much as protons, which is typical for solar wind shocks.
  (Reproduced from Hull et al.\ 2001.)}

\label{hull}
\end{figure}

\subsubsection{Comparison with other astrophysical plasmas}

To our knowledge, the above result is the first direct indication for any
astrophysical plasma that after a shock, electrons reach the equilibrium
temperature on a timescale shorter than collisional.  However, it cannot
distinguish between two interesting possibilities: electrons being heated to
the ``correct'' temperature right at the shock, or electrons and protons
equilibrating soon after via a mechanism unrelated to the shock. This is
where the solar wind and supernova remnants (SNR) provide complementary
evidence.

A large amount of data has been gathered by space probes for the
heliospheric shocks, especially for the Earth's bow shock (for recent
reviews see, e.g., Russell 2005 and references therein).  Many of these
shocks have high Alfv\'en Mach numbers $M_A$, moderate sonic Mach numbers,
and a ratio of pre-shock thermal to magnetic pressure $\beta>1$, which
should make them qualitatively similar to merger shocks in clusters
(although $M_A$ and $\beta$ in clusters are typically much higher; e.g.,
upstream of the \1e\ shock, $M_A=70\; (B/1\,\mu{\rm G})^{-1}$ and
$\beta=650\; (B/1\,\mu{\rm G})^{-2}$, so the similarity is not guaranteed).
In most observed heliospheric shocks, and in all the stronger ones,
electrons were heated much less than protons, barely above the adiabatic
compression temperature (e.g., Schwartz et al.\ 1988).  Figure \ref{hull}
gives a typical example of a shock crossing by the {\em Wind}\/ satellite
(Hull et al.\ 2001).  Both $T_p$ and $T_e$ quickly reach their respective
asymptotic values, which are $T_e < T_p$.  These in-situ measurements cannot
tell us what happens after that on timescales comparable to $\tau_{\rm c}$
(which corresponds to a linear scale of several A.U.), but they do show that
equilibration at least does not occur on scales $\sim 10^{-7} \tau_{\rm c}$.

Another site of actively studied shocks is SNRs. The typical $\beta$ in SNR
plasmas is of the same order of that in clusters, much higher than that in
the solar wind. Outer shocks in SNRs are typically very strong and may be
modified by cosmic ray acceleration (see, e.g., Vink 2004 for a review).
Nevertheless, these very different shocks also exhibit $T_e/T_i<1$ in most
cases (e.g., Rakowski 2005; Raymond \& Korreck 2005), although with
considerable uncertainty due to the difficulty of estimating the velocity
and $T_i$ for a strong shock.  (One cannot use the density jump as we did,
as it is at its asymptotic value; on the other hand, in some Galactic SNRs,
the shock velocity can be directly measured using high-resolution X-ray
images taken several years apart, which, of course, is impossible for
clusters.)  The typical timescale $\tau_{\rm c}$ for an SNR plasma is of
order $10^3$ yr, which in many cases is comparable to the SNR's age.
Furthermore, the region for which the temperatures are derived is usually
close to the shock, so the timescales sampled by the temperature
measurements in the SNR plasma are usually $\ll \tau_{\rm c}$.  Thus,
similarly to the solar wind results, the SNR data indicate that
collisionless shocks with a wide range of parameters produce $T_e/T_i<1$,
but, to our knowledge, do not constrain the timescale of the subsequent
equilibration.

The solar wind and SNR results on one hand, and our \1e\ measurement on the
other, appear to leave only one of the two possibilities mentioned above,
namely, that shocks produce unequal electron and ion temperatures, which
quickly equalize outside the shock layer. Proposing an equilibration
mechanism that works faster than Coulomb collisions in plasma is beyond the
scope of this review (and our expertise).

\subsection{Shocks and cluster cosmic ray population}
\label{sec:halos}

Cluster mergers convert kinetic energy of the gas in colliding subclusters
into thermal energy by driving shocks and turbulence. A fraction of this
energy may be diverted into nonthermal phenomena, such as magnetic field
amplification and the acceleration of ultrarelativistic particles. Such
particles manifest themselves via synchrotron radio halos (recently reviewed
by, e.g., Feretti 2002, 2004; Kempner et al.\ 2004) and IC hard X-ray
emission (e.g., Fusco-Femiano et al.\ 2005; Rephaeli \& Gruber 2002).  This
energy fraction may be significant and depends on the exact mechanism of the
production of halo-emitting electrons, which is not yet understood. However,
a consensus emerges that the underlying cause and energy source should
indeed be the subcluster mergers (e.g., Feretti 2002; Buote 2001; G04).

The radio halo generating electrons are relatively short-lived ($10^7-10^8$
yr) due to IC and synchrotron energy losses (e.g., Sarazin 1999). This is
much shorter than the diffusion time across a cluster-sized halo, so a
mechanism is required by which these electrons are locally and
simultaneously (re)accelerated over the halo volume.  Several possibilities
were proposed, including radio galaxies, so-called ``secondary electron''
production by interactions of the long-lived cosmic ray protons with the ICM
protons, and ``primary electron'' mechanisms, in which electrons are
accelerated directly by merger-driven turbulence or shocks (see, e.g.,
Brunetti 2003 for a review).

Shock fronts provide a unique experimental tool in this area. They are those
rare locations in clusters where we can determine gas velocities in the sky
plane and the gas compression ratio (\S\ref{sec:1e_M}). They also create
high-contrast features in the cluster X-ray images and, as we will see, in
the radio images. Simply by comparing the X-ray and radio images of clusters
with shock fronts, one may be able to determine the contribution of shocks
to the halo production.

\begin{figure}[t]
\centering
\includegraphics[width=0.7\textwidth]%
{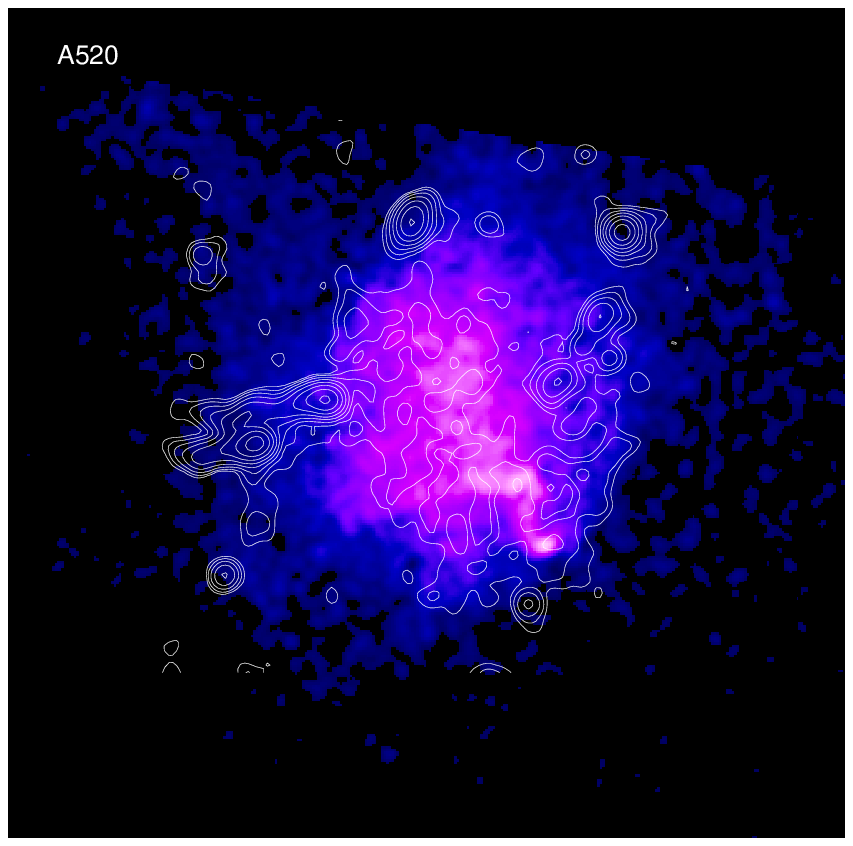}

\caption{Radio halo brightness (contours; Govoni et al.\ 2001) overlaid on
  the \chandra\ X-ray image of A520 (shown without this overlay in Fig.\ 
  \ref{a520_profs}{\em a}). This 1.4 GHz map has an angular resolution of
  15\as; unrelated compact radio sources are left in the image.  The halo is
  a low-brightness mushroom-like structure with a ``cap'' that coincides
  with the X-ray bow shock southwest of center and a wide ``stem'' extending
  along the NE-SW axis of the cluster. See Fig.\ \ref{1e}{\em d}\/ for a
  similar overlay for \1e.}

\label{a520_halo}
\end{figure} 

Both clusters with known shock fronts, \1e\ and A520, have prominent radio
halos (Liang et al.\ 2000; Govoni et al.\ 2001). Their radio brightness
contours are overlaid on the X-ray images in Figs.\ \ref{1e}{\em d}\/ and
\ref{a520_halo}.  One immediately notices a striking coincidence of the SW
edge of the radio halo in A520 with the shock front (M05). A similar
extension of the radio halo edge to the bow shock is seen in the less
well-resolved halo in \1e\ (M02; G04).  In another merging cluster, A665, a
``leading'' edge of the radio halo also corresponds to a region of hot gas
that's probably behind a bow shock (Markevitch \& Vikhlinin 2001; G04),
although the X-ray image of A665 does not show a gas density edge at the
putative shock (\S\ref{sec:shockwidth}). The cold front cluster A3667
exhibits two radio relics at the opposite sides of the cluster (R\"ottgering
et al.\ 1997), along the merger axis suggested by the cold front.  They are
often mentioned as evidence of bow shocks at those locations; however, they
are located too far in the cluster periphery, where X-ray imaging is not
currently feasible.

The overall structure of the radio halo in A520 suggests two distinct
components, a mushroom with a stem and a cap, where the main stem component
goes across the cluster and the cap ends at the bow shock.  The main halo
component is located in the region of the cluster where one expects
relatively strong turbulence (G04), but about which we cannot say much
observationally at present.  However, the part of the halo that coincides
with the shock front can already be used for interesting tests.  Below we
discuss two main possibilities for its origin, how to distinguish between
them, and propose some future tests that can be done using this shock (M05).

\subsubsection{Shock acceleration}
\label{sec:accel}

One explanation for the radio edge is acceleration of electrons to
ultrarelativistic energies by the shock.  The shock with a density jump
$r$\/ should generate electrons with a Lorentz factor $\gamma$ and energy
spectrum $dN/d\gamma=N_0 \gamma^{-p}$ with 
\begin{equation}
p=\frac{r+2}{r-1}
\label{eq:p}
\end{equation}
via first-order Fermi acceleration (see, e.g., Blandford \& Eichler 1987 for
a review). For the A520 shock with $r\simeq 2.3$, $p\simeq 3.3$.  The
synchrotron emission should have a spectrum $I_\nu \propto \nu^{-\alpha}$
with $\alpha=(p-1)/2 \simeq 1.2$ right behind the shock front.  However,
these electrons are short-lived because of IC and synchrotron energy losses
(e.g., Rybicki \& Lightman 1979), and their spectrum will quickly steepen.
The respective electron lifetimes, $t_{\rm\,IC}$ and $t_{\rm syn}$, are
\begin{equation}
t_{\rm\,IC} = 2.3\times 10^{8}\; \left(\frac{\gamma}{10^4}\right)^{-1}
              (1+z)^{-4}\; {\rm yr}
\label{eq:tic}
\end{equation}
and
\begin{equation}
t_{\rm syn} = 2.4\times 10^{9}\; \left(\frac{\gamma}{10^4}\right)^{-1}
      \left(\frac{B}{1\,\mu G}\right)^{-2}\; {\rm yr}
\label{eq:tsyn}
\end{equation}
(e.g., Sarazin 1999). IC losses dominate for $B<3\,\mu G$; other losses are
negligible for our range of energies and fields.  For a power-law electron
spectrum with $p=2-4$ (expected for shocks with $M>1.7$), the contribution
of different $\gamma$ at a given synchrotron frequency has a peak at
\begin{equation}
\gamma_{\rm peak} \approx 10^4\, \left(\frac{\nu}{1\,{\rm GHz}}\right)^{1/2}
   \left(\frac{B}{1\,\mu G}\right)^{-1/2}.
\label{eq:gmax}
\end{equation}
Assuming $B\sim 1\,\mu G$, the lifetime for electrons with $\gamma\sim
1.2\times 10^4$ that emit at our radio image frequency of 1.4 GHz in A520 is
$\sim 10^8$ yr.  Thus, given the 1000 \kms\ velocity of the downstream flow
(\S\ref{sec:1e_M}) that carries these electrons away from the shock, the
width of the synchrotron-emitting region should only be about 100 kpc,
beyond which the electrons cool out of the 1.4 GHz band.  This scale is an
order of magnitude smaller than the size of the halo, so the whole halo
cannot be produced by particles accelerated at this shock. A similar
conclusion was reached for the front in \1e\ by Siemieniec-Ozieb{\l}o
(2004), and this is expected for merger shocks in general (e.g., Brunetti
2003).  However, the cap-like part of the radio halo appears to have just
the right width, $\Delta R\lax 100$ kpc (considering the finite angular
resolution).  Thus, with the available data, this region is not inconsistent
with shock acceleration.  While the relativistic electrons in this structure
cool down soon after the shock passage, some may later be picked up and
re-accelerated as they reach the turbulent region behind the subcluster
core, where the stem-like halo component forms.

Because the bow shock is spatially separated from the turbulent area further
downstream (except for the region around the small dense core fragments) and
there is no reason to expect significant turbulence and additional
acceleration in that intermediate region, the cap-like structure is likely
to exhibit a measurable spectral difference from the main halo. Within the
100 kpc-wide strip along the shock, the spectrum should quickly steepen
starting from $\alpha=1.2$.  If the region is unresolved, the resulting
mixture would have a volume-averaged slope $\bar{\alpha}\approx \alpha+1/2$
(Ginzburg \& Syrovatskii 1964) which is significantly steeper than
$\alpha\simeq 1-1.2$ observed on average in most halos (e.g., Feretti 2004),
the bulk of which is probably continuously powered by turbulence.
Interestingly, Feretti et al.\ (2004) found that the presumed post-shock
region in A665 indeed exhibits the steepest radio spectrum in the spectral
index map of the cluster, which is consistent with the above two-component
cap + stem picture.

\subsubsection{Compression of fossil electrons}
\label{sec:compr}

The efficiency with which collisionless shocks can accelerate relativistic
particles is unknown, and may be insufficient to generate the observed radio
brightness.  The radio edges in A520 and \1e\ offer an interesting prospect
for constraining it. If the acceleration efficiency is low, the observed
radio edge may alternatively be explained by an increase in the magnetic
field strength and the energy density of the {\em pre-existing}\/
relativistic electrons, simply due to the gas compression at the shock. Such
pre-existing electrons with energies below those required to emit at our
radio frequencies (eq.\ \ref{eq:gmax}) may accumulate from past mergers
(e.g., Sarazin 1999).  In this model, the pre-existing electrons must
produce low-brightness diffuse radio emission {\em in front of}\/ the bow
shock, whose intensity and spectrum may be predicted from the shock
compression factor and the post-shock radio spectrum.

M05 made such a prediction under the assumption that, as the plasma crosses
the shock surface, each volume element, with its frozen-in tangled magnetic
field, is compressed isotropically.  Indeed, observations at the Earth's bow
shock show that a shock passage strengthens the field whether it is parallel
or perpendicular to the shock (e.g., Wilkinson 2003), so this assumption is
adequate.  The average field strength $B$\/ then increases by a factor
\begin{equation}
B\propto r^{\,2/3}.
\label{eq:b}
\end{equation}
This increase would cause the particles to spin up as \mbox{$\gamma \propto
  B^{1/2}$} (the adiabatic invariant). In addition, the number density of
relativistic electrons increases by another factor of $r$\/ due to the
compression. As a result, for a power-law fossil electron spectrum of the
form $dN/d\gamma=N_0\, \gamma^{-\delta}$, the synchrotron brightness at a
given radio frequency should exhibit a jump at the shock
\begin{equation}
I_\nu \propto r\rlap{\phantom{I}}^{%
               \frac{\scriptstyle 2\delta}{\scriptstyle 3}+1}. 
\label{eq:icompr}
\end{equation}
The power-law slope of the spectrum is preserved. Thus, for the A520 shock
with $r=2.3$, if the radio edge is due to the compression only, there must
be a pre-shock radio emission with the same spectrum as post-shock, but
fainter by a factor of $7-20$ for $\delta=2-4$, respectively. (In practice,
$\delta$\/ can be determined from the post-shock radio spectrum; this
measurement is currently feasible but has not yet been made.)  Of course,
projection of a spheroidal shape of the shock should be taken into account,
as is done for deriving the amplitude of the X-ray brightness jump.

Although the low surface brightness within the radio halos is already near
the limit of detectability, an improvement of sensitivity by an order of
magnitude required to look for emission from pre-shock fossil electrons is
not completely out of reach.  If future sensitive measurements do not detect
such a pre-shock emission at the level predicted for simple compression, it
would mean that the shock generates relativistic electrons and/or a magnetic
field, as opposed to simply compressing them.  Observations of solar wind
shocks of a similar strength and $\beta>1$ as in cluster plasma, as well as
lower-$\beta$ shocks, seem to be consistent with a simple compression of the
field (e.g., Russell \& Greenstadt 1979; Hull et al.\ 2001), so significant
magnetic field generation is unlikely.  Thus, such a non-detection would
provide a lower limit on the shock's particle acceleration efficiency, a
quantity that is interesting for a wide range of astrophysical problems.

\begin{figure}[t]
\centering

\includegraphics[width=0.65\linewidth]%
{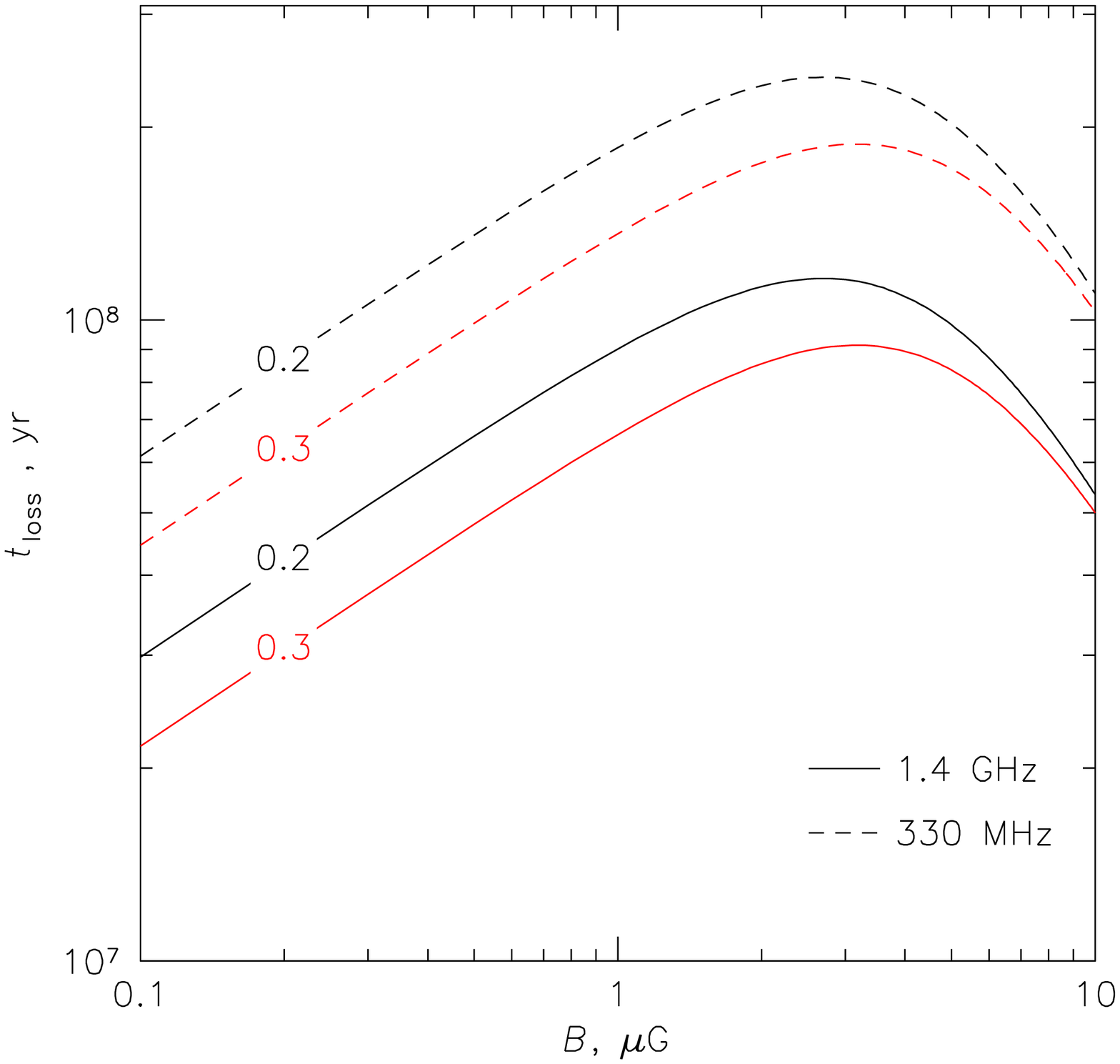}

\caption{Lifetime of the relativistic electrons  that contribute the most to
  the synchrotron emission at two frequencies, as a function of the magnetic
  field strength (for a $p=3-4$ electron energy spectrum). Two frequencies
  are shown as solid and dashed lines.  Black and red lines correspond to
  two different redshifts as labeled ($z=0.2$ for A520 and $z=0.3$ for \1e).
  This timescale can be determined from the width of the radio-bright strip
  at the shock, and gives an estimate of $B$.  (Reproduced from M05.)}

\label{a520_B}
\end{figure} 

\subsubsection{Yet another method to measure intracluster magnetic field}
\label{sec:b}

The strength of the magnetic field in the ICM is an important quantity, but
its measurements are still very limited in accuracy and often restricted to
certain non-representative locations in clusters (e.g., Carilli \& Taylor
2002).  The shock-related edge of the radio halo in A520, and perhaps a
similar region in \1e, may allow an independent estimate of the magnetic
field strengths behind the shock, regardless of the exact origin of the
relativistic electrons. As discussed above, IC and synchrotron cooling cause
the electron spectrum to steepen and the electrons to drop out of the radio
image on timescales of order $10^8$ yr.  This timescale depends on the field
strength as shown in Fig.\ \ref{a520_B}, which combines eqs.\ 
(\ref{eq:tic}--\ref{eq:gmax}). It gives the lifetime of the electrons that
contribute the most emission at a given frequency, for an electron energy
spectrum with the slope $p=3-4$ and an interesting range of $B$.  Assuming
that the bow shock causes a momentary increase in electron energy and $B$,
and that diffusion of the relativistic particles is negligible (see M05 for
a discussion), the flow of the post-shock gas (that carries the relativistic
electrons with it) will spread the time evolution of the electron spectrum
along the spatial coordinate --- something we have already used for another
test in \S\ref{sec:tei}.  Thus, the width of the cap-like region of the halo
can give us the magnetic field strength. Of course, the magnetic field
behind the shock will be amplified from its pre-shock value as discussed in
\S\ref{sec:compr}. This method can distinguish among the values in the
currently controversial range of $B\sim 0.1-3\,\mu G$, although, as seen
from Fig.\ \ref{a520_B}, it cannot give a unique value of $B$, because
$t_{\rm loss}$ is not a monotonic function of $B$.  In practice, such a
measurement will need to be done at more than one frequency, in order to
determine the spectrum of the electrons and to verify that they cool as
predicted at different frequencies, that is, no acceleration occurs after
the shock has passed.  The available single-frequency radio data (Fig.\ 
\ref{a520_halo}; Govoni et al.\ 2001) do not have the needed signal to noise
ratio or angular resolution, but are not inconsistent with $B\sim 1\,\mu$G.

\subsection{Constraints on the nature of dark matter}
\label{sec:dm}

\begin{figure}
\centering

\includegraphics[width=0.7\textwidth,angle=-90]%
{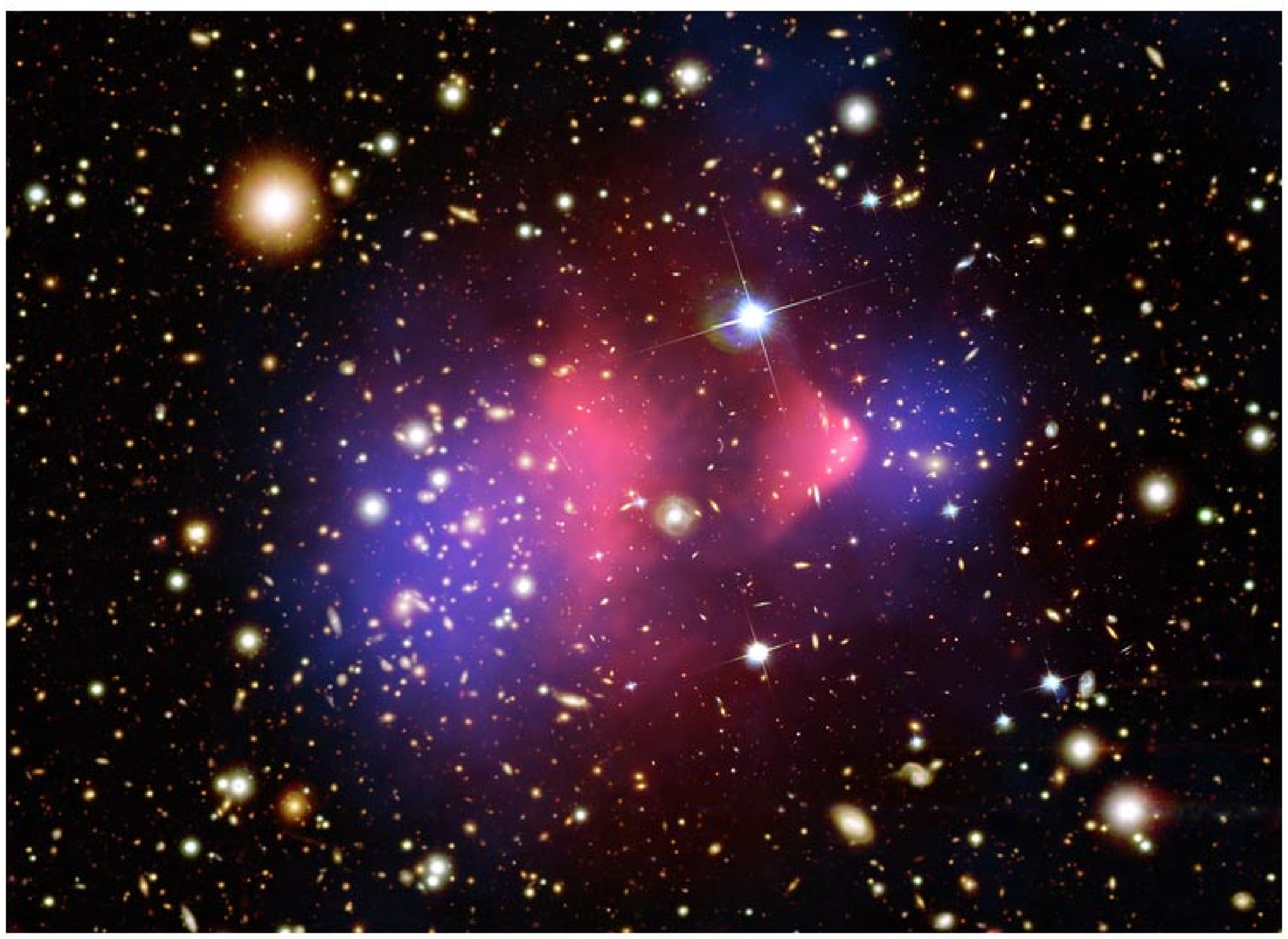}

\caption{\chandra\ X-ray image of \1e\ (pink, same as in Fig.\ \ref{1e}{\em
    c}) and the projected total mass map from weak lensing (blue; Clowe et
  al.\ 2006; same as in Fig.\ \ref{1e}{\em b}) overlaid on the optical
  image. The gas bullet lags behind its host dark matter subcluster, while
  the two mass peaks coincide with centroids of the respective galaxy
  concentrations. This overlay proves the existence of dark matter, as
  opposed to alternative gravity theories (Clowe et al.\ 2004, 2006). (Image
  created by the \chandra\ press group.)}

\label{1e_lens}
\end{figure}

We finally mention two interesting results that rely on the determination of
the velocity and the direction of motion of the subcluster in \1e\ based on
its shock front. The fact that we observe a cold front or a shock front
tells that the subcluster responsible for the feature moves very nearly in
the plane of the sky, because otherwise, the sharp gas density jump would be
smeared by projection. From the shapes of either feature, usually the
direction of motion in the sky plane is also clear, as seen in the images of
A3667 (Fig.\ \ref{2142_3667}), NGC\,1404 (Fig.\ \ref{n1404}), and \1e\ 
(Fig.\ \ref{1e}{\em c}). For the latter cluster, the shock and the cold
front (the bullet) in the X-ray image strongly suggest that the X-ray gas
distribution has axial symmetry, at least in its western half containing the
shock and the bullet.  This allows a reasonably accurate estimate of the gas
mass in that region. As expected from observations of other clusters,
invariably showing that gas is the dominant baryonic mass component, the gas
mass of the bullet is several times higher than the stellar mass in galaxies
in a comparable aperture centered on the brightest galaxy of that
subcluster.  At the same time, comparison of the gravitational lensing mass
map and the X-ray image (Fig.\ \ref{1e_lens}; Clowe et al.\ 2006; see also
Brada\v{c} et al.\ 2006) shows that the subcluster's total mass peak is
offset from the baryonic mass peak (the X-ray bullet). The same is true for
the bigger eastern subcluster. Clowe et al.\ (2004, 2006) interpret this as
the first direct evidence for the existence of dark matter, as opposed to
alternative theories that avoid the notion of dark matter by, e.g.,
modifying the gravitational force law on cluster scales (e.g., MOND: Milgrom
1983; Brownstein \& Moffat 2006).  If cluster had contained only visible
matter (i.e., mostly the X-ray emitting gas), regardless of the form of the
gravity law, the gravitational lensing would have to show a mass peak
coincident with the gas bullet (see Clowe et al.\ 2006 for a more detailed
discussion, and Angus et al.\ 2007 for a MOND response admitting the need
for dark matter).

Furthermore, the current orientation of the shock and the bullet let us
reconstruct the trajectory of the merging subclusters back in time at least
to the core passage. It is clear that the small halo has passed near the
center of the bigger halo. The shock front also gives an estimate of the
current velocity of the subcluster (\S\ref{sec:1e_M}).  Combined with the
mass measurements from lensing and the distribution of the member galaxies
from the optical images, this can be used to derive a limit on the
self-interaction cross-section of the dark matter particles. Although the
common wisdom is that dark matter is collisionless, a nonzero cross-section
was proposed as a solution for several observational problems (Spergel \&
Steinhardt 2000).  \1e\ offers an interesting opportunity to test this
hypothesis in a relatively robust and model-independent way (Markevitch et
al.\ 2004).  If the dark matter particles experienced scattering by elastic
collisions, it would have several observable consequences. At large
cross-sections, the dark matter would behave as a fluid --- just like the
intracluster gas --- and there would be no offset between the gas peak and
the dark matter peak.  This extreme is excluded simply by comparing the
X-ray and mass images (Fig.\ \ref{1e_lens}). Smaller values would lead to
subtler effects, such as an anomalously low mass-to-light ratio for the
subcluster caused by loss of dark matter particles by the subcluster, and a
lag of the subcluster mass peak from the centroid of the collisionless
galaxy distribution. The absence of these effects gives a conservative upper
limit on the cross-section of $\sigma/m<0.7$ cm$^{2}\,$g$^{-1}$, where $m$
is the (unknown) mass of the dark matter particle (Randall et al.\ 2007).
This excludes almost the entire range of values (0.5--5 cm$^{2}\,$g$^{-1}$)
required to solve the problems for which the self-interacting dark matter
has been invoked.

\section{SUMMARY}

Recent observations of sharp gas density edges in galaxy clusters provided
us with novel ways to study the cluster mergers, and even to perform remote
sensing of some otherwise inaccessible physical properties of the
intracluster plasma.  Classic bow shocks have been observed in two clusters
undergoing violent mergers at the right stage and with the right orientation
for us to see the front discontinuities unobstructed.  They have been used
to determine the velocities and trajectories of the infalling subclusters.

The velocity and the unique geometry of the subcluster in \1e, whose gas is
clearly stripped from its dark matter by ram pressure, were combined with
the gravitational lensing and optical data to place an upper limit on the
dark matter's self-interaction cross-section, and to directly demonstrate
the dark matter existence.

A jump of the electron temperature at the shock in \1e\ provides evidence
that electron-ion equilibration in the ICM occurs on a timescale shorter
than the collisional timescale. To our knowledge, this is the first such
test in any astrophysical plasma. Combined with finding based on the shocks
in solar wind and SNRs that electrons are not heated to equilibrium
temperature immediately in the shock, the \1e\ result indicates the presence
of a fast electron-ion equilibration mechanism unrelated to shocks (although
the uncertainty of the \1e\ result is quite large).

The shock front in A520 coincides with a distinct edge-like feature in the
radio halo observed in this cluster. This indicates that ultrarelativistic
electrons responsible for this part of the radio halo emission are either
accelerated by this weak shock, or that there is a pre-existing relativistic
population that is being compressed by the shock.  More sensitive radio
observations will be able to distinguish among these interesting
possibilities and constrain the efficiency of electron acceleration by
shocks.  The A520 shock also provides a setup for a measurement of the
strength of the magnetic field in this cluster by determining the cooling
time of the relativistic electrons.

In addition to observing shock fronts, \chandra\ has discovered a new kind
of transient features in the intracluster gas, named ``cold fronts.'' These
sharp density edges are contact discontinuities between gas phases with
different specific entropies in approximate pressure equilibrium. They form
as a result of bulk motion of a dense gas cloud through the hotter ICM.  In
merging clusters, these clouds are the surviving remnants of the cores of
the infalling subclusters. In cooling flow clusters, such edges are produced
by displacement and subsequent sloshing of the low-entropy central gas.
Cold fronts are much more common than shock fronts.  Just as shock fronts,
they can be used to determine the velocity and the direction of motion of
the gas flow.

Cold fronts are remarkably sharp and stable. The observed temperature jumps
across cold fronts require thermal conduction across the front to be
severely suppressed. Furthermore, the density jump in the best-observed
front in A3667 is narrower than the collisional mean free path in the
plasma. These observations demonstrate that transport processes in the ICM
can be easily suppressed.  The KH stability of the front in A3667 also
suggests the presence of a layer of the compressed magnetic field oriented
along the front surface, such as the one expected to form as a result of
magnetic draping.  Such a layer is exactly what is needed to completely
suppress thermal conduction and diffusion across the front surface.  In
addition, Rayleigh-Taylor stability of the front in A3667 reveals the
presence of an underlying massive dark matter subcluster.

This area of cluster research is relatively new (although much of the
underlying hydrodynamics is textbook). It is not yet completely clear what
these findings mean for the more general questions of how the energy is
dissipated in a cluster merger, how much of it goes into relativistic
particles and magnetic fields, how prevalent and long-lived is turbulence,
how effective is mixing of the various gas phases, what is the effect of the
central gas sloshing and turbulence on cooling flows and on the hydrostatic
mass estimates, how representative of the hydrostatic temperatures are the
average temperatures derived from the X-ray data, etc.  Our goal for this
mostly observational review was to encourage detailed theoretical and
numerical studies of these interesting phenomena that can be used as novel
diagnostic tools in clusters.

\ack{The phenomena discussed in this review owe their discovery to the
  sharpness of the \chandra\ X-ray mirror, whose development was led by the
  late Leon VanSpeybroeck. These subtle X-ray features in clusters may have
  gone unnoticed without the convenience of the software tool SAOimage
  written by M. VanHilst, and its later versions. We thank Y.~Ascasibar,
  P.~Blasi, G.~Brunetti, E.~Churazov, W.~Forman, L.~Hernquist, N.~Inogamov,
  C.~Jones, A.~Loeb, M.~Lyutikov, P.~Mazzotta, P.~Nulsen, A.~Petrukovich,
  C.~Sarazin, P.~Slane, and A.~Schekochikhin for stimulating discussions,
  and the referee, A. Kravtsov, for helpful comments. This work was
  supported by NASA contract NAS8-39073 and grants NAG5-9217, GO4-5152X and
  GO5-6121A.}



\end{document}